\documentclass[a4paper,12pt]{article}
\usepackage{hyperref}
\usepackage{amsmath}
\usepackage{amssymb}
\usepackage{graphicx, color}
\usepackage[dvipsnames]{xcolor}
\usepackage{physics}
\usepackage{MnSymbol}
\usepackage{float}
\usepackage{comment}
\usepackage{bbm}
\usepackage{ascmac}
\usepackage{cite}

\newlength{\abstractwidth}
\makeatletter
\g@addto@macro\normalsize{
  \setlength\abovedisplayskip{10pt}
  \setlength\belowdisplayskip{20pt}
  \setlength\abovedisplayshortskip{10pt}
  \setlength\belowdisplayshortskip{20pt}
}
\makeatother

\def\YY{\mathcal{Y}}
\def\OO{\mathcal{O}}
\def\KK{\mathcal{K}}
\def\NN{\mathcal{N}}
\def\EE{\mathcal{E}}
\def\PP{\mathcal{P}}
\def\QQ{\mathcal{Q}}
\def\RRR{\mathcal{R}}
\def\id{\mathbbm{1}}
\def\eq{\equiv}
\def\fft#1#2{_{\left [\substack{{#1}\\~\\ {#2}}\right ]}}
\newcommand{\be}{\begin{equation}}
\newcommand{\ee}{\end{equation}}
\newcommand{\ba}{\begin{align}}
\newcommand{\bml}{\begin{multline}}
\newcommand{\eml}{\end{multline}}
\allowdisplaybreaks

\hypersetup{
setpagesize=false,
colorlinks=true,
citecolor=OliveGreen,
linkcolor=Mulberry,
urlcolor=BrickRed,
}
\hypersetup{linktocpage}

\begin{document}
\renewcommand{\theequation}{\thesection.\arabic{equation}}
\renewcommand{\thefigure}{\arabic{figure}}
\begin{titlepage}
\rightline{}
\bigskip
\bigskip\bigskip\bigskip\bigskip
\bigskip
\centerline{\Large \bf {Conformal bootstrap in momentum space at finite volume}}
\bigskip

\bigskip
\begin{center}
\bf Kanade Nishikawa  \rm

\bigskip

\bigskip

\it{Kavli Institute for the Physics and Mathematics of the Universe (WPI), \\
the University of Tokyo
Kashiwa-no-ha, Chiba 277-8583, Japan}
\bigskip
\bigskip
\vspace{2cm}
\end{center}
\bigskip\bigskip
\bigskip\bigskip
\begin{abstract}

  In this paper, we Fourier transform the Wightman function concerning energy and angular momentum on the $S^{D-1}$ spatial slice in radial quantization
  in $D=2,3$ dimensions. In each case, we use the conformal Ward Identities to solve systematically for the Fourier components. We then use these
  Fourier components to build conformal blocks for the four-point function in momentum space, giving a finite-volume version of the momentum-space conformal blocks. We check that this construction is consistent with the known result in infinite volume. Our construction may help to find bootstrap equations that can give nontrivial constraints that do not appear in analysis in infinite volume. We show some examples of bootstrap equations and their nontriviality.
\medskip
\noindent
\end{abstract}
\end{titlepage}

\setcounter{tocdepth}{2}
\tableofcontents

\setcounter{equation}{0}

\section{Introduction}
The conformal bootstrap method was applied by the paper \cite{Belavin:1984vu} to solve an infinite class of two-dimensional conformal field theories (CFTs). Now, it is known as one of the most influential and fascinating tools to analyze CFT from the point of nonperturbative aspects of the theory. It carves out the space of consistent CFTs by imposing physical conditions such as symmetry, causality, and unitarity. It leads to nontrivial conditions that cannot be obtained from the algebraic structure of conformal algebra and perturbative analysis. About thirty years after their work  \cite{Belavin:1984vu}, the paper \cite{Rattazzi:2008pe} introduced the numerical method for conformal bootstrap equations to get constraints for higher-dimensional CFTs. Using crossing symmetry for the four-point function, they showed the upper bound on the conformal dimension of the first scalar appearing in the Operator Product Expansion (OPE) of two identical scalars. After their work, much progress has been made, such as fixing the critical exponents of the Ising model in three-dimensional CFT \cite{El-Showk:2012cjh,El-Showk:2014dwa,Kos:2013tga,Kos:2015mba,Kos:2014bka,Kos:2016ysd,Simmons-Duffin:2015qma} and getting bounds on the scaling dimensions of the theory with global $O(N)$ symmetry \cite{Kos:2013tga,Kos:2015mba,Kos:2016ysd,Nakayama:2014yia}. Please see the reviews \cite{Poland:2016chs,Poland:2018epd,Chester:2019wfx} to check the recent works in conformal bootstrap.\par
As numerical bootstrap methods developed, there was a growing body of research investigating the analytical properties of bootstrap. One of the current research directions on conformal bootstrap is large N expansion. This analysis method is robust in AdS/CFT correspondence \cite{Maldacena:1997re,Gubser:1998bc,Witten:1998qj,Heemskerk:2009pn}, in which we have a good correspondence between weakly coupled gravity in AdS and its dual CFTs. “Weakly coupled” translates to the CFT with large N, and in addition, we demand that CFT on the boundary is strongly coupled to make the gravity theories on the bulk without light particles of spin greater than two. In the analysis, we perturb the CFT by $1/N$ and $1/\Delta$, where $\Delta$ is the conformal dimension of the lightest operator in the CFT. Many researchers have studied the relationship between weakly coupled bulk theory and CFT data on its boundary \cite{Aharony:2016dwx,Aprile:2017bgs,Alday:2017vkk,Alday:2017xua,Rastelli:2016nze,Rastelli:2017udc,Caron-Huot:2018kta}. The conformal bootstrap helps obtain consistent theories of gravity in AdS from the effective field theories, the swampland program \cite{Vafa:2005ui,Arkani-Hamed:2006emk,Palti:2019pca}. In AdS/CFT correspondence, the AdS cylinder in global coordinates corresponds to the boundary CFT in radial quantization. With these physical motivations, formulating conformal bootstrap at finite volume on the cylinder is naturally needed to understand the bulk theory.\par
Recently, CFT in momentum space has been developed formally in \cite{Gillioz:2019lgs,Coriano:2013jba,Bzowski:2013sza,Isono:2019ihz,Bautista:2019qxj}. Those papers define correlation functions in momentum space as a Fourier transform of that in position space. CFT in momentum space has physical applications such as the study of anomalies \cite{Giannotti:2008cv,Mottola:2010gp,Armillis:2009im,Armillis:2009pq,Armillis:2010qk,Coriano:2011ti,Coriano:2011zk}, the determination of the form of conformal invariance in the non-Gaussian features of the cosmic microwave background \cite{Antoniadis:1996dj,Antoniadis:2011ib}, and an investigation for inflation \cite{Maldacena:2011nz,Kehagias:2012pd}.\par
One of the difficulties of analysis in momentum space is that we cannot expect the time-ordered correlation function to behave well because the integral calculation in the Fourier transform involves the position where the operators are not time-ordered. Instead, the Wightman function is a good function for the analysis in momentum space. The Wightman function has an operator algebra, but we have yet to study its structure so much. The conclusion and discussion summarize the future direction that surveys algebraic construction in momentum space.\par
When studying CFT in radial quantization, we can expect that the analysis in momentum space is the most suitable because the feature of CFT in radial quantization is that the energy and momentum are quantized. So, studying CFT in momentum space is natural if we try to find the difference between CFT in $\mathbb{R}^{1,d-1}$, and CFT in $\mathbb{R}\times S^{d-1}$.\par
From those considerations, we formulate the basis for the conformal bootstrap in momentum space using the Wightman function at finite volume in this paper. We expect that the conformal bootstrap at finite volume gives constraints for CFT data that cannot be obtained in CFT at infinite volume. In this paper, we show that the two- and three-point function at finite volume leads to that at infinite volume under the “Large volume limit.” It implies that the information in conformal bootstrap at finite volume is richer than that at infinite volume. As an analysis method that is particular for CFT at finite volume, we consider expansion by $1/R$, where $R$ is a compactification radius. We leave this kind of analysis to future work.\par
This paper is organized as follows. We summarize the result in two-dimensional CFT in section 2. We compute two- and three-point functions in momentum space with Ward Identities (WIs). This method is powerful because it applies to general operators and is helpful for computational calculation. In section 3, we study three-dimensional CFT by using WIs. Though we cannot get a closed form for the three-point function, ODEs obtained from WIs may be helpful for calculation with a computer. Finally, we show one of the applications of our calculation result. We develop bootstrap equations by using improved microcausality conditions. It was designed in infinite volume in the paper \cite{Gillioz:2019iye}. We resolve some problems and get a proper bootstrap equation in finite volume. Though our development is incomplete, it implies that the bootstrap method in momentum space might be helpful. We leave the problem for future work.

\section{Main results in \texorpdfstring{$D=2$}{TEXT}}
\setcounter{equation}{0}
This section summarizes the derivation of two- and three-point Wightman functions in two-dimensional momentum space. We deal with CFT quantized on a cylinder.\par
For two-dimensional CFT, the factorization method is potent as we can independently calculate holomorphic and antiholomorphic parts. After that, we can construct complete correlation functions for general operators.\par
There are three ways to get them. The first is a direct integral calculation, which is easy for a simple case, but an analytic continuation for general situations is not trivial. \par
On the other hand, the Ward Identity (WI) method is helpful for the general case, which is the second procedure. Though solving the ODEs obtained from WI is a little tricky for three-point functions, this method emphasizes that we can determine all of the correlation functions of the mode operator from ones including only primaries.\par
The third procedure is an algebraic construction, which helps construct a completeness relation when calculating a four-point function.\par
The most crucial difference between two-dimensional CFT and higher-dimensional CFT is that there are Virasoro descendants in the two-dimensional CFT. In this paper, we call the descendants made by acting $L_{-1}$s on primaries as “descendants” and the descendants made by working $L_{-n}\ (n\geq2)$s on primaries as “Virasoro descendants.” When considering four-point functions in momentum space, we do not have to consider descendants, but we cannot ignore the contributions from Virasoro descendants. We will see the calculation in the latter chapter. We summarize the results briefly, so please visit the appendix for a more detailed analysis and supplements.
\subsection{Conformal generators and their action}
We use cylinder coordinates and complex plane coordinates. In the complex plane, we use $z$ and $\bar{z}$ as coordinates, and $\omega=\sigma_1+i\sigma_2=\sigma-i\tau$ and $\bar{\omega}=\sigma_1-i\sigma_2=\sigma+i\tau$ for the cylinder frame. They are related to each other by conformal transformations with radius $R$.
\begin{equation}
z=re^{i\theta}=Re^{\frac{i\omega}{R}}=Re^{\frac{i\sigma+\tau}{R}},\hspace{1cm}\bar{z}=re^{-i\theta}=Re^{-\frac{i\bar{\omega}}{R}}=Re^{\frac{-i\sigma+\tau}{R}}
\end{equation}
We define two “Spatially-integrated mode” types for primary operators. The first one is
\begin{equation}
\tilde{\OO}_n(r)\eq\int_{C_r}\frac{\dd{\theta}}{2\pi}\OO(r,\theta)e^{-in\theta},
\end{equation}
where the integral path $C_r$ is the circumference of radius $r$. And the second one is defined for holomorphic operators $\OO(z)$ and antiholomorphic operators $\bar{\OO}(\bar{z})$, respectively.
\begin{equation}
\OO_n\eq\frac{1}{2\pi i}\oint\frac{\dd{z}}{z^{1+n}}\OO(z),\hspace{1cm}\bar{\OO}_n\eq\frac{1}{2\pi i}\oint\frac{\dd{\bar{z}}}{\bar{z}^{1+n}}\bar{\OO}(\bar{z})
\end{equation}
For example, for holomorphic operators, $\tilde{\OO}_n(r)=r^n\OO_n$. For operators with spin $s$, we also use other types of representation. Define $J\eq n+s$ and $\bar{J}\eq-n-s$, then
\begin{equation}
\OO_{[J]}\eq\OO_n,\hspace{1cm}\bar{\OO}_{[\bar{J}]}\eq\bar{\OO}_n.
\end{equation}
We obtain the action of the conformal generators on these modes. For example,
\begin{equation}
P_z\cdot\tilde{\OO}_{n}=\frac{1}{2}\int\frac{\dd{\theta}}{2\pi}\left(\partial_r\OO-\frac{i}{r}\partial_{\theta}\OO\right)e^{-i(n+1)\theta}=\frac{1}{2}\left(\partial_r+\frac{n+1}{r}\right)\tilde{\OO}_{n+1}=\frac{1}{2r}\left(\hat{E}+n+1\right)\tilde{\OO}_{n+1},
\end{equation}
where $\hat{E}=r\partial_r$. Similarly, we get the following equations.
\begin{align}
P_{\bar{z}}\cdot\tilde{\OO}_n&=\frac{1}{2}\left(\partial_r-\frac{n-1}{r}\right)\tilde{\OO}_{n-1}=\frac{1}{2r}\left(\hat{E}-n+1\right)\tilde{\OO}_{n-1}\\
K^z\cdot\tilde{\OO}_n&=\frac{r^2}{2}\left(\partial_r+\frac{n-1}{r}\right)\tilde{\OO}_{n-1}+2hr\tilde{\OO}_{n-1}=\frac{r}{2}\left(\hat{E}+n+4h-1\right)\tilde{\OO}_{n-1}\\
K^{\bar{z}}\cdot\tilde{\OO}_n&=\frac{r^2}{2}\left(\partial_r-\frac{n+1}{r}\right)\tilde{\OO}_{n+1}+2\tilde{h}r\tilde{\OO}_{n+1}=\frac{r}{2}\left(\hat{E}-n+4\tilde{h}-1\right)\tilde{\OO}_{n+1}\\
L_0\cdot\tilde{\OO}_n&=\frac{1}{2}\left(r\partial_r+n+2h\right)\tilde{\OO}_n=\frac{1}{2}\left(\hat{E}+n+2h\right)\tilde{\OO}_{n}\\
\tilde{L}_0\cdot\tilde{\OO}_n&=\frac{1}{2}\left(r\partial_r-n+2\tilde{h}\right)\tilde{\OO}_n=\frac{1}{2}\left(\hat{E}-n+2\tilde{h}\right)\tilde{\OO}_n
\end{align}

\subsection{Two-point function}
\subsubsection{Two-point function for primaries}
Define the complete Wightman two-point function of primaries as
\begin{equation}
C(n_1,n_2,r_1,r_2)\eq\langle\tilde{\OO}_{n_2}^{(2)}(r_2)\tilde{\OO}_{n_1}^{(1)}(r_1)\rangle\eq\delta(n_1+n_2+s_1+s_2)r_1^{-\Delta_1}r_2^{-\Delta_2}F_{n_1}(y),
\end{equation}
where $y\eq r_1/r_2$ is a ratio of radius.
We used $L_0$ and $\tilde{L}_0$ WIs to get the reduced form.
Usually, the above delta function means Dirac's delta function, but in this paper, it often means $0$ or $1$ function when its content has a discrete value.
\begin{equation}
  \delta(x)\eq
  \begin{cases}
    1& \text{if $x=0$}\\
    0& \text{otherwise}
  \end{cases}
\end{equation}
For this reduced form, we get WIs, which give ODEs for a two-point function. Four are left, as we used two of six WIs to get the reduced form.
\begin{align}
L_{-1}&:\left(y\dv{y}-\Delta_1+n+1\right)F_{n+1}(y)+y\left(-y\dv{y}-\Delta_2-n-s_1-s_2\right)F_{n}(y)=0\label{Ward Id2-1}\\
  \tilde{L}_{-1}&:y\left(-y\dv{y}-\Delta_2+n+1+s_1+s_2\right)F_{n+1}(y)+\left(y\dv{y}-\Delta_1-n\right)F_n(y)=0\label{Ward Id2-2}\\
  L_1&:\left(-y\dv{y}+\Delta_2-n-1-s_1+s_2\right)F_{n+1}(y)+y\left(y\dv{y}+\Delta_1+2s_1+n\right)F_n(y)=0\label{Ward Id2-3}\\
  \tilde{L}_1&:y\left(y\dv{y}+\Delta_1-2s_1-n-1\right)F_{n+1}(y)+\left(-y\dv{y}+\Delta_2+n+s_1-s_2\right)F_n(y)=0\label{Ward Id2-4}
\end{align}
The solution of these ODEs is as follows.
\begin{align}
G_{n+s}&\eq F_n=K_sy^{\Delta+|n|}(\sum_{l=0}c_{n+s|l}y^{2l})\\
c_{n+s|l}&=\frac{\Gamma(\Delta+s+l+n)\Gamma(\Delta-s+l)}{\Gamma(\Delta+s)\Gamma(\Delta-s)\Gamma(l+n+1)\Gamma(l+1)}
\end{align}
$K_s$ is an undetermined constant depending on the normalization of operators. The appendix summarizes how to solve these WIs and checks that this solution matches that obtained by a direct integral.\par
In the two-point function, $\Delta+s=h$ and $\Delta-s=\tilde{h}$ appear independently. In other words, we can factorize it into holomorphic and antiholomorphic parts. It is natural, but we summarize it to clarify the situation. Let us perform the Fourier transform of the two-point function of scalar operators with the above solution.
\begin{equation}
  \int\frac{\dd\theta_1}{2\pi}\frac{\dd\theta_2}{2\pi}e^{-in_1\theta_1-in_2\theta_2}\langle\OO^{(2)}\OO^{(1)}\rangle=\delta(n_1+n_2)r_2^{-2\Delta}\sum_{l=0}\frac{\Gamma(\Delta+l+n_1)\Gamma(\Delta+l)}{\Gamma(l+n_1+1)\Gamma(l+1)\Gamma(\Delta)^2}\left(\frac{r_1}{r_2}\right)^{2l+n_1}
\end{equation}
The operator normalisation has been chosen to set $K_s$ to unity.  The coordinate transformation from the complex plane to the cylinder, inverse Wick rotation, and complete Fourier transform give
\begin{multline}
 \int\frac{\dd\sigma_1}{2\pi R}\int\frac{\dd\sigma_2}{2\pi R}\int\frac{\dd t_1}{2\pi}\int\frac{\dd t_2}{2\pi}e^{-iE_1t_1}e^{-iE_2t_2}e^{-i\sigma_1\frac{n_1}{R}}e^{-i\sigma_2\frac{n_2}{R}}\langle\OO^{(2)}(\sigma_2,\tau_2)\OO^{(1)}(\sigma_1,\tau_1)\rangle\\
 =\delta(k_1+k_2)\delta(E_1+E_2)\frac{R^{-2\Delta}}{(\Gamma(\Delta))^2}\frac{\Gamma\left(\frac{R(E_1-k_1)+\Delta}{2}\right)\Gamma\left(\frac{R(E_1+k_1)+\Delta}{2}\right)}{\Gamma\left(\frac{R(E_1-k_1)-\Delta}{2}+1\right)\Gamma\left(\frac{R(E_1+k_1)-\Delta}{2}+1\right)}. \label{full}
\end{multline}
The point is that we can factorize the two-point function in momentum space into $E_1\pm k_1$ parts. It indicates that we can calculate it by multiplying the results in holomorphic and antiholomorphic parts, but what is the procedure? For holomorphic operators,
\begin{equation}
\int\frac{d\theta_1}{2\pi}\frac{d\theta_2}{2\pi}e^{-in_1\theta_1-in_2\theta_2}\langle\OO^{(2)}\OO^{(1)}\rangle\propto r_1^{-h}r_2^{-h}\delta(n_1+n_2+2h)\frac{\Gamma(n_1+2h)}{\Gamma(n_1+1)\Gamma(2h)}\left(\frac{r_1}{r_2}\right)^{n_1+h}.\label{holo}
\end{equation}
And for antiholomorphic operators,
\begin{equation}
  \int\frac{d\theta_1}{2\pi}\frac{d\theta_2}{2\pi}e^{i\bar{n}_1\theta_1+i\bar{n}_2\theta_2}\langle\mathcal{\bar{O}}^{(2)}\mathcal{\bar{O}}^{(1)}\rangle\propto r_1^{-\tilde{h}}r_2^{-\tilde{h}}\delta(\bar{n}_1+\bar{n}_2+2\tilde{h})\frac{\Gamma(\bar{n}_1+2\tilde{h})}{\Gamma(\bar{n}_1+1)\Gamma(2\tilde{h})}\left(\frac{r_1}{r_2}\right)^{\bar{n}_1+\tilde{h}}.\label{antiholo}
\end{equation}
Multiplying them doesn't give (\ref{full}). The physical states created by $\tilde{O}_{n_1}$ are ones with spin $n_1$, so they have the form $(P_z)^{n_1+l_1}(P_{\bar{z}})^{l_1}\ket{\Phi}$ with some integer $l_1$, where $\ket{\Phi}$ is a primary state created by inserting primary operator with conformal weight $(h,\tilde{h})$.\par
For this state, $n_1$ in (\ref{holo}) should be replaced by $n_1+l_1$, and $\bar{n}_1$ in (\ref{antiholo}) should be replaced by $l_1$. From here, we only concentrate on the Gamma function dependent part in the two-point function since the delta functional part and $r_1,r_2$ dependent part only give conservation law.\par
The multiplication of the holomorphic part and the antiholomorphic part gives
\begin{equation}
\frac{\Gamma(n_1+2h+l_1)}{\Gamma(n_1+l_1+1)\Gamma(2h)}\frac{\Gamma(l_1+2h)}{\Gamma(l_1+1)\Gamma(2h)}=\frac{\Gamma(\Delta+n_1+l_1)}{\Gamma(n_1+l_1+1)\Gamma(\Delta)}\frac{\Gamma(\Delta+l_1)}{\Gamma(l_1+1)\Gamma(\Delta)}.
\end{equation}
Energy for the state $(P_z)^{n_1+l_1}(P_{\bar{z}})^{l_1}\ket{\Phi}$ is $E_1=(h+\tilde{h}+n_1+l_1+l_1)/R=(\Delta+n_1+2l_1)/R$ and momentum for the state is $k_1=n_1/R$. So, the Gamma function dependent part is
\begin{equation}
  \frac{1}{(\Gamma(\Delta))^2}\frac{\Gamma\left(\frac{R(E_1-k_1)+\Delta}{2}\right)\Gamma\left(\frac{R(E_1+k_1)+\Delta}{2}\right)}{\Gamma\left(\frac{R(E_1-k_1)-\Delta}{2}+1\right)\Gamma\left(\frac{R(E_1+k_1)-\Delta}{2}+1\right)}.\label{two-point-func in two-dim CFT}
\end{equation}
It is the same as (\ref{full}).\par
In the above example, we calculated the two-point function of scalar operators: direct integral calculation and factorization. Their results are the same, and we can generalize this method for other two-point functions. We will show more clearly that we can use the factorization method to calculate three-point functions later, and the logic is the same.

\subsubsection{Two-point function for descendants}
As well known, inserting descendant fields in the correlation function means acting differential operators on others.
\begin{equation}
  \langle(L_{-m}\phi)(z)\phi_1(z_1)\cdots\phi_N(z_N)\rangle=\sum_{i=1}^N\left[\frac{(m-1)h_i}{(z_i-z)^m}-\frac{1}{(z_i-z)^{m-1}}\pdv{z_i}\right]\langle\phi(z)\phi_1(z_1)\cdots\phi_N(z_N)\rangle
\end{equation}
So, for example,
\begin{align}
  \langle\phi_{[-J]}(L_{-m}\phi)_{[J]}\rangle&=\oint\frac{dz_2}{z_2^{1-J-h}}\oint\frac{dz_1}{z_1^{1+J-h-m}}\left[\frac{(m-1)h}{(z_2-z_1)^m}-\frac{1}{(z_2-z_1)^{m-1}}\pdv{z_2}\right]\langle\phi(z_2)\phi(z_1)\rangle\notag\\
  &=(m+1)h\frac{\Gamma(J+h)}{\Gamma(1+J-h-m)\Gamma(2h+m)}.
\end{align}
In general,
\begin{align}
\langle\phi_{[-J]}(L_{-m_k}\cdots L_{-m_2}L_{-m_1}\phi)_{[J]}\rangle=\frac{\prod_{i=1}^k\left[(m_i+1)h+\sum_{j=1}^{i-1}m_j\right]\Gamma(J+h)}{\Gamma(1+J-h-\sum_{i=1}^km_i)\Gamma(2h+\sum_{i=1}^km_i)}.
\end{align}
For more general calculations such as $\langle (L_{-2}\phi)_{[-J]}(L_{-3}\phi)_{[J]}\rangle$, we also need to consider more complicated calculation including OPEs between energy-momentum tensors. Although one can compute an arbitrary correlation function in momentum space by a straightforward analysis, we can understand it more simply by using the algebraic method described next.
\subsubsection{Algebraic construction}
Like CFT in position space, we can calculate the two-point function by algebraic calculation. In the procedure, we consider in-state and out-state naturally, which is useful when constructing the completeness relation to calculate a four-point function.\par
As we can calculate the holomorphic and antiholomorphic parts independently, we only focus on the holomorphic part. First, the ket $\ket{\phi_{[J]}}$ is defined as
\begin{equation}
  \ket{\phi_{[J]}}=\oint\frac{dz}{z^{1+J-h}}\phi(z)\ket{0}=\oint\frac{dz}{z^{1+J-h}}e^{iL_{-1}z}\phi(0)e^{-iL_{-1}z}\ket{0}=\frac{i^{J-h}}{(J-h)!}L_{-1}^{J-h}\ket{h}.
\end{equation}
The ket for descendant field $(L_{-m_k}\cdots L_{-m_2}L_{-m_1}\phi)(z)$ is
\begin{equation}
  \ket{L_{-m_k}\cdots L_{-m_2}L_{-m_1}\phi_{[J]}}=\frac{i^{J-h}}{(J-h)!}L_{-1}^{J-h}L_{-m_k}\cdots L_{-m_2}L_{-m_1}\ket{h}.
\end{equation}
In this paper, we arrange the Virasoro generators so that $m_1\geq m_2\geq\cdots\geq m_k$ is satisfied.\par
We define the bra for the primary field and for the descendant field as
\begin{align}
  &\bra{\phi_{[J]}}=\oint\frac{dz}{z^{1-J-h}}\frac{1}{z^{2h}}\bra{0}\phi(\infty)e^{-iL_1\frac{1}{z}}=\frac{(-i)^{J-h}}{(J-h)!}\bra{h}L_{1}^{J-h}\\
  &\bra{L_{-m_k}\cdots L_{-m_2}L_{-m_1}\phi_{[J]}}=\frac{(-i)^{J-h}}{(J-h)!}\bra{h}L_{m_1}L_{m_2}\cdots L_{m_k}L_{1}^{J-h}.
\end{align}
The phases are chosen so that $\bra{\phi_{[J]}}\ket{\phi_{[J]}}\geq0$ and
$\bra{L_{-m_k}\cdots L_{-m_1}\phi_{[J]}}\ket{L_{-m_k}\cdots L_{-m_1}\phi_{[J]}}\geq0$ are satisfied. With those definitions, we can recover the previous result, for example,
\begin{equation}
  \langle\phi_{[-h-2]}\phi_{[h+2]}\rangle=\frac{\Gamma(2h+2)}{\Gamma(3)\Gamma(2h)}=\bra{\phi_{[h+2]}}\ket{\phi_{[h+2]}}=\frac{1}{2!}\frac{1}{2!}\bra{h}L_1^2L_{-1}^2\ket{h}=h(2h+1).
\end{equation}
This algebraic construction helps make a completeness relation needed to calculate the four-point function by conformal block decomposition. We will deal with it later.
\subsubsection{Large volume limit for two-point function}
In Gillioz's paper \cite{Gillioz:2021sce}, they calculated the two-point function of scalars in momentum space in $d$-dimensional Minkowski space. They defined the double bracket notation as
\begin{equation}
  \bra{0}\phi_1(k_1)\cdots\phi_n(k_n)\ket{0}=(2\pi)^d\delta^d(k_1+\cdots+k_n)\llangle\phi_1(k_1)\cdots\phi_n(k_n)\rrangle.
\end{equation}
Here, Fourier transform of operator $\tilde{\phi}(x)$ is $\phi(k)\eq\int d^dxe^{ik\cdot x}\tilde{\phi}(x)$. So,
\begin{equation}
 \llangle\phi_1(k_1)\cdots\phi_n(k_n)\rrangle=\int d^dx_1\cdots d^dx_{n-1}e^{i(k_1\cdot x_1+\cdots+k_{n-1}\cdot x_{n-1})}\bra{0}\tilde{\phi}_1(x_1)\cdots\tilde{\phi}_n(0)\ket{0}.
\end{equation}
The two-point function of the scalar operators for momentum $k$ lying in a forward light cone is
\begin{equation}
  \llangle\phi(-k)\phi(k)\rrangle=\frac{(4\pi)^{d/2+1}}{2^{2\Delta+1}\Gamma(\Delta)\Gamma(\Delta-\frac{d}{2}+1)}(k^2)^{\Delta-d/2}.\label{Gillioz's result2}
\end{equation}
In two-dimensional spacetime,
\begin{equation}
  \langle\phi_1(k_1)\phi_2(k_2)\rangle=(2\pi)^2\delta^2(k_1+k_2)\frac{(4\pi)^2}{2^{2\Delta+1}(\Gamma(\Delta))^2}(k^2)^{\Delta-1}.\label{Gillioz's result}
\end{equation}
We want this result by taking the limit of $R\rightarrow\infty$ for our result in finite volume. We call it ``Large Volume Limit.'' Our result is
\begin{multline}
  \tilde{C}(k_1,k_2,E_1,E_2)\eq\int\frac{dt_1}{2\pi}\frac{dt_2}{2\pi}\int\frac{d\sigma_1}{2\pi R}\frac{d\sigma_2}{2\pi R}e^{i\sigma_1k_1}e^{i\sigma_2k_2}e^{-it_1E_1}e^{-it_2E_2}\langle\phi_2(t_2,\sigma_2)\phi_1(t_1,\sigma_1)\rangle\\
  =\delta(k_1+k_2)\delta(E_1+E_2)\frac{R^{-2\Delta}}{(\Gamma(\Delta))^2}\frac{\Gamma\left(\frac{R(E_1-k_1)+\Delta}{2}\right)\Gamma\left(\frac{R(E_1+k_1)+\Delta}{2}\right)}{\Gamma\left(\frac{R(E_1-k_1)-\Delta}{2}+1\right)\Gamma\left(\frac{R(E_1+k_1)-\Delta}{2}+1\right)}.
\end{multline}
With Stirling's approximation, we get
\begin{equation}
  \lim_{R\rightarrow\infty}\text{(LHS)}\propto\delta(k_1+k_2)^2\frac{R^{-2\Delta}}{\Gamma(\Delta)^2}\frac{(k_1-E_1)^{\Delta-1}R^{\Delta}}{2^{\Delta-1}}\frac{(k_1+E_1)^{\Delta-1}R^{\Delta}}{2^{\Delta-1}}\propto\delta(k_1+k_2)^2\frac{(k^2)^{\Delta-1}}{(\Gamma(\Delta))^2}.
\end{equation}
This is identical to the result (\ref{Gillioz's result}) at infinite volume calculated in the paper \cite{Gillioz:2021sce}.
\subsection{Three-point function}
Solving WIs for the three-point function is complicated. So first, we Fourier transform the three-point function of holomorphic operators directly and check that the solution satisfies the ODEs obtained from WIs. The three-point function of holomorphic operators is
\begin{equation}
  \langle\OO_3(z_3)\OO_2(z_2)\OO_1(z_1)\rangle\eq\lambda_{321}z_{21}^{h_3-h_1-h_2}z_{32}^{h_1-h_3-h_2}z_{31}^{h_2-h_1-h_3}\eq \lambda_{321}z_{21}^{-b_3}z_{32}^{-b_1}z_{31}^{-b_2}.
\end{equation}
Here, $h_1,h_2,h_3$ are conformal weight of operator $\OO_1,\OO_2,\OO_3$. And we define $z_{ij}\eq z_i-z_j$, $b_1\eq h_2+h_3-h_1$, $b_2\eq h_1+h_3-h_2$ and $b_3\eq h_1+h_2-h_3$. We assume that conformal weights are all integers, but we can analytically continue the region of conformal weights by taking $n_i\ (i=1,2,3)$ non-integers. Then,
\begin{multline}
\langle\OO_{n_3}\OO_{n_2}\OO_{n_1}\rangle=\frac{1}{(2\pi i)^3}\oint\frac{dz_3}{z_3^{1+n_3}}\oint\frac{dz_2}{z_2^{1+n_2}}\oint\frac{dz_1}{z_1^{1+n_1}}\langle\OO_{3}(z_3)\OO_{2}(z_2)\OO_1(z_1)\rangle\\
=\tfrac{\lambda_{321}\delta(\sum_{i=1}^3J_i)}{\Gamma(b_1)\Gamma(b_2)\Gamma(b_3)}\sum_{q=0}^{\mathrm{Min}\{n_1,m_3\}}\tfrac{\Gamma[b_3+q+\mathrm{Max}\{{0,n_1-m_3}\}]\Gamma[b_2-q+\mathrm{Min}\{{n_1,m_3}\}]\Gamma[b_1+q+\mathrm{Max}\{{m_3-n_1,0}\}]}{\Gamma[1+q+\mathrm{Max}\{{0,n_1-m_3}\}]\Gamma[1-q+\mathrm{Min}\{{n_1,m_3}\}]\Gamma[1+q+\mathrm{Max}\{{m_3-n_1,0}\}]},
\end{multline}
where $m_3=-(n_3+2h_3)$. In the appendix, we summarize the detail of the derivation and check that the solution for the three-point function satisfies the ODEs obtained from WIs.\par
In the paper \cite{Gillioz:2019lgs} by Gillioz, they calculated the three-point function in infinite volume in the fully factorized form
\begin{equation}
\llangle\phi_f(p_f)\phi_0(p_0)\phi_i(p_i)\rrangle=\lambda_{f0i}W(p_f^+,p_0^+,p_i^+)W(p_f^-,p_0^-,p_i^-),
\end{equation}
where $p^\pm=p_0\pm p_1$ and
\begin{multline}
  W(p_f^\pm,p_0^\pm,p_i^\pm)=\frac{(2\pi)^2}{2^{(\Delta_f+\Delta_0+\Delta_i-2)/2}}\frac{(p_f^\pm)^{\Delta_f-d/2}(p_i^\pm)^{\Delta_i-d/2}}{|p_0^\pm|^{(\Delta_f-\Delta_0+\Delta_i)/2}}\\
  \left[\frac{\Theta(p_0^\pm)}{\Gamma(\Delta_i)\Gamma\left(\frac{\Delta_f+\Delta_0-\Delta_i}{2}\right)}{}_2F_1\left(\frac{\Delta_f-\Delta_0+\Delta_i}{2},\frac{\Delta_f-\tilde{\Delta}_0+\Delta_i}{2};\Delta_i;-\frac{p^\pm_i}{p^\pm_0}\right)\right.\\
  \left.+\frac{\Theta(-p_0^\pm)}{\Gamma(\Delta_f)\Gamma\left(\frac{\Delta_i+\Delta_0-\Delta_f}{2}\right)}{}_2F_1\left(\frac{\Delta_f-\Delta_0+\Delta_i}{2},\frac{\Delta_f-\tilde{\Delta}_0+\Delta_i}{2};\Delta_f;-\frac{p^\pm_f}{p^\pm_0}\right)\right].
\end{multline}
Here, ${}_2F_1$ is a hypergeometric function
\begin{equation}
  {}_2F_1(a,b:c:z)=\sum_{n=0}^{\infty}\frac{(a)_n(b)_n}{(c)_n}\frac{z^n}{n!},
\end{equation}
where $(a)_n$ is the Pochhammer symbol defined as follows.
\begin{equation}
  (a)_n=\begin{cases}
  1 & (n=0) \\
  a(a+1)(a+2)\cdots(a+n-1) & (n>0)
\end{cases}
\end{equation}
We can see that the Wightman 3-point function is factorized into holomorphic and anti-holomorphic pieces, or equivalently into left- and right-movers. \par
We want to get this result by taking a large volume limit for our result at a finite volume. We use a time-reversal symmetric form for three-point functions. $\KK_{\mathrm{red}}$ is defined as
\begin{equation}
  \KK(n_1,n_2,n_3)\eq\langle\OO_{n_3}(z_3)\OO_{n_2}(z_2)\OO_{n_1}(z_1)\rangle\eq\delta(n_1+n_2+n_3+h_1+h_2+h_3)\KK_{\mathrm{red}}(n_1,n_3).
\end{equation}
And remember that this is equal to
\begin{multline}
  \KK_{\mathrm{red}}(n_1,n_3)\\
  =\tfrac{\lambda_{321}}{\Gamma(b_1)\Gamma(b_2)\Gamma(b_3)}\sum_{q=0}^{\mathrm{Min}\{n_1,m_3\}}\tfrac{\Gamma[b_3+q+\mathrm{Max}\{{0,n_1-m_3}\}]\Gamma[b_2-q+\mathrm{Min}\{{n_1,m_3}\}]\Gamma[b_1+q+\mathrm{Max}\{{m_3-n_1,0}\}]}{\Gamma[1+q+\mathrm{Max}\{{0,n_1-m_3}\}]\Gamma[1-q+\mathrm{Min}\{{n_1,m_3}\}]\Gamma[1+q+\mathrm{Max}\{{m_3-n_1,0}\}]}.
\end{multline}
To take a large volume limit, define $\NN[x,y]\eq\Gamma(1+x+y)/\Gamma(1+x)\Gamma(1+y)$. Under the $y\rightarrow\infty$ limit, $\NN[x,y]\simeq y^x/\Gamma(1+x)$. With it, we can write a three-point function as
\begin{multline}
\KK_{\mathrm{red}}(n_1,n_3)=\lambda_{321}\sum_{q=0}^{\mathrm{Min}\{n_1,m_3\}}\NN[b_3-1,q+\mathrm{Max}\{{0,n_1-m_3}\}]\\
\NN[b_2-1,-q+\mathrm{Min}\{n_1,m_3\}]\NN[b_1-1,q+\mathrm{Max}\{m_3-n_1,0\}].
\end{multline}
Under a large volume limit, $n_1$ and $m_3$ also go to infinity. We define $\hat{q}\eq q/R$, $\hat{n}_1\eq n_1/R$, and $\hat{m}_3\eq m_3/R$. Then, we get
\begin{align}
  &\NN[b_1-1,q+\mathrm{Max}\{m_3-n_1,0\}]\simeq\frac{1}{\Gamma(b_1)}(q+\mathrm{Max}\{m_3-n_1,0\})^{b_1-1}\\
  &\NN[b_2-1,-q+\mathrm{Min}\{n_1,m_3\}]\simeq\frac{1}{\Gamma(b_2)}(-q+\mathrm{Min}\{n_1,m_3\})^{b_2-1}\\
  &\NN[b_3-1,q+\mathrm{Max}\{{0,n_1-m_3}\}]\simeq\frac{1}{\Gamma(b_3)}(q+\mathrm{Max}\{{0,n_1-m_3}\})^{b_3-1}.
\end{align}
The product is
\begin{align}
&\NN[b_1-1,q+\mathrm{Max}\{m_3-n_1,0\}]\NN[b_2-1,-q+\mathrm{Min}\{n_1,m_3\}]\NN[b_3-1,q+\mathrm{Max}\{{0,n_1-m_3}\}]\notag\\
&=\left(\prod_{i=1}^3\tfrac{1}{\Gamma(b_i)}\right)(q+\mathrm{Max}\{m_3-n_1,0\})^{b_1-1}(-q+\mathrm{Min}\{n_1,m_3\})^{b_2-1}(q+\mathrm{Max}\{{0,n_1-m_3}\})^{b_3-1}\notag\\
&=\left(\prod_{i=1}^3\tfrac{R^{b_i-1}}{\Gamma(b_i)}\right)(\hat{q}+\mathrm{Max}\{\hat{m}_3-\hat{n}_1,0\})^{b_1-1}(-\hat{q}+\mathrm{Min}\{\hat{n}_1,\hat{m}_3\})^{b_2-1}(\hat{q}+\mathrm{Max}\{{0,\hat{n}_1-\hat{m}_3}\})^{b_3-1}
\end{align}
Using $\sum_q\simeq\int \dd q=R\int \dd\hat{q}$, we get
\begin{align}
&\KK_{\mathrm{red}}(n_1,n_3)\notag\\
&\simeq\frac{A}{R}\sum_{q=0}^{\mathrm{Min}\{\hat{n}_1,\hat{m}_3\}}(\hat{q}+\mathrm{Max}\{\hat{m}_3-\hat{n}_1,0\})^{b_1-1}(-\hat{q}+\mathrm{Min}\{\hat{n}_1,\hat{m}_3\})^{b_2-1}(\hat{q}+\mathrm{Max}\{{0,\hat{n}_1-\hat{m}_3}\})^{b_3-1}\notag\\
&\simeq A\int_0^{\mathrm{Min}\{\hat{n}_1,\hat{m}_3\}}d\hat{q}(\hat{q}+\mathrm{Max}\{\hat{m}_3-\hat{n}_1,0\})^{b_1-1}\notag\\
&\hspace{5cm}(-\hat{q}+\mathrm{Min}\{\hat{n}_1,\hat{m}_3\})^{b_2-1}(\hat{q}+\mathrm{Max}\{{0,\hat{n}_1-\hat{m}_3}\})^{b_3-1},
\end{align}
where $A\eq\lambda_{321}R^{b_1+b_2+b_3-2}/\Gamma(b_1)\Gamma(b_2)\Gamma(b_3)$. Now define $x\eq\hat{q}/\mathrm{Min}\{\hat{n}_1,\hat{m}'_3\}$ to change the range of integration from $[0,\mathrm{Min}\{\hat{n}_1,\hat{m}_3\}]$ to $[0,1]$.
\begin{align}
  &\KK_{\mathrm{red}}(n_1,n_3)A^{-1}\notag\\
  &\simeq [\mathrm{Min}\{\hat{n}_1,\hat{m}_3\}]^{b_1+b_2+b_3-2}\int_0^1dx\left(x+\tfrac{\mathrm{Max}\{\hat{m}_3-\hat{n}_1,0\}}{\mathrm{Min}\{\hat{n}_1,\hat{m}_3\}}\right)^{b_1-1}\left(1-x\right)^{b_2-1}\left(x+\tfrac{\mathrm{Max}\{0,\hat{n}_1-\hat{m}_3\}}{\mathrm{Min}\{\hat{n}_1,\hat{m}_3\}}\right)^{b_3-1}\notag\\
  &=
  \begin{cases}
    \hat{m}_3^{h_1+h_2+h_3-2}\int_0^1dx\ x^{b_1-1}\left(1-x\right)^{b_2-1}\left(x+\frac{\hat{n}_1-\hat{m}_3}{\hat{m}_3}\right)^{b_3-1} & (\hat{n}_1\geq\hat{m}_3)\\
    \hat{n}_1^{h_1+h_2+h_3-2}\int_0^1dx\left(x+\frac{\hat{m}_3-\hat{n}_1}{\hat{m}_3}\right)^{b_1-1}\left(1-x\right)^{b_2-1}\left(x+\frac{\hat{n}_1-\hat{m}_3}{\hat{m}_3}\right)^{b_3-1} & (\hat{n}_1<\hat{m}_3)\\
  \end{cases}
\end{align}
To make this expression simpler, we use a hypergeometric function.
\begin{equation}
  \int_0^1dx\ x^{\alpha}(1-x)^{\beta}(x+t)^{\gamma}=t^\gamma\tfrac{\Gamma(1+\alpha)\Gamma(1+\beta)}{\Gamma(2+\alpha+\beta)}{}_2F_1\left[1+\alpha,-\gamma;2+\alpha+\beta;-\tfrac{1}{t}\right]
\end{equation}
Let us consider the case of $\hat{n}_1\geq\hat{m}_3$. In the above expression, this case corresponds to $\alpha=b_1-1$, $\beta=b_2-1$, $\gamma=b_3-1$, $t=(\hat{n}_1-\hat{m}_3)/\hat{m}_3$.
\begin{multline}
  \int_0^1dx\ x^{b_1-1}(1-x)^{b_2-1}\left(x+\tfrac{\hat{n}_1-\hat{m}_3}{\hat{m}_3}\right)^{b_3-1}\\
  =\left(\tfrac{\hat{n}_1-\hat{m}_3}{\hat{m}_3}\right)^{b_3-1}\tfrac{\Gamma(b_1)\Gamma(b_2)}{\Gamma(b_1+b_2)}{}_2F_1\left[b_1,1-b_3;b_1+b_2;-\tfrac{\hat{m}_3}{\hat{n}_1-\hat{m}_3}\right]
\end{multline}
So, the three-point function is
\begin{align}
  \left.\KK_{\mathrm{red}}(n_1,m_3)\right|_{n_1\geq m_3}\simeq A\hat{m}_3^{2h_3-1}(\hat{n}_1-\hat{m}_3)^{h_1+h_2+h_3-1}{}_2F_1\left[b_1,1-b_3;2h_3;-\tfrac{\hat{m}_3}{\hat{n}_1-\hat{m}_3}\right].
\end{align}
In chiral theory, $P^0=E=P^1$ and $E=P^1=J/R$, so under large volume limit,
\begin{align}
  P^+_i&=P^+_1=\frac{2J_1}{R}=\frac{2(n_1+h_1)}{R}\simeq2\hat{n}_1\\
  P^+_o&=P^+_2=\frac{2J_2}{R}=\frac{2(n_2+h_2)}{R}\simeq2(\hat{m}_3-\hat{n}_1)\\
  P^+_f&=P^+_3=\frac{2J_3}{R}=\frac{2(n_3+h_3)}{R}=-\frac{2(m_3+h_3)}{R}\simeq-2\hat{m}_3.
\end{align}
In this case, $P_i\geq0$, $P_o\leq0$, $P_f\leq0$. And $h_i=\tilde{h}_i=\Delta_i/2$ for $i=1,2,3$. We can write the three-point function in terms of $P_i$, $P_o$, $P_f$ as
\begin{multline}
  \left.\KK_{\mathrm{red}}(n_1,m_3)\right|_{P^+_i\geq -P^+_f}\\
  \simeq\tfrac{\lambda_{321}(R/2)^{h_1+h_2+h_3-2}}{\Gamma(\Delta_f)\Gamma(h_1+h_2-h_3)}(|P^+_f|)^{\Delta_f-1}(|P^+_0|)^{\frac{\Delta_i+\Delta_0-\Delta_f}{2}-1}{}_2F_1\left[\tfrac{\Delta_f+\Delta_o-\Delta_i}{2},\tfrac{\Delta_f+\tilde{\Delta}_o-\Delta_i}{2};\Delta_f;-\tfrac{P^+_f}{P^+_o}\right].
\end{multline}
Here, $\tilde{\Delta}_o=D-\Delta_o=2-\Delta_o$. With Euler Identity
${}_2F_1[A,B;C;z]=(1-z)^{C-A-B}{}_2F_1[C-A,C-B;C;z]$, we get
\begin{multline}
  \left.\KK_{\mathrm{red}}(n_1,m_3)\right|_{P^+_o\leq0}\simeq\tfrac{\lambda_{321}(R/2)^{h_1+h_2+h_3-2}}{\Gamma(\Delta_f)\Gamma\left(\frac{\Delta_i+\Delta_o-\Delta_f}{2}\right)}(|P^+_f|)^{\Delta_f-1}(|P^+_i|)^{\Delta_i-1}(|P^+_0|)^{\frac{\Delta_o-\Delta_i-\Delta_f}{2}}\\
  {}_2F_1\left[\tfrac{\Delta_f-\Delta_o+\Delta_i}{2},\tfrac{\Delta_f-\tilde{\Delta}_o+\Delta_i}{2};\Delta_f;-\tfrac{P^+_f}{P^+_o}\right].
\end{multline}
Compare it to the $P^+_o\leq0$ component of the chiral factor of the three-point function in $D=2$ of Gillioz's formula. They are identical modulo the overall momentum-independent and time-reversal-invariant prefactor. Our formula is consistent with results in infinite volume under a large volume limit.
\subsection{Some comments}
\paragraph{Analytic continuation}
We derived two-point functions by the WI method and direct integral calculation. We derived three-point functions by straightforward Fourier transform. Whether the conformal weights of operators are an integer or not is not essential in the WI method, but it is vital in the direct integral calculation because cuts can appear when the conformal weights are not an integer. Therefore, we must show that we can analytically extend the three-point function formula to the region where the conformal weights are not integers. We summarize the detail of the discussion in the appendix. Here, we only deal with an outline of the debate.\par
We can show that two and three-point functions are polynomials in conformal weight, and the polynomial order is limited to a finite value by the number of excitations. Polynomials of finite degree that agree at an infinite number of points are equal to each other, so the three-point function formula, which we obtained by direct integral calculation, can be extended analytically to the case where the conformal weights are not integers. For more details, please see the appendix.
\paragraph{Four point function}
We can construct a four-point function from the two- and three-point functions in momentum space as in position space. One of the most important features is that decomposing four-point functions into two- and three-point functions must satisfy the conservation laws. We summarize the construction briefly.\par
As we can calculate the holomorphic and antiholomorphic parts independently, we only focus on the four-point function of holomorphic operators. In usual CFT in position space, we can calculate four-point functions by inserting an “intermediate state” between the second and the third operators. It enables us to calculate the whole four-point function with the data of three-point functions.\par
The procedure is the same in momentum space, but the conservation rule of $J\eq n+s$ restricts the intermediate state. Inserting the completeness relation (\ref{Completeness relation in infinite volume}) gives
\begin{equation}
\langle\OO_{[J_4]}^{(4)}\OO_{[J_3]}^{(3)}\OO_{[J_2]}^{(2)}\OO_{[J_1]}^{(1)}\rangle=\sum_{(5)}\frac{\langle\OO_{[J_4]}^{(4)}\OO_{[J_3]}^{(3)}\OO_{[J_1+J_2]}^{(5)}\rangle\langle\OO_{[-J_1-J_2]}^{(5)}\OO_{[J_2]}^{(2)}\OO_{[J_1]}^{(1)}\rangle}{\langle\OO_{[-J_1-J_2]}^{(5)}\OO_{[J_1+J_2]}^{(5)}\rangle}.
\end{equation}
Here, $J_1+J_2+J_3+J_4=0$ must be satisfied for the four-point function not to vanish. The sum is about conformal families. We will summarize its explicit calculation later.
\section{Main results in \texorpdfstring{$D=3$}{TEXT}}
\setcounter{equation}{0}
Calculations in three-dimensional CFT proceed the same way as in two-dimensional CFT, but some difficulties were absent in two-dimensional CFT. In two-dimensional CFT, factorization made the analysis easier. In higher dimensional CFT, on the other hand, we must perform a highly complicated calculation. In particular, the form of the three-point function is complicated, and there is no closed description. We must calculate the three-point function with recursion relations obtained from WI term by term.
Nevertheless, the momentum space description for higher dimensional CFT is compelling. One of the reasons is that we only have to consider the contributions from primaries when we calculate a four-point function. So, the structure of conformal blocks is straightforward.
\subsection{Conformal generators and their action}
We use a spherical coordinate system.
\begin{equation}
x=r\sin\theta\cos\phi,\hspace{1cm}y=r\sin\theta\sin\phi,\hspace{1cm}z=r\cos\theta\hspace{1cm}(0\leq\phi<2\pi,0\leq\theta\leq\pi)
\end{equation}
We now turn our attention to scalar operators to make the discussion simple. We can make the same argument for operators with general spin. The actions of the conformal generators on the scalar operator are as follows.
\begin{align}
P_x\cdot \OO&\eq\left(\sin\theta\cos\phi\pdv{r}+\frac{\cos\theta\cos\phi}{r}\pdv{\theta}-\frac{\sin\phi}{r\sin\theta}\pdv{\phi}\right)\OO\\
P_y\cdot\OO&\eq\left(\sin\theta\sin\phi\pdv{r}+\frac{\cos\theta\sin\phi}{r}\pdv{\theta}+\frac{\cos\phi}{r\sin\theta}\pdv{\phi}\right)\OO\\
P_z\cdot\OO&\eq\left(\cos\theta\pdv{r}-\frac{\sin\theta}{r}\pdv{\theta}\right)\OO\\
K_x\cdot\OO&\eq\left(r^2\sin\theta\cos\phi\pdv{r}-r\cos\theta\cos\phi\pdv{\theta}+r\frac{\sin\phi}{\sin\theta}\pdv{\phi}+2\Delta r\sin\theta\cos\phi\right)\OO\\
K_y\cdot\OO&\eq\left(r^2\sin\theta\sin\phi\pdv{r}-r\cos\theta\sin\phi\pdv{\theta}-r\frac{\cos\phi}{\sin\theta}\pdv{\phi}+2\Delta r\sin\theta\sin\phi\right)\OO\\
K_z\cdot\OO&\eq\left(r^2\cos\theta\pdv{r}+r\sin\theta\pdv{\theta}+2\Delta r\cos\theta\right)\OO\\
J_x\cdot\OO&\eq\left(-\sin\phi\pdv{\theta}-\frac{\cos\phi}{\tan\theta}\pdv{\phi}\right)\OO\\
J_y\cdot\OO&\eq\left(\cos\phi\pdv{\theta}-\frac{\sin\phi}{\tan\theta}\pdv{\phi}\right)\OO\\
J_z\cdot\OO&\eq\pdv{\phi}\OO\\
D\cdot\OO&\eq\left(r\pdv{r}+\Delta\right)\OO
\end{align}
Spatially-integrated mode is defined as
\begin{equation}
O_{l,m}\eq\int \dd\Omega Y^{*}_{l,m}(\theta,\phi)O(r,\theta,\phi)\hspace{1cm}(0\leq\theta\leq\pi,0\leq\phi<2\pi,\dd\Omega=\dd\theta\dd\phi\sin\theta).
\end{equation}
We removed coefficient factors such as $i$ to simplify the calculations.
$Y_{l,m}$ is a spherical harmonics, and $Y^{*}_{l,m}=(-1)^{m}Y_{l,-m}$ is its complex conjugate. The actions of conformal generators on the spatially integrated mode are as follows.
\begin{flalign}
&(P_x+iP_y)\cdot\OO_{l,m}=D_r^{P_+^+}\OO_{l+1,m-1}+D_r^{P_+^-}\OO_{l-1,m-1}&\\
&D_r^{P_+^+}=(-1)^{m-1}\sqrt{(2l+1)(2l+3)}\mqty(l&1&l+1\\0&0&0)\left[\sqrt{2}\mqty(l&1&l+1\\-m&1&m-1)\left(\pdv{r}-\frac{m-2}{r}\right)\right.&\notag\\
&\left.\hspace{6cm}+\mqty(l&1&l+1\\-m+1&0&m-1)\frac{\sqrt{(l+m)(l-m+1)}}{r}\right]&\\
&D_r^{P_+^-}=(-1)^{m-1}\sqrt{(2l+1)(2l-1)}\mqty(l&1&l-1\\0&0&0)\left[\sqrt{2}\mqty(l&1&l-1\\-m&1&m-1)\left(\pdv{r}-\frac{m-2}{r}\right)\right.&\notag\\
&\left.\hspace{6cm}+\mqty(l&1&l-1\\-m+1&0&m-1)\frac{\sqrt{(l+m)(l-m+1)}}{r}\right]&
\end{flalign}
\begin{flalign}
&(P_x-iP_y)\cdot\OO_{l,m}=D_r^{P_-^+}\OO_{l+1,m+1}+D_r^{P_-^-}\OO_{l-1,m+1}&\\
&D_r^{P_-^+}=(-1)^{m}\sqrt{(2l+1)(2l+3)}\mqty(l&1&l+1\\0&0&0)\left[\sqrt{2}\mqty(l&1&l+1\\-m&-1&m+1)\left(\pdv{r}+\frac{m+2}{r}\right)\right.&\notag\\
&\left.\hspace{6cm}+\mqty(l&1&l+1\\-m-1&0&m+1)\frac{\sqrt{(l-m)(l+m+1)}}{r}\right]&\\
&D_r^{P_-^-}=(-1)^{m}\sqrt{(2l+1)(2l-1)}\mqty(l&1&l-1\\0&0&0)\left[\sqrt{2}\mqty(l&1&l-1\\-m&-1&m+1)\left(\pdv{r}+\frac{m+2}{r}\right)\right.&\notag\\
&\left.\hspace{6cm}+\mqty(l&1&l-1\\-m-1&0&m+1)\frac{\sqrt{(l-m)(l+m+1)}}{r}\right]&
\end{flalign}
\begin{flalign}
&P_z\cdot\OO_{l,m}=D_r^{P_z^+}\OO_{l+1,m}+D_r^{P_z^-}\OO_{l-1,m}&\\
&D_r^{P_z^+}=(-1)^{m}\sqrt{(2l+1)(2l+3)}\mqty(l&1&l+1\\0&0&0)\left[\mqty(l&1&l+1\\-m&0&m)\left(\pdv{r}-\frac{m-2}{r}\right)\right.&\notag\\
&\left.\hspace{6cm}+\sqrt{2}\mqty(l&1&l+1\\-m+1&-1&m)\frac{\sqrt{(l+m)(l-m+1)}}{r}\right]&\\
&D_r^{P_z^-}=(-1)^{m}\sqrt{(2l+1)(2l-1)}\mqty(l&1&l-1\\0&0&0)\left[\mqty(l&1&l-1\\-m&0&m)\left(\pdv{r}-\frac{m-2}{r}\right)\right.&\notag\\
&\left.\hspace{6cm}+\sqrt{2}\mqty(l&1&l-1\\-m+1&-1&m)\frac{\sqrt{(l+m)(l-m+1)}}{r}\right]&
\end{flalign}
\begin{flalign}
&(K_x+iK_y)\cdot\OO_{l,m}=D_r^{K_+^+}\OO_{l+1,m-1}+D_r^{K_+^-}\OO_{l-1,m-1}&\\
&D_r^{K_+^+}=(-1)^{m-1}r^2\sqrt{(2l+1)(2l+3)}\mqty(l&1&l+1\\0&0&0)\left[\sqrt{2}\mqty(l&1&l+1\\-m&1&m-1)\left(\pdv{r}+\frac{2\Delta+m-2}{r}\right)\right.&\notag\\
&\left.\hspace{6cm}-\mqty(l&1&l+1\\-m+1&0&m-1)\frac{\sqrt{(l+m)(l-m+1)}}{r}\right]&\\
&D_r^{K_+^-}=(-1)^{m-1}r^2\sqrt{(2l+1)(2l-1)}\mqty(l&1&l-1\\0&0&0)\left[\sqrt{2}\mqty(l&1&l-1\\-m&1&m-1)\left(\pdv{r}+\frac{2\Delta+m-2}{r}\right)\right.&\notag\\
&\left.\hspace{6cm}-\mqty(l&1&l-1\\-m+1&0&m-1)\frac{\sqrt{(l+m)(l-m+1)}}{r}\right]&
\end{flalign}
\begin{flalign}
&(K_x-iK_y)\cdot\OO_{l,m}=D_r^{K_-^+}\OO_{l+1,m+1}+D_r^{K_-^-}\OO_{l-1,m+1}&\\
&D_r^{K_-^+}=(-1)^{m}r^2\sqrt{(2l+1)(2l+3)}\mqty(l&1&l+1\\0&0&0)\left[\sqrt{2}\mqty(l&1&l+1\\-m&-1&m+1)\left(\pdv{r}-\frac{-2\Delta+m+2}{r}\right)\right.&\notag\\
&\left.\hspace{6cm}-\mqty(l&1&l+1\\-m-1&0&m+1)\frac{\sqrt{(l-m)(l+m+1)}}{r}\right]&\\
&D_r^{K_-^-}=(-1)^{m}r^2\sqrt{(2l+1)(2l-1)}\mqty(l&1&l-1\\0&0&0)\left[\sqrt{2}\mqty(l&1&l-1\\-m&-1&m+1)\left(\pdv{r}-\frac{-2\Delta+m+2}{r}\right)\right.&\notag\\
&\left.\hspace{6cm}-\mqty(l&1&l-1\\-m-1&0&m+1)\frac{\sqrt{(l-m)(l+m+1)}}{r}\right]&
\end{flalign}
\begin{flalign}
&K_z\cdot\OO_{l,m}=D_r^{K_z^+}\OO_{l+1,m}+D_r^{K_z^-}\OO_{l-1,m}&\\
&D_r^{K_z^+}=(-1)^{m}r^2\sqrt{(2l+1)(2l+3)}\mqty(l&1&l+1\\0&0&0)\left[\mqty(l&1&l+1\\-m&0&m)\left(\pdv{r}+\frac{2\Delta+m-2}{r}\right)\right.&\notag\\
&\left.\hspace{6cm}-\sqrt{2}\mqty(l&1&l+1\\-m+1&-1&m)\frac{\sqrt{(l+m)(l-m+1)}}{r}\right]&\\
&D_r^{K_z^-}=(-1)^{m}r^2\sqrt{(2l+1)(2l-1)}\mqty(l&1&l-1\\0&0&0)\left[\mqty(l&1&l-1\\-m&0&m)\left(\pdv{r}+\frac{2\Delta+m-2}{r}\right)\right.&\notag\\
&\left.\hspace{6cm}-\sqrt{2}\mqty(l&1&l-1\\-m+1&-1&m)\frac{\sqrt{(l+m)(l-m+1)}}{r}\right]&
\end{flalign}
\begin{align}
&(J_x+iJ_y)\cdot\OO_{l,m}=i\sqrt{(l+m)(l-m+1)}\OO_{l,m-1}\\
&(J_x-iJ_y)\cdot\OO_{l,m}=i\sqrt{(l-m)(l+m+1)}\OO_{l,m+1}\\
&J_z\cdot\OO_{l,m}=-im\OO_{l,m}
\end{align}
The 2-by-3 matrix in the above is the Wigner $3j$ symbol. We summarize it and the derivation of the above expressions in the appendix.
\subsection{Two-point function}
The action of $J_z$ on the operator mode implies the conservation law of $m$. So, we have the following reduced form for the two-point function.
\begin{equation}
\langle\OO_{l_2,m_2}(r_2)\OO_{l_1,m_1}(r_1)\rangle=\delta(m_1+m_2)r_1^{-\Delta_1}r_2^{-\Delta_2}F_{m_1}^{l_1,l_2}(y)\hspace{1cm}\left(y=\frac{r_1}{r_2}\right)
\end{equation}
We want to get the exact form of it. First, start from $J_x\pm iJ_y$ WIs.
\begin{align}
&(J_x+iJ_y)\cdot\langle\OO_{l_2,m_2}(r_2)\OO_{l_1,m_1}(r_1)\rangle\\
&=i\sqrt{(l_2+m_2)(l_2-m_2+1)}\langle\OO_{l_2,m_2-1}\OO_{l_1,m_1}\rangle+i\sqrt{(l_1+m_1)(l_1-m_1+1)}\langle\OO_{l_2,m_2}\OO_{l_1,m_1-1}\rangle\notag\\
&=i\delta(m_1+m_2-1)r_1^{-\Delta_1}r_2^{-\Delta_2}[\sqrt{(l_2+m_2)(l_2-m_2+1)}F_{m_1}^{l_1,l_2}+\sqrt{(l_1+m_1)(l_1-m_1+1)}F_{m_1-1}^{l_1,l_2}]\notag
\end{align}
When $m_2=-m_1+1$,
\begin{equation}
\sqrt{(l_2-m_1+1)(l_2+m_1)}F_{m_1}^{l_1,l_2}+\sqrt{(l_1+m_1)(l_1-m_1+1)}F_{m_1-1}^{l_1,l_2}=0.\label{J+ Ward Identity}
\end{equation}
\begin{align}
&(J_x-iJ_y)\cdot\langle\OO_{l_2,m_2}(r_2)\OO_{l_1,m_1}(r_1)\rangle\\
&=i\sqrt{(l_2-m_2)(l_2+m_2+1)}\langle\OO_{l_2,m_2+1}\OO_{l_1,m_1}\rangle+i\sqrt{(l_1-m_1)(l_1+m_1+1)}\langle\OO_{l_2,m_2}\OO_{l_1,m_1+1}\rangle\notag\\
&=i\delta(m_1+m_2+1)r_1^{-\Delta_1}r_2^{-\Delta_2}[\sqrt{(l_2-m_2)(l_2+m_2+1)}F_{m_1}^{l_1,l_2}+\sqrt{(l_1-m_1)(l_1+m_1+1)}F_{m_1+1}^{l_1,l_2}]\notag
\end{align}
When $m_2=-m_1-1$,
\begin{equation}
\sqrt{(l_2+m_1+1)(l_2-m_1)}F_{m_1}^{l_1,l_2}+\sqrt{(l_1-m_1)(l_1+m_1+1)}F_{m_1+1}^{l_1,l_2}=0.\label{J- Ward Identity}
\end{equation}
Shifting (\ref{J+ Ward Identity}) by $m_1\rightarrow m_1+1$, we get
\begin{equation}
\sqrt{(l_1+m_1+1)(l_1-m_1)}F_{m_1}^{l_1,l_2}+\sqrt{(l_2-m_1)(l_2+m_1+1)}F_{m_1+1}^{l_1,l_2}=0.\label{J+ Ward Identity2}
\end{equation}
From (\ref{J- Ward Identity}) and (\ref{J+ Ward Identity2}), for $-l_1\leq m_1\leq l_1-1$ $-l_2\leq m_2\leq l_2-1$, we get
\begin{equation}
(l_1-m_1)(l_1+m_1+1)=(l_2-m_1)(l_2+m_1+1).
\end{equation}
As $l_1\geq0$ and $l_2\geq0$, we get $l_1=l_2$ and
\begin{equation}
  F_{m_1}^{l_1,l_2}+F_{m_1+1}^{l_1,l_2}=0.\label{Jpm Ward Identity result}
\end{equation}
These relations say that the two-point function has the following structure.
\begin{equation}
\langle\OO_{l_2,m_2}(r_2)\OO_{l_1,m_1}(r_1)\rangle=\delta(m_1+m_2)\delta(l_1-l_2)r_1^{-\Delta_1}r_2^{-\Delta_2}(-1)^{m_1}G^{l_1}(y)
\end{equation}
where $G^{l_1}=F_{0}^{l_1,l_2}$. From $P_z$ and the $K_z$ WIs, we can find that it vanishes unless the conformal dimensions of the two operators are the same. We summarize the detail of the calculation in the appendix to get the exact form of $G^{l_1}$. From the WIs, we get the following second-order differential equation.
\begin{equation}
\dv[2]{y}G^l(y)+\frac{2\Delta y^2+2(\Delta-1)}{y^3-y}\dv{y}G^l(y)+\frac{(\Delta+l)(\Delta-l-1)}{y^2}G^l(y)=0\label{ODE for two pt func in 3 dim}
\end{equation}
The solution for it is
\begin{align}
G^l&\eq K_ly^{\Delta+l}(\sum_{n=0}b_{l|n}y^{2n})=K_0y^{\Delta+l}(\sum_{n=0}c_{l|n}y^{2n})\\
K_{l}&=\frac{\Gamma(\frac{3}{2})}{\Gamma(l+\frac{3}{2})}\frac{\Gamma(\Delta+l)}{\Gamma(\Delta)}K_0\\
b_{l|n}&=\frac{\Gamma(\Delta+l+n)\Gamma(\Delta-\frac{1}{2}+n)\Gamma(1)\Gamma(\frac{3}{2}+l)}{\Gamma(\Delta+l)\Gamma(\Delta-\frac{1}{2})\Gamma(1+n)\Gamma(\frac{3}{2}+l+n)}\\
c_{l|n}&=\frac{\Gamma(\Delta+l+n)\Gamma(\Delta-\frac{1}{2}+n)\Gamma(1)\Gamma(\frac{3}{2})}{\Gamma(\Delta)\Gamma(\Delta-\frac{1}{2})\Gamma(1+n)\Gamma(\frac{3}{2}+l+n)}.\label{cln}
\end{align}
The normalization of operators determines the value of $K_0$. We can always set it unity.\par
To see if the solution for the two-point function of scalar primaries is consistent with the result in \cite{Gillioz:2021sce}, we only have to focus on the form of (\ref{cln}). As we calculated in two-dimensional CFT, with Stirling's approximation, we get
\begin{equation}
  c_{l|n}=\frac{1}{\Gamma(\Delta)\Gamma(\Delta-\frac{1}{2})}\frac{\Gamma\left(\frac{R(E_1-k_1)+\Delta}{2}-\frac{1}{2}\right)\Gamma\left(\frac{R(E_1+k_1)+\Delta}{2}\right)}{\Gamma\left(\frac{R(E_1-k_1)-\Delta}{2}+1\right)\Gamma\left(\frac{R(E_1+k_1)-\Delta}{2}+\frac{3}{2}\right)}\xrightarrow{R\rightarrow\infty}\frac{(k^2)^{\Delta-3/2}}{\Gamma(\Delta)\Gamma(\Delta-\frac{3}{2}+1)},
\end{equation}
where $k=(E_1,k_1)=\left(\frac{\Delta+l+2n}{R},\frac{l}{R}\right)$. This is identical to the result (\ref{Gillioz's result2}) with $d=3$ at infinite volume calculated in the paper \cite{Gillioz:2021sce}.
\subsection{Three-point function}
Finding the explicit closed formula for the three-point function is complicated. Of course, we can perform direct integral calculations honestly, but the number of terms for that expression is too large to handle. \par
On the other hand, the WI method implies that the number of terms in the formula should be manageable. However, the WIs are too complicated to find a definitive solution. We must get the exact solution term by term with recursion relations obtained from WIs.\par
First, consider the reduction of the three-point function. As the $J_z$ WI implies the conservation of $m$, we can write the three-point function as
\begin{align}
\langle\OO_{l_3,m_3}\OO_{l_2,m_2}\OO_{l_1,m_1}\rangle&\eq\int\dd\Omega_1\int\dd\Omega_2\int\dd\Omega_3Y^*_{l_1,m_1}Y^*_{l_2,m_2}Y^*_{l_3,m_3}\langle\OO_3\OO_2\OO_1\rangle\notag\\
&\eq\delta(m_1+m_2+m_3)r_1^{-\Delta_1}r_2^{-\Delta_2}r_3^{-\Delta_3}F^{l_1,l_2,l_3}_{m_1,m_3}(y_1,y_3),
\end{align}
where $y_1\eq r_1/r_2$ and $y_3\eq r_3/r_2$. Next, consider the $J_x\pm iJ_y$ WIs.
\begin{multline}
J_x+iJ_y:\sqrt{(l_3+m_3)(l_3-m_3+1)}F_{m_1,m_3-1}^{l_1,l_2,l_3}+\sqrt{(l_2+m_1+m_3)(l_2-m_1-m_3+1)}F_{m_1,m_3}^{l_1,l_2,l_3}\\
+\sqrt{(l_1+m_1)(l_1-m_1+1)}F_{m_1-1,m_3}^{l_1,l_2,l_3}=0\label{Jx+iJy Ward Identity}
\end{multline}
\begin{multline}
J_x-iJ_y:\sqrt{(l_3-m_3)(l_3+m_3+1)}F_{m_1,m_3+1}^{l_1,l_2,l_3}+\sqrt{(l_2-m_1-m_3)(l_2+m_1+m_3+1)}F_{m_1,m_3}^{l_1,l_2,l_3}\\
+\sqrt{(l_1-m_1)(l_1+m_1+1)}F_{m_1+1,m_3}^{l_1,l_2,l_3}=0\label{Jx-iJy Ward Identity}
\end{multline}
They are related by conjugate transformation $m_1\rightarrow-m_1$ and $m_3\rightarrow-m_3$.\par
The equations (\ref{Jx+iJy Ward Identity}) and (\ref{Jx-iJy Ward Identity}) imply that $F_{m_1,m_3}^{l_1,l_2,l_3}$ and $F_{m'_1,m'_3}^{l_1,l_2,l_3}$ differ only by a constant factor. The relation between $F_{m_1,m_3}^{l_1,l_2,l_3}$ and $F_{l_1,-l_3}^{l_1,l_2,l_3}$ is
\begin{equation}
F_{l_1-n,-l_3+k}^{l_1,l_2,l_3}\eq(-1)^{n+k}\sqrt{\frac{(2l_1-n)!(2l_3-k)!(l_2-l_1+l_3)!(l_2+l_1-l_3+k-n)!}{(2l_1)!n!(2l_3)!k!(l_2-l_1+l_3-k+n)!(l_2+l_1-l_3)!}}f_{n,k}F_{l_1,-l_3}^{l_1,l_2,l_3}\label{reduction form1}
\end{equation}
with
\begin{equation}
f_{n,k}=\sum_{l=0}^{k}(-1)^{l}\frac{k!n!(2l_3-k+l)!(l_2+l_1-l_3+k-l)!(l_2-l_1+l_3+n-k)!}{l!(k-l)!(n-l)!(2l_3-k)!(l_2+l_1-l_3-n+k)!(l_2-l_1+l_3-k+l)!}.\label{reduction form2}
\end{equation}
We summarize its derivation in the appendix.
Now, we only have to consider $F^{l_1,l_2,l_3}\eq F^{l_1,l_2,l_3}_{l_1,-l_3}$ because we can derive all $F^{l_1,l_2,l_3}_{m_1,m_3}$s from $F^{l_1,l_2,l_3}$ by using the above relation.\par
The WIs for a general $F_{m_1,m_3}^{l_1,l_2,l_3}$ are very complicated. For example, the $P_x+iP_y$ WI for a three-point function is
\begin{align}
  0=&\sqrt{\tfrac{(l_3-m_3+1)(l_3-m_3+2)}{(2l_3+1)(2l_3+3)}}\frac{1}{y_3}\left[y_3\dv{y_3}-\Delta_3+l_3+2\right]F_{m_1,m_3-1}^{l_1,l_2,l_3}\notag\\
  &-\sqrt{\tfrac{(l_3+m_3-1)(l_3+m_3)}{(2l_3-1)(2l_3+1)}}\frac{1}{y_3}\left[y_3\dv{y_3}-\Delta_3-l_3+1\right]F_{m_1,m_3-1}^{l_1,l_2,l_3-1}\notag\\
  &-\sqrt{\tfrac{(l_2+m_1+m_3)(l_1+m_1+m_3+1)}{(2l_2+1)(2l_2+3)}}\left[y_1\dv{y_1}+y_3\dv{y_3}+\Delta_2-l_2-2\right]F_{m_1,m_3}^{l_1,l_2+1,l_3}\notag\\
  &+\sqrt{\tfrac{(l_2-m_1-m_3)(l_2-m_1-m_3+1)}{(2l_2-1)(2l_2+1)}}\left[y_1\dv{y_1}+y_3\dv{y_3}+\Delta_2+l_2-1\right]F_{m_1,m_3}^{l_1,l_2-1,l_3}\notag\\
  &+\sqrt{\tfrac{(l_1-m_1+1)(l_1-m_1+2)}{(2l_1+1)(2l_1+3)}}\frac{1}{y_1}\left[y_1\dv{y_1}-\Delta_1+l_1+2\right]F_{m_1-1,m_3}^{l_1+1,l_2,l_3}\notag\\
  &-\sqrt{\tfrac{(l_1+m_1-1)(l_1+m_1)}{(2l_1-1)(2l_1+1)}}\frac{1}{y_1}\left[y_1\dv{y_1}-\Delta_1-l_1+1\right]F_{m_1-1,m_3}^{l_1-1,l_2,l_3}.
\end{align}
We have similar WIs for $P_x-iP_y$, $P_z$, $K_x\pm iK_y$ and $K_z$. They are too complicated to handle, but the reduction (\ref{reduction form1}) helps us a little. For example, the $P_x+iP_y$ WI becomes
\begin{align}
  0=&\sqrt{\tfrac{2l_3+2}{2l_3+3}}\frac{1}{y_3}\left[y_3\dv{y_3}+\Delta_3-l_3-2\right]F^{l_1,l_2,l_3+1}\notag\\
  &-\sqrt{\tfrac{(l_2+l_1-l_3)(l_2+l_1-l_3+1)}{(2l_2+1)(2l_2+3)}}\left[y_1\dv{y_1}+y_3\dv{y_3}-\Delta_2+l_2+2\right]F^{l_1,l_2+1,l_3}\notag\\
  &+\sqrt{\tfrac{(l_2-l_1+l_3)(l_2-l_1+l_3+1)}{(2l_2-1)(2l_2+1)}}\left[y_1\dv{y_1}+y_3\dv{y_3}-\Delta_2-l_2+1\right]F^{l_1,l_2-1,l_3}\notag\\
  &+\sqrt{\tfrac{(2l_1+2)(l_2+l_1-l_3)(l_2+l_1-l_3+1)(l_2-l_1+l_3)(l_2-l_1+l_3+1)}{2(2l_1+1)(2l_1+3)}}\frac{1}{y_1}\left[y_1\dv{y_1}+\Delta_1-l_1-2\right]F^{l_1+1,l_2,l_3}\notag\\
  &-\sqrt{\tfrac{2l_1}{2l_1+1}}\frac{1}{y_1}\left[y_1\dv{y_1}+\Delta_1+l_1-1\right]F^{l_1-1,l_2,l_3}.
\end{align}
We have similar WIs for $P_x-iP_y$, $P_z$, $K_x\pm iK_y$, and $K_z$. However, they are still difficult to handle as they are two-variable differential equations. Therefore, we use new notation to get one-variable differential equations.
\begin{align}
F^{l_1,l_2,l_3}=\sum_{p,q\in\mathbb{Z}_{\geq0}}F_{(p,q)}^{l_1,l_2,l_3}y_1^{\Delta_1+l_1+2p}y_3^{-\Delta_3-l_3-2q}=\sum_{p\in\mathbb{Z}_{\geq0}}K_p^{l_1,l_2,l_3}(y_3)y_1^{\Delta_1+l_1+2p}
\end{align}
Substituting it for the WIs and comparing the coefficients for $y_1^k$, we get one-variable differential equations for $K_p^{l_1,l_2,l_3}$. Here, we focus on getting $F^{0,0,0}$. We list the differential equations related to $K^{0,0,0}_p$. $K_p^{l_1,l_2,l_3}$ vanishes when $p<0$.
\begin{align}
  0&=y_3K_p^{'000}+K_p^{'011}+(\Delta_1+\Delta_2+2p)K_p^{000}+\frac{-\Delta_3+2}{y_3}K_p^{011}+(2p+3)K_p^{110}\label{ODE1}\\
  0&=y_3K_p^{'000}+y_3^2K_p^{'011}+(\Delta_1-\Delta_2+2p)K_p^{000}+(\Delta_3-2)y_3K_p^{011}+(2\Delta_1+2p-3)K_{p-1}^{110}\label{ODE2}\\
  0&=K_p^{'000}+y_3K_p^{'011}-\frac{\Delta_3}{y_3}K_p^{000}+(\Delta_1+\Delta_2+2p-2)K_p^{011}-(2p+3)K_p^{101}\label{ODE3}\\
  0&=y_3^2K_p^{'000}+y_3K_p^{'011}+\Delta_3y_3K_p^{000}+(\Delta_1-\Delta_2+2p+2)K_p^{011}-(2\Delta_1+2p-3)K_{p-1}^{101}\label{ODE4}\\
  0&=y_3K_p^{'110}-K_p^{'101}+(2p+2)K_{p+1}^{000}+\frac{\Delta_3-2}{y_3}K_p^{101}+(\Delta_1+\Delta_2+2p-1)K_p^{110}\label{ODE5}\\
  0&=y_3K_p^{'110}-y_3^2K_p^{'101}+2(\Delta_1+p)K_p^{000}-(\Delta_3-2)y_3K_p^{101}+(\Delta_1-\Delta_2+2p+3)K_p^{110}\label{ODE6}
\end{align}
From (\ref{ODE2}) with $p=0$ and (\ref{ODE4}) with $p=0$, we get a second order differential equation.
\begin{multline}
y_3^2(y_3^2-1)K_0^{''000}+2[\Delta_3y_3^3+(\Delta_2-\Delta_1-1)y_3]K_0^{'000}+[\Delta_3(\Delta_3-1)y_3^2-(\Delta_1-\Delta_2)(\Delta_1-\Delta_2+1)]K_0^{000}\\=0
\end{multline}
We can get the solution for this ODE assuming that $K_0^{000}$ is an analytic function at $y_3=0$.
\begin{equation}
K_0^{000}=\sum_{q=0}^{\infty}\frac{(\Delta_1-\Delta_2-\Delta_3)!}{(2q+1)!(\Delta1-\Delta_2-\Delta_3+1)!}y_3^{-\Delta_3-2q}\label{K_0^000}
\end{equation}
Substituting it for six ODEs (\ref{ODE1})-(\ref{ODE6}), we get all $K_p^{000}$, $K_p^{011}$, $K_p^{101}$ and $K_p^{110}$.
The procedure is as follows.
First, substituting (\ref{K_0^000}) for (\ref{ODE2}) with $p=0$ and (\ref{ODE4}) with $p=0$, we get $K_0^{011}$.
Substituting $K_0^{000}$ and $K_0^{011}$ for (\ref{ODE1}), (\ref{ODE3}) and (\ref{ODE6}) gives $K_0^{101}$ and $K_0^{110}$.
Next, we can get $K_1^{000}$ by substituting $K_0^{101}$ and $K_0^{110}$ for (\ref{ODE5}).
$K_1^{001}$ can be obtained from (\ref{ODE2}) or (\ref{ODE4}).
$K_0^{101}$ and $K_0^{110}$ can be obtained from (\ref{ODE1}), (\ref{ODE3}) and (\ref{ODE6}).
In this way, we can get all $K_p^{000}$, $K_p^{011}$, $K_p^{101}$ and $K_p^{110}$.
We can guess from the procedure that not all ODEs are needed to get the solutions.\par
The direct integral calculation is less helpful in this case. We summarize the detail of the calculation in the appendix. Here, we only show the result for $\OO^{(3)}_{0,0}\OO^{(2)}_{0,0}\OO^{(1)}_{0,0}$ and compare it to that in the WI method.
\begin{equation}
\int\frac{\dd\Omega_1}{4\pi}\int\frac{\dd\Omega_2}{4\pi}\int\frac{\dd\Omega_3}{4\pi}\langle\OO_3\OO_2\OO_1\rangle=\lambda_{321}r_1^{-\Delta_1}r_2^{-\Delta_2}r_3^{-\Delta_3}\sum_{p,q\in\mathbb{Z}_{\geq0}}F_{(p,q)}^{000}y_1^{\Delta_1+2p}y_3^{-\Delta_3-2q}
\end{equation}
$F_{(p,q)}^{000}$ has the following form.
\begin{align}
  F_{(p,q)}^{000}=&\sum_{k=0}^{\min(p,q)}\sum_{j=0}^{k}\sum_{n=0}^{p-k}\sum_{l=0}^{q-k}A_{k,j,n,l}\int\frac{\dd\Omega_1}{4\pi}\int\frac{\dd\Omega_2}{4\pi}\int\frac{\dd\Omega_3}{4\pi}\Phi_{12}^{2n}\Phi_{23}^{2l}\Phi_{31}^{2j}\notag\\
  &-\sum_{k=0}^{\min(p-1,q-1)}\sum_{j=0}^{k}\sum_{n=0}^{p-k-1}\sum_{l=0}^{q-k-1}B_{k,j,n,l}\int\frac{\dd\Omega_1}{4\pi}\int\frac{\dd\Omega_2}{4\pi}\int\frac{\dd\Omega_3}{4\pi}\Phi_{12}^{2n+1}\Phi_{23}^{2l+1}\Phi_{31}^{2j+1}
\end{align}
\begin{align}
  A_{k,j,n,l}&=\tfrac{2^{2(j+n+l)}}{(2j)!(2n)!(2l)!(k-j)!(p-k-n)!(q-k-l)!}\tfrac{g_{12}!g_{23}!g_{31}!}{(g_{31}-k-j)!(g_{12}-p+k-n)!(g_{23}-q+k-l)!}\\
  B_{k,j,n,l}&=\tfrac{2^{2(j+n+l)+3}}{(2j+1)!(2n+1)!(2l+1)!(k-j)!(p-k-n-1)!(q-k-l-1)!}\tfrac{g_{12}!g_{23}!g_{31}!}{(g_{31}-k-j-1)!(g_{12}-p+k-n)!(g_{23}-q+k-l)!}
\end{align}
where $g_{12}\eq(\Delta_3-\Delta_1-\Delta_2)/2$ and $\Phi_{12}\eq\sin\theta_1\sin\theta_2\cos(\phi_1-\phi_2)+\cos\theta_1\cos\theta_2$.
The problem is that this configuration has too many terms. At least, the number of the terms is of the order of $pq[\min(p,q)]^2$. Moreover, we have not yet found a simple expression for $\int\dd\Omega_1\int\dd\Omega_2\int\dd\Omega_3\Phi_{12}^{2n}\Phi_{23}^{2l}\Phi_{31}^{2j}$ and $\int\dd\Omega_1\int\dd\Omega_2\int\dd\Omega_3\Phi_{12}^{2n+1}\Phi_{23}^{2l+1}\Phi_{31}^{2j+1}$. We would like to get a valuable expression for three-point functions because we need to calculate more for conformal bootstrap in momentum space. This direct integral calculation result is not helpful in that sense.

\section{Applications}
\setcounter{equation}{0}
In this section, we show some examples of applications of the calculations in previous sections. The ultimate goal is to compare the results of conformal bootstrap at finite volume with those at infinite volume to see how the $1/R$ term comes into play, but this has not been fully explored yet and will be left as an issue to be resolved in the future.\par
In this paper, we only deal with two types of test functions. The first is the delta function, also used in the previous article \cite{Gillioz:2019iye}. The other one is the step-like function, which helps us understand how to handle the step function.\par
In those cases, it is easy to derive bootstrap equations in momentum space. However, we must consider contributions equally from all intermediate states, from the primary state to the state with high energy. \par
On the other hand, it may be possible to suppress contributions from the intermediate state with high energy if we choose a better test function. However, there is currently a challenge in that it is difficult to do Fourier transforms for all but delta and step-like functions. If we could succeed in finding a test function that is calculable and well-behaved for summing up the intermediate state, that would be a better test function.\par
Though the delta function and step-like function are interesting enough to be studied, searching for a better test function will be left as an issue to resolve.
\subsection{Conformal bootstrap in momentum space for infinite volume}
\subsubsection{Conformal bootstrap equation}
We review the Momentum-Space Bootstrap for Wightman functions at infinity volume in brief. For more details, please check the paper \cite{Gillioz:2019iye}. The paper seems they use a test function with support at coincident point $\sigma_2=\sigma_3, \tau_2=\tau_3$. It makes a problem when the commutator of the second and the third operators have a singular term at the coincident point. If we try to Fourier transform the commutation relation, including the singular point, we must thoroughly consider the commutation relation of the second and third operators. We leave this problem to future work.\par
So instead of that, we use a test function with support only at spacelike points. Though we deal with two-dimensional CFT in the following, the logic is parallel for general dimensional CFT in infinite volume. In this part, the delta function represents Dirac's delta function.\par
First, let us consider the four-point function in infinite volume. We use $\sigma$ and $\tau$ to represent space and time coordinates. The microcausality condition says
\begin{equation}
\bra{0}\OO^{(4)}(x_4)[\OO^{(3)}(x_3),\OO^{(2)}(x_2)]\OO^{(1)}(x_1)\ket{0}=0\hspace{1cm}\text{for $x_3-x_2$ spacelike.}
\end{equation}
It is inconvenient for Fourier transform since we have to integrate it over where $x_3-x_2$ is timelike. Instead, we adopt the following improved microcausality condition.
\begin{equation}
\bra{0}\OO^{(4)}(\sigma_4,\tau_4)[\OO^{(3)}(\sigma_3,\tau_3),\OO^{(2)}(\sigma_2,\tau_2)]\OO^{(1)}(\sigma_1,\tau_1)\ket{0}f(x_3-x_2)=0
\end{equation}
Here, $f(x)$ has support only for spacelike $x$. For example,
\begin{equation}
f(x_3-x_2)=\delta(\sigma_3-\sigma_2-a)\delta(\tau_3-\tau_2)\hspace{1cm}(a\neq0).
\end{equation}
With this test function, we get
\begin{flalign}
  &\left(\prod_{j=1}^4\int\dd^2x_je^{-iP_j\cdot x_j}\right)\bra{0}\OO^{(4)}[\OO^{(3)},\OO^{(2)}]\OO^{(1)}\ket{0}\delta(\sigma_3-\sigma_2-a)\delta(\tau_3-\tau_2)&\notag\\
  &=\int\frac{\dd^2k}{(2\pi)^2}\dd^2x_2\dd^2x_3e^{-i(P_2\cdot x_2+P_3\cdot x_3)} e^{ik\cdot(x_3-x_2)}e^{-iak_1}\bra{0}\tilde{\OO}^{(4)}(P_4)[\OO^{(3)},\OO^{(2)}]\tilde{\OO}^{(1)}(P_1)\ket{0}&\notag\\
  &\propto\int\dd^2k\dd^2x_2\dd^2x_3e^{-iak_1-i(P_2+k)\cdot x_2-i(P_3-k)\cdot x_3}\bra{0}\tilde{\OO}^{(4)}(P_4)[\OO^{(3)}(x_3),\OO^{(2)}(x_2)]\tilde{\OO}^{(1)}(P_1)\ket{0}&\notag\\
  &=\int\dd^2ke^{-iak_1}\bra{0}\tilde{\OO}^{(4)}(P_4)[\tilde{\OO}^{(3)}(P_3-k),\tilde{\OO}^{(2)}(P_2+k)]\tilde{\OO}^{(1)}(P_1)\ket{0}&\notag\\
  &\propto\delta^2(\sum_iP_i)\int\dd^2Qe^{-iaQ_1}\llangle\tilde{\OO}^{(4)}(P_4)[\tilde{\OO}^{(3)}(-Q-\tfrac{P_1+P_4}{2}),\tilde{\OO}^{(2)}(Q-\tfrac{P_1+P_4}{2})]\tilde{\OO}^{(1)}(P_1)\rrangle&\notag\\
  &\equiv\delta^2(\sum_iP_i)\int\dd^2Qe^{-iaQ_1}[W(P_4,P_1|Q)-W(P_4,P_1|-Q)]&\notag\\
  &=0.&\label{BE for Identical scalar operators}
\end{flalign}
The last equation does not automatically hold, and this gives non-trivial information as the bootstrap equation. We defined $Q=(Q_0,Q_1)\equiv(P_2-P_3+2k)/2$.\par
Let us check whether this is satisfied by the four-point function in generalized free field theory. In that case, the four-point function is given by multiplying two-point functions.
\begin{flalign}
  &\llangle\tilde{\phi}(P_4)\tilde{\phi}(P_3)\tilde{\phi}(P_2)\tilde{\phi}(P_1)\rrangle&\notag\\
  &=(2\pi)^2\sum_{(i,j,k)=(2,3,4)(3,2,4)(4,2,3)}\delta^2(P_1+P_i)\llangle\tilde{\phi}(P_k)\tilde{\phi}(P_j)\rrangle\llangle\tilde{\phi}(P_i)\tilde{\phi}(P_1)\rrangle&\notag\\
  &=\frac{(2\pi)^6}{2^{4\Delta_\phi-2}[\Gamma(\Delta_\phi)]^4}(\prod_{j=1}^4P_j^2)^{\frac{\Delta_\phi-1}{2}}[\sum_{j=2,3}\delta^2(P_1+P_j)+\delta^2(P_1+P_4)\Theta(P_{2+})\Theta(P_{2-})]&\label{Four point func for identical scalar operators}
\end{flalign}
So,
\begin{align}
  &W(P_4,P_1|Q)-W(P_4,P_1|-Q)\notag\\
  &=\frac{(2\pi)^6}{2^{4\Delta_\phi-2}[\Gamma(\Delta_\phi)]^4}(P_1^2)^{\Delta_\phi-1}\delta^2(P_1+P_4)(Q^2)^{\Delta_\phi-1}[\Theta(Q_+)\Theta(Q_-)-\Theta(-Q_+)\Theta(-Q_-)].
\end{align}
As this is an odd function of $Q_0$, (\ref{BE for Identical scalar operators}) is actually satisfied.\par
As we substituted the form (\ref{Four point func for identical scalar operators}) to the Ward Identity, we do not get any constraint for CFT data. This is only a check to ensure the validity of the bootstrap equations we have obtained. In general, to get a constraint for CFT data, we should describe the four-point function with OPE coefficients.
\subsubsection{Other test functions for infinite volume}
Let us consider other test functions. We define a step function as follows.
\begin{equation}
  H(x)\equiv
  \begin{cases}
    0& \text{if $x<0$}\\
    \frac{1}{2}& \text{if $x=0$}\\
    1& \text{if $x>0$}
  \end{cases}
\end{equation}
This function has the following representation.
\begin{equation}
H(x)=1-\int_x^\infty\delta(y)\dd y
\end{equation}
In this paper, we define the delta function as having the following properties.
\begin{equation}
  \int_0^\epsilon\delta(x)=\int_{-\epsilon}^0\delta(x)=\frac{1}{2}\hspace{1cm}(0<\epsilon)
\end{equation}
With this delta function, we have
\begin{equation}
  H(x)=1-\int_x^{\infty}\dd y\frac{1}{2\pi}\int_{-\infty}^\infty\dd k e^{iky}=1-\frac{1}{2\pi}\int_{-\infty}^\infty\dd k\int_x^{\infty}\dd ye^{iky}.
\end{equation}
Introduce the convergence parameter $\epsilon$.
\begin{equation}
  H(x)=1-\frac{1}{2\pi}\lim_{\epsilon\rightarrow0}\int_{-\infty}^{\infty}\dd k\int_x^{\infty}\dd ye^{i(k+i\epsilon)y}=1+\frac{1}{2\pi i}\lim_{\epsilon\rightarrow0}\int_{-\infty}^{\infty}\dd k\frac{e^{i(k+i\epsilon)x}}{k+i\epsilon}
\end{equation}
The second term is $-1$ when $x<0$, $-1/2$ when $x=0$, and $0$ when $x>0$. We can check it by complex analysis.
When $x>0$, we take the integral path in the upper semicircle. As the contour surrounds no pole, the second term above is 0.
When $x<0$, we take the integral path in the lower semicircle. In this case, the pole exists at $k=-i\epsilon$, which gives the residue $1$. Since we take the contour clockwise, a negative sign appears, and the second term above is $-1$.
When $x=0$, we can close the contour in either the upper or lower semicircle. In the former, the contour doesn't enclose any poles, so only the integral contribution of the arc portion exists.
In total, the second term above is $(-1)(\pi i)/(2\pi i)=-1/2$. In the latter, the contour encloses a pole at $k=-i\epsilon$. In total, the second term above is $(-1)\{(2\pi i)/(2\pi i)-(\pi i)/(2\pi i)\}=-1/2$. Both calculations yield the same result.\par
\begin{figure}[tb]
  \begin{center}
  \includegraphics[width=8cm]{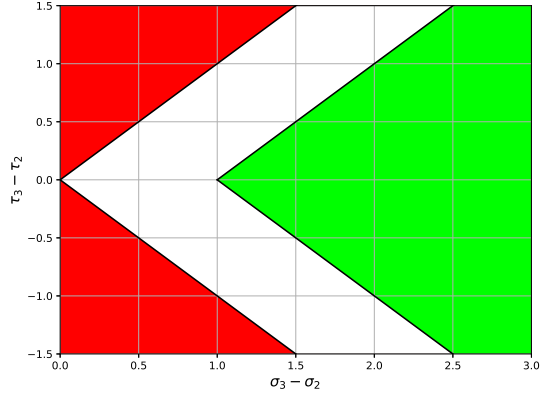}
  \caption{The test function has support at the green region, including the boundary. The red area represents the timelike region.}
  \label{support1}
\end{center}
\end{figure}

With this function, we can formulate various test functions. For example,
\begin{align}
f(x_3-x_2)&=H((\sigma_3-\sigma_2)-(\tau_3-\tau_2)-1)H((\sigma_3-\sigma_2)+(\tau_3-\tau_2)-1)\notag\\
&=\left[1+\frac{1}{2\pi i}\lim_{\epsilon_1\rightarrow0}\int_{-\infty}^\infty\dd k_1\frac{e^{i(k_1+i\epsilon_1)(\sigma_3-\sigma_2-\tau_3+\tau_2-1)}}{k_1+i\epsilon_1}\right]\notag\\
&\hspace{4cm}\left[1+\frac{1}{2\pi i}\lim_{\epsilon_2\rightarrow0}\int_{-\infty}^\infty\dd k_2\frac{e^{i(k_2+i\epsilon_2)(\sigma_3-\sigma_2+\tau_3-\tau_2-1)}}{k_2+i\epsilon_2}\right].
\end{align}
This test function has support for $-(\sigma_3-\sigma_2)+1\leq\tau_3-\tau_2\leq(\sigma_3-\sigma_2)-1$, which is shown in the Fig. \ref{support1}.
With this test function, we get
\begin{flalign}
  &\left(\prod_{j=1}^4\int\dd^2 x_je^{-iP_j\cdot x_j}\right)\bra{0}\OO^{(4)}[\OO^{(3)},\OO^{(2)}]\OO^{(1)}\ket{0}f(x_3-x_2)&\notag\\
  &=\left(\prod_{j=1}^4\int\dd^2 x_je^{-iP_j\cdot x_j}\right)\left[1+\frac{1}{2\pi i}\lim_{\epsilon_1\rightarrow0}\int_{-\infty}^\infty\dd k_1\frac{e^{i(k_1+i\epsilon_1)(\sigma_3-\sigma_2-\tau_3+\tau_2-1)}}{k_1+i\epsilon_1}\right]&\notag\\
  &\hspace{2cm}\left[1+\frac{1}{2\pi i}\lim_{\epsilon_2\rightarrow0}\int_{-\infty}^\infty\dd k_2\frac{e^{i(k_2+i\epsilon_2)(\sigma_3-\sigma_2+\tau_3-\tau_2-1)}}{k_2+i\epsilon_2}\right]\bra{0}\OO^{(4)}[\OO^{(3)},\OO^{(2)}]\OO^{(1)}\ket{0}&\notag\\
  &=\delta^2(P_1+P_2+P_3+P_4)\left[\llangle\tilde{\OO}^{(4)}(P_4)[\tilde{\OO}^{(3)}(P_3),\tilde{\OO}^{(2)}(P_2)]\tilde{\OO}^{(1)}(P_1)\rrangle\right.&\notag\\
  &+\frac{1}{2\pi i}\lim_{\epsilon_1\rightarrow0}\int_{-\infty}^\infty\dd k_1\frac{e^{-i(k_1+i\epsilon_1)}}{k_1+i\epsilon_1}\llangle\tilde{\OO}^{(4)}[\tilde{\OO}^{(3)}(E_3-k_1,p_3-k_1),\tilde{\OO}^{(2)}(E_2+k_1,p_2+k_1)]\tilde{\OO}^{(1)}\rrangle&\notag\\
  &+\frac{1}{2\pi i}\lim_{\epsilon_2\rightarrow0}\int_{-\infty}^\infty\dd k_2\frac{e^{-i(k_2+i\epsilon_2)}}{k_2+i\epsilon_2}\llangle\tilde{\OO}^{(4)}[\tilde{\OO}^{(3)}(E_3+k_2,p_3-k_2),\tilde{\OO}^{(2)}(E_2-k_2,p_2+k_2)]\tilde{\OO}^{(1)}\rrangle&\notag\\
  &+\left(\frac{1}{2\pi i}\right)^2\lim_{\epsilon_1\rightarrow0}\lim_{\epsilon_2\rightarrow0}\int_{-\infty}^\infty\dd k_1\frac{e^{-i(k_1+i\epsilon_1)}}{k_1+i\epsilon_1}\int_{-\infty}^\infty\dd k_2\frac{e^{-i(k_2+i\epsilon_2)}}{k_2+i\epsilon_2}&\notag\\
  &\llangle\tilde{\OO}^{(4)}(P_4)[\tilde{\OO}^{(3)}(E_3-k_1+k_2,p_3-k_1-k_2),\tilde{\OO}^{(2)}(E_2+k_1-k_2,p_2+k_1+k_2)]\tilde{\OO}^{(1)}(P_1)\rrangle].&\label{long}
\end{flalign}
We omitted $\epsilon_1$ and $\epsilon_2$ in the double brackets. Let us consider the four-point function of generalized free field theory. Using the previous result (\ref{Four point func for identical scalar operators}), we get
\begin{align}
  &\delta(E_2+E_3)\delta(p_2+p_3)\left[(\prod_{j=1}^4P_j^2)^{\frac{\Delta_\phi-1}{2}}[\Theta(P_{2+})\Theta(P_{2-})-\Theta(P_{3+})\Theta(P_{3-})]\right.\notag\\
  &+\frac{1}{2\pi i}\lim_{\epsilon_1\rightarrow0}\int_{-\infty}^\infty\dd k_1\frac{e^{-i(k_1+i\epsilon_1)}}{k_1+i\epsilon_1}(P_4^2(P_3^2+2k_1(E_3-p_3))(P_2^2-2k_1(E_2-p_2))P_1^2)^{\frac{\Delta_\phi-1}{2}}\notag\\
  &\hspace{3cm}[\Theta(E_2+p_2+2k_1)\Theta(E_2-p_2)-\Theta(E_3+p_3-2k_1)\Theta(E_3-p_3)]&\notag\\
  &+\frac{1}{2\pi i}\lim_{\epsilon_2\rightarrow0}\int_{-\infty}^\infty\dd k_2\frac{e^{-i(k_2+i\epsilon_2)}}{k_2+i\epsilon_2}(P_4^2(P_3^2-2k_2(E_3+p_3))(P_2^2+2k_2(E_2+p_2))P_1^2)^{\frac{\Delta_\phi-1}{2}}\notag\\
  &\hspace{3cm}[\Theta(E_2+p_2)\Theta(E_2-p_2-2k_2)-\Theta(E_3+p_3)\Theta(E_3-p_3+2k_2)]\notag\\
  &+\left(\frac{1}{2\pi i}\right)^2\lim_{\epsilon_1\rightarrow0}\lim_{\epsilon_2\rightarrow0}\int_{-\infty}^\infty\dd k_1\frac{e^{-i(k_1+i\epsilon_1)}}{k_1+i\epsilon_1}\int_{-\infty}^\infty\dd k_2\frac{e^{-i(k_2+i\epsilon_2)}}{k_2+i\epsilon_2}(P_4^2P_1^2)^{\frac{\Delta_\phi-1}{2}}\notag\\
  &\hspace{1cm}\{P_3^2+2k_1(E_3-p_3)-2k_2(E_3+p_3)+4k_1k_2\}^{\frac{\Delta_\phi-1}{2}}\notag\\
  &\hspace{1cm}\{P_2^2-2k_1(E_2-p_2)+2k_2(E_2+p_2)+4k_1k_2)\}^{\frac{\Delta_\phi-1}{2}}\notag\\
  &\left.\hspace{1cm}[\Theta(E_2+p_2+2k_1)\Theta(E_2-p_2-2k_2)-\Theta(E_3+p_3-2k_1)\Theta(E_3-p_3+2k_2)]\right].
\end{align}
Considering the delta functions, we get
\begin{align}
  &P_2^{2(\Delta_\phi-1)}[\Theta(P_{2+})\Theta(P_{2-})-\Theta(-P_{2+})\Theta(-P_{2-})]\notag\\
  &+\frac{1}{2\pi i}\lim_{\epsilon_1\rightarrow0}\int_{-\infty}^\infty\dd k_1\frac{e^{-i(k_1+i\epsilon_1)}}{k_1+i\epsilon_1}(P_2^2-2k_1(E_2-p_2))^{\Delta_\phi-1}\notag\\
  &\hspace{4cm}[\Theta(E_2+p_2+2k_1)\Theta(E_2-p_2)-\Theta(-E_2-p_2-2k_1)\Theta(-E_2+p_2)]\notag\\
  &+\frac{1}{2\pi i}\lim_{\epsilon_2\rightarrow0}\int_{-\infty}^\infty\dd k_2\frac{e^{-i(k_2+i\epsilon_2)}}{k_2+i\epsilon_2}(P_2^2+2k_2(E_2+p_2))^{\Delta_\phi-1}\notag\\
  &\hspace{4cm}[\Theta(E_2+p_2)\Theta(E_2-p_2-2k_2)-\Theta(-E_2-p_2)\Theta(-E_2+p_2+2k_2)]\notag\\
  &+\left(\frac{1}{2\pi i}\right)^2\lim_{\epsilon_1\rightarrow0}\lim_{\epsilon_2\rightarrow0}\int_{-\infty}^\infty\dd k_1\frac{e^{-i(k_1+i\epsilon_1)}}{k_1+i\epsilon_1}\int_{-\infty}^\infty\dd k_2\frac{e^{-i(k_2+i\epsilon_2)}}{k_2+i\epsilon_2}\notag\\
  &\hspace{1cm}(P_2^2-2k_1(E_2-p_2)+2k_2(E_2+p_2)+4k_1k_2)^{\Delta_\phi-1}\notag\\
  &\hspace{1cm}[\Theta(E_2+p_2+2k_1)\Theta(E_2-p_2-2k_2)-\Theta(-E_2-p_2-2k_1)\Theta(-E_2+p_2+2k_2)].
\end{align}
Taking each integral path in the lower semicircle, we have residues at $k_1=-i\epsilon_1$ and $k_2=-i\epsilon_2$. It vanishes as expected. Though we checked that the exact four-point function in generalized field theory satisfies the bootstrap equation, it can constrain OPE data. We skip the analysis here and leave the problem to future work.
\subsection{Conformal bootstrap in momentum space for finite volume}
\subsubsection{Conformal bootstrap equation}
Consider the following four-point function.
\begin{equation}
\bra{0}\OO^{(4)}(\sigma_4,\tau_4)[\OO^{(3)}(\sigma_3,\tau_3),\OO^{(2)}(\sigma_2,\tau_2)]\OO^{(1)}(\sigma_1,\tau_1)\ket{0}
\end{equation}
$\sigma$ is a coordinate for spatial direction $(-\pi R\leq\sigma<\pi R)$ and $\tau$ is a coordinate for time direction $-\infty<\tau<\infty$. We summarize them as $x=(x^0,x^1)=(\tau,\sigma)$.\par
\begin{figure}[tb]
  \begin{center}
  \includegraphics[width=5cm]{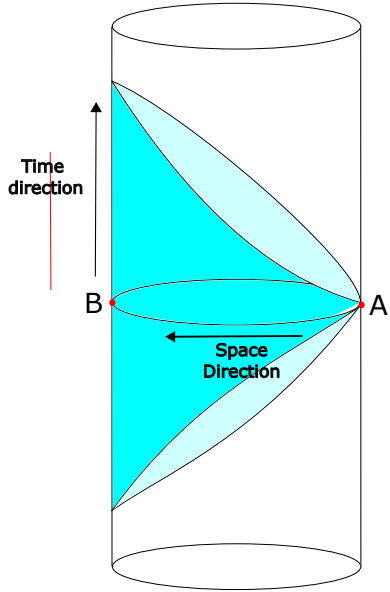}
  \caption{When we set the second operator at A and the third operator at B, they commute because of the microcausality condition.}
  \label{cylinder}
\end{center}
\end{figure}

The microcausality condition says,
\begin{equation}
\bra{0}\OO^{(4)}(\sigma_4,\tau_4)[\OO^{(3)}(\sigma_3,\tau_3),\OO^{(2)}(\sigma_2,\tau_2)]\OO^{(1)}(\sigma_1,\tau_1)\ket{0}=0\hspace{1cm}\text{for $x_3-x_2$ spacelike.}
\end{equation}
As in the case of infinite volume, we adopt the following improved microcausality condition.
\begin{equation}
\bra{0}\OO^{(4)}(\sigma_4,\tau_4)[\OO^{(3)}(\sigma_3,\tau_3),\OO^{(2)}(\sigma_2,\tau_2)]\OO^{(1)}(\sigma_1,\tau_1)\ket{0}f(x_3-x_2)=0
\end{equation}
Here, $f(x)$ has support only for spacelike $x$ (Fig. \ref{cylinder}). For example,
\begin{equation}
f(x_3-x_2)=\delta(\sigma_3-\sigma_2-a)\delta(\tau_3-\tau_2)\hspace{1cm}(0<a<2\pi R).
\end{equation}
For this test function, we get
\begin{align}
  &\left(\prod_{j=1}^4\int\dd x_je^{-iP_j\cdot x_j}\right)\bra{0}\OO^{(4)}[\OO^{(3)},\OO^{(2)}]\OO^{(1)}\ket{0}\delta(\sigma_3-\sigma_2-a)\delta(\tau_3-\tau_2)\notag\\
  &\propto\int\dd^2x_2\dd^2x_3e^{-i(P_2\cdot x_2+P_3\cdot x_3)}\int\dd k_0e^{-ik_0(\tau_3-\tau_2)}\notag\\
  &\hspace{6.5cm}\sum_{k_1=\frac{\mathbb{Z}}{R}}e^{ik_1(\sigma_3-\sigma_2-a)}\bra{0}\tilde{\OO}^{(4)}[\OO^{(3)},\OO^{(2)}]\tilde{\OO}^{(1)}\ket{0}\notag\\
  &\propto\int\dd k_0\left(\prod_{j=2,3}\int\dd x_j^2\right)\sum_{k_1=\frac{\mathbb{Z}}{R}}e^{-i[ak_1+(p_2+k_1)\sigma_2+(p_3-k_1)\sigma_3-(E_2+k_0)\tau_2-(E_3-k_0)\tau_3]}\notag\\
  &\hspace{9.4cm}\bra{0}\tilde{\OO}^{(4)}[\OO^{(3)},\OO^{(2)}]\tilde{\OO}^{(1)}\ket{0}\notag\\
  &=\delta^2(\sum_iP_i)\sum_{k_0,k_1}e^{-iak_1}\llangle\tilde{\OO}^{(4)}(P_4)\tilde{\OO}^{(3)}\left(E_3-k_0,p_3-k_1\right),\tilde{\OO}^{(2)}\left(E_2+k_0,p_2+k_1\right)]\tilde{\OO}^{(1)}(P_1)\rrangle\notag\\
  &=\delta^2(\sum_iP_i)\sum_{k_0,k_1}e^{-ia k_1}\llangle\tilde{\OO}^{(4)}(P_4)[\tilde{\OO}^{(3)}(P_3-k),\tilde{\OO}^{(2)}(P_2+k)]\tilde{\OO}^{(1)}(P_1)\rrangle\notag\\
  &\propto\delta^2(\sum_iP_i)\sum_{Q_0,Q_1}e^{-ia Q_1}\llangle\tilde{\OO}^{(4)}(P_4)[\tilde{\OO}^{(3)}(-Q-\tfrac{P_1+P_4}{2}),\tilde{\OO}^{(2)}(Q-\tfrac{P_1+P_4}{2})]\tilde{\OO}^{(1)}(P_1)\rrangle\notag\\
  &\equiv\delta^2(\sum_iP_i)\sum_{Q_0,Q_1}e^{-iaQ_1}[W(P_4,P_1|Q)-W(P_4,P_1|-Q)]\notag\\
  &=0.\label{finite BE for Identical operators}
\end{align}
The last equation does not automatically hold, and this gives non-trivial information as the bootstrap equation.
We defined $Q=(Q_0,Q_1)\equiv(P_2-P_3+2k)/2$. The value of $Q_0$ and $Q_1$ depends on the theory and the boundary conditions of “in-state” and “out-state.” Let us see the situation next.
\subsubsection{Completeness relation}
We need to insert a completeness relation between the second and the third operators to calculate the four-point function. In previous papers, it is constructed in momentum space in an infinite volume \cite{Gillioz:2019iye,Gillioz:2016jnn,Gillioz:2018kwh,Karateev:2018oml}
\begin{align}
  \id=\ket{0}\bra{0}+\sum_{\psi\neq\id}\int\frac{\dd^dp}{(2\pi)^d}\Theta(p^0)\Theta(p^2)\frac{\ket{\tilde{\psi}(p)}\bra{\tilde{\psi}(p)}}{\llangle\tilde{\psi}(-p)\tilde{\psi}(p)\rrangle},\label{Completeness relation in infinite volume}
\end{align}
where
\begin{align}
  \ket{\tilde{\psi}(p)}=\int\dd^dxe^{-ip\cdot x}\psi(x)\ket{0},\hspace{1cm}\bra{\tilde{\psi}(p)}=\bra{\tilde{\psi}(p)}^\dagger=\bra{0}\tilde{\psi}(-p).
\end{align}
The point is that in higher dimensional CFT, we only have to consider the contributions from primaries because we can make all descendants acting $P_\mu$ on the primary state, which are eigenstates of the momentum operator. So, only primary states can expand the completeness relation in momentum space.\par
On the other hand, in two-dimensional CFT, we also have to consider Virasoro descendants because they include $L_{-m}\ (m\geq2)$ in general.\par
So, our goal is to construct completeness relations from the primary state and Virasoro descendants. Moreover, we need to study the summation for momentum because it is discrete in finite volume. The integral for momentum in (\ref{Completeness relation in infinite volume}) is replaced by $\sum_p$.\par
As correlation functions in two-dimensional CFT are factorized into the holomorphic and antiholomorphic parts, we only have to consider the completeness relation for the former.
First, make a linear combination of almost orthogonal states with coefficients.
\begin{align}
  &\sum_{J=h+\mathbb{Z}_{\geq0}}\sum_{[\{k\}],[\{k'\}]}C^{[\{k\}][\{k'\}]}_J\ket{L_{\{k\}}\phi_{[J]}}\bra{L_{\{k'\}}\phi_{[J]}}\notag\\
  &=C^{[0][0]}_{h}\dyad{\phi_{[h]}}+C^{[0][0]}_{h+1}\dyad{\phi_{[h+1]}}\notag\\
  &\hspace{1cm}+C^{[0][0]}_{h+2}\dyad{\phi_{[h+2]}}+C^{[2][2]}_{h+2}\dyad{L_{-2}\phi_{[h+2]}}\notag\\
  &\hspace{2cm}+C^{[2][0]}_{h+2}\dyad{L_{-2}\phi_{[h+2]}}{\phi_{[h+2]}}+C^{[0][2]}_{h+2}\dyad{\phi_{[h+2]}}{L_{-2}\phi_{[h+2]}}+\cdots
\end{align}
Here, $\{k\}=k_i,\cdots,k_1\ (k_i\geq\cdots\geq k_1)$ means that the Virasoro operators act in the form $L_{-k_i}\cdots L_{-k_1}\ket{h}$. Please note that $L_{-1}$ does not appear in the above expression. They are absorbed in the basis $\dyad{\phi_{[J]}}$ as $L_{-1}$ corresponds to a momentum operator.\par
Let us see how we can determine the coefficients $C^{[\{k\}][\{k'\}]}_J$ from the demand that it should correspond to a completeness relation.
Before that, we need some preparation.
\begin{align}
  &\braket{\phi_{[J]}}=\frac{1}{(J-h)!^2}\bra{h}L_1^{J-h}L_{-1}^{J-h}\ket{h}\\
  &\braket{L_{-m_k}\cdots L_{-m_1}\phi_{[J]}}{\phi_{[J]}}=\frac{i^{\sum_{j=1}^km_j}}{(J-h-\sum_{i=1}^km_i)!(J-h)!}\bra{h}L_{m_1}\cdots L_{m_k}L_1^{J-h-\sum_{j=1}^km_j}L_{-1}^{J-h}\ket{h}\\
  &\braket{\phi_{[J]}}{L_{-m_k}\cdots L_{-m_1}\phi_{[J]}}=\frac{(-i)^{\sum_{j=1}^km_j}}{(J-h)!(J-h-\sum_{j=1}^km_j)!}\bra{h}L_1^{J-h}L_{-1}^{J-h-\sum_{j=1}^km_j}L_{-m_k}\cdots L_{-m_1}\ket{h}\\
  &\braket{L_{-m_k}\cdots L_{-m_1}\phi_{[J]}}{L_{-n_l}\cdots L_{-n_1}\phi_{[J]}}\notag\\
  &=\frac{i^{\sum_{j=1}^km_j}(-i)^{\sum_{j=1}^ln_j}}{(J-h-\sum_{j=1}^km_j)!(J-h-\sum_{j=1}^ln_j)!}\bra{h}L_{m_1}\cdots L_{m_k}L_1^{J-h-\sum_{j=1}^km_j}L_{-1}^{J-h-\sum_{j=1}^ln_j}L_{-n_l}\cdots L_{-n_1}\ket{h}
\end{align}
Now we have everything, let us start from the level $J=h$. The demand that the above expression should be the completeness relation means that the expression above is identity. So, multiplying $\bra{\phi_{[h]}}$ from the left and $\ket{\phi_{[h]}}$ from the right, we get
\begin{align}
  \braket{\phi_{[h]}}=C_h^{[0][0]}\braket{\phi_{[h]}}^2.
\end{align}
For $h\neq0$, it gives
\begin{align}
  C_h^{[0][0]}=1.
\end{align}
Next, consider the level $J=h+1$. Sandwich the above expression between $\bra{\phi_{[h+1]}}$ and $\ket{\phi_{[h+1]}}$.
\begin{align}
  \braket{\phi_{[h+1]}}=C_{h+1}^{[0][0]}\braket{\phi_{[h+1]}}^2
\end{align}
It gives
\begin{align}
  C_{h+1}^{[0][0]}=\frac{1}{2h}.
\end{align}
When the primary operator is identity, the level $J=h+1=1$ does not appear because the vacuum is invariant under the action of $L_{-1}$.\par
Nontriviality appears from the next level $J=h+2$. We have four parameters $C_{h+2}^{[0][0]}$, $C_{h+2}^{[2][2]}$, $C_{h+2}^{[2][0]}$, $C_{h+2}^{[0][2]}$ and four equations obtained by sandwiching it between $\bra{\phi_{[h+2]}}$ or $\bra{L_{-2}\phi_{[h+2]}}$ and $\ket{\phi_{[h+2]}}$ or $\ket{L_{-2}\phi_{[h+2]}}$. We get the following four equations.
\begin{align}
  &h(2h+1)=h^2(2h+1)^2C_{h+2}^{[0][0]}-3h^2(2h+1)C_{h+2}^{[2][0]}-3h^2(2h+1)C_{h+2}^{[0][2]}+(3h)^2C_{h+2}^{[2][2]}\\
  &4h+\tfrac{c}{2}=(3h)^2C_{h+2}^{[0][0]}-3h(4h+\tfrac{c}{2})C_{h+2}^{[2][0]}-3h(4h+\tfrac{c}{2})C_{h+2}^{[0][2]}+(4h+\tfrac{c}{2})^2C_{h+2}^{[2][2]}\\
  &-3h=-3h^2(2h+1)C_{h+2}^{[0][0]}+(4h+\tfrac{c}{2})h(2h+1)C_{h+2}^{[2][0]}+(3h)^2C_{h+2}^{[0][2]}-3h(4h+\tfrac{c}{2})C_{h+2}^{[2][2]}\\
  &-3h=-3h^2(2h+1)C_{h+2}^{[0][0]}+(3h)^2C_{h+2}^{[2][0]}+(4h+\tfrac{c}{2})h(2h+1)C_{h+2}^{[0][2]}-3h(4h+\tfrac{c}{2})C_{h+2}^{[2][2]}
\end{align}
They only sometimes have a solution. For example, when $h=1/2,\ c=1/2$, they become the same equation.
\begin{align}
  1=C_{h+2}^{[0][0]}-\frac{3}{2}C_{h+2}^{[2][0]}-\frac{3}{2}C_{h+2}^{[0][2]}+\frac{9}{4}C_{h+2}^{[2][2]}
\end{align}
It is all constraint for them so that we can set $C_{h+2}^{[0][0]}=1, C_{h+2}^{[0][2]}=C_{h+2}^{[2][0]}=C_{h+2}^{[2][2]}=0$.\par
This corresponds to a energy operator $\epsilon$ in Ising model ($\mathcal{M}_3,\ c=1/2$). The fact that the above four equations are not linearly independent corresponds to the fact that the Kac determinant \cite{Kac:1601455} vanishes at level 2 for $h=1/2,\ c=1/2$.\par
Next, consider the case $h=3/2,\ c=7/10$. It is one of the operators in the Tricritical Ising model \cite{Zamolodchikov:1986db}. We get $C_{h+2}^{[0][0]}=127/357$, $C_{h+2}^{[2][0]}=30/119$, $C_{h+2}^{[0][2]}=30/119$, $C_{h+2}^{[2][2]}=40/119$. Of course, we can get eigenvectors, but it is not so useful for calculating the four-point function.
\begin{multline}
  \id=\cdots+\left[\frac{20}{119}\left(1+\frac{128}{\sqrt{32449}}\right)\right]^2\left|\frac{7+\sqrt{32449}}{180}\ket{\phi_{[h+2]}}+\ket{L_{-2}\phi_{[h+2]}}\right|^2\\
  +\left[\frac{20}{119}\left(1-\frac{128}{\sqrt{32449}}\right)\right]^2\left|\frac{7-\sqrt{32449}}{180}\ket{\phi_{[h+2]}}-\ket{L_{-2}\phi_{[h+2]}}\right|^2+\cdots
\end{multline}
Let us summarize the result. We can construct the completeness relation by determining each coefficient $C_{J}^{[\{k\}][\{k'\}]}$ by sandwiching it between in-states and out-states. As well known, the number of linearly independent states is determined by the number of a singular vector. So we need to study the Kac determinant to construct the completeness relation.\par
We can write the four-point function as
\begin{align}
  \langle\OO^{(4)}_{[J_4]}\OO^{(3)}_{[J_3]}\OO^{(2)}_{[J_2]}\OO^{(2)}_{[J_1]}\rangle=\sum_{\OO^{(5)}}\frac{\langle\OO^{(4)}_{[J_4]}\OO^{(3)}_{[J_3]}\OO^{(5)}_{[J_5]}\rangle\langle\OO^{(5)}_{[-J_5]}\OO^{(4)}_{[J_2]}\OO^{(4)}_{[J_1]}\rangle}{\langle\OO^{(5)}_{[-J_5]}\OO^{(5)}_{[J_5]}\rangle}
\end{align}
where $J_5=J_1+J_2=-J_3-J_4$ and $\ket{\OO^{(5)}_{[J_5]}}$ is an orthogonal states at the level $J_5$.\par
In the bootstrap calculation, we start by considering intermediate states at $J=0$ and increase the number of $J$. By viewing the intermediate states at a higher level, we can improve the accuracy of constraint for CFT.
\subsubsection{Nontriviality}
The question is whether or not the bootstrap equation (\ref{finite BE for Identical operators}) can restrict the OPE data for some CFT. In other words, is (\ref{finite BE for Identical operators}) satisfied regardless of the details of OPE data? Let us consider this issue.\par
As an example, consider the four-point function of the identical operator $\phi$ with conformal weight $(h_\phi,\tilde{h}_\phi)=((\Delta_\phi+s_\phi)/2, (\Delta_\phi-s_\phi)/2)$. When we compute the four-point function, many contributions from conformal families are inserted between the second and the third operators. We call the contribution from the OPE $\OO\in\phi\times\phi$ “$\OO$ conformal block.” And we write the $\OO$ conformal block in $W(P_4,P_1|Q)$ as $W_\OO(P_4,P_1|Q)$. \par
The $\OO$ conformal block includes all contributions from the conformal family of $\OO$. So, there are Virasoro descendants in the intermediate state. We call the contribution from the intermediate state $\dyad{\OO_{[J]}}$ “$\OO$ primary conformal block” and write it as $W_{\OO}^P(P_4,P_1|Q)$. We call the contribution from other intermediate states such as $\dyad{L_{-m_k}\cdots\OO_{[J]}}$ “$\OO$ Virasoro descendants conformal block” and write it as $W_{\OO}^V(P_4,P_1|Q)$. As $\dyad{\OO_{[J]}}$ and $\dyad{L_{-m_k}\cdots L_{-m_1}\OO_{[J]}}$ are not orthogonal basis at level $J$, $W_{\OO}^P(P_4,P_1|Q)+W_{\OO}^V(P_4,P_1|Q)\neq W_{\OO}(P_4,P_1|Q)$, in general.\par
If the sum of the difference of $W_\OO$ is 0 for each $\OO$ contribution, the bootstrap equation is trivially satisfied.
\begin{align}
  &\text{Bootstrap equation is trivially satisfied.}\notag\\
  &\Leftrightarrow \sum_{Q_0,Q_1}e^{-iaQ_1}[W_\OO(P_4,P_1|Q)-W_\OO(P_4,P_1|-Q)]=0 \text{\ \ for each $\OO\in\phi\times\phi$}\notag
\end{align}
We focus on the difference of $\OO$ primary conformal block to see whether it is true.
\begin{equation}
  \sum_{Q_0,Q_1}e^{-iaQ_1}[W_\OO^P(P_4,P_1|Q)-W_\OO^P(P_4,P_1|-Q)]=0 \text{\ \ is satisfied or not?}\notag
\end{equation}
Though there are many other contributions from $\OO$ conformal block, to check whether this is correct or not is enough because we can give the same logic for other contributions, including Virasoro descendants. We can write $\OO$ primary conformal block as follows. $\OO$ has conformal weight $(h_\OO,\tilde{h}_\OO)=((\Delta_\OO+s_\OO)/2,(\Delta_\OO-s_\OO)/2)$.
\begin{equation}
  W_\OO^P(P_4,P_1|Q)=C(h,c)
  \llangle\tilde{\phi}\fft{\PP_4}{\EE_4}\tilde{\phi}\fft{\PP_3}{\EE_3}\tilde{\OO}\fft{\PP_5}{\EE_5}\rrangle\llangle\tilde{\OO}\fft{-\PP_5}{-\EE_5}\tilde{\phi}\fft{\PP_2}{\EE_2}\tilde{\phi}\fft{\PP_1}{\EE_1}\rrangle
\end{equation}
The notation is defined by (\ref{def of new mode}).
The coefficient $C(h,c)$ is determined by the conformal weight of $\OO$ and the central charge. It has a nontrivial configuration when some nonnegative integers $n_L,\ n_R,\ n'_L,\ n'_R,\ N_L,\ N_R$ exist such that the following relations hold.
\begin{align}
  \PP_1&=n_L-n_R+s_\phi, & \EE_1&=n_L+n_R+\Delta_\phi\\
  \PP_4&=-(n'_L-n'_R+s_\phi), & \EE_4&=-(n'_L+n'_R+\Delta_\phi)\\
  \PP_5&=N_L-N_R+s_\OO, & \EE_5&=N_L+N_R+\Delta_\OO\\
  \PP_2&=-n_L+n_R+N_L-N_R-s_\phi+s_\OO, & \EE_2&=-n_L-n_R+N_L+N_R-\Delta_\phi+\Delta_\OO\\
  \PP_3&=n'_L-n'_R-N_L+N_R+s_\phi-s_\OO, & \EE_3&=n'_L+n'_R-N_L-N_R+\Delta_\phi-\Delta_\OO
\end{align}
We can rewrite the last four equations as
\begin{align}
  \QQ_0&\equiv\frac{\EE_2-\EE_3}{2}=-\frac{n_L+n_R+n'_L+n'_R}{2}+N_L+N_R-\Delta_\phi+\Delta_\OO\\
  \QQ_1&\equiv\frac{\PP_2-\PP_3}{2}=-\frac{n_L-n_R+n'_L-n'_R}{2}+N_L-N_R-s_\phi+s_\OO.
\end{align}
$\PP_2+\PP_3$ and $\EE_2+\EE_3$ equations are satisfied if $\PP_1,\ \EE_1,\ \PP_4,\ \EE_4$ equations are confident because of the conservation law of energy and momentum.\par
We ignore the condition as we set $\PP_1,\ \EE_1,\ \PP_4,\ \EE_4$ conditions as “boundary” conditions for each bootstrap equation. And $\PP_5,\ \EE_5$ conditions are satisfied if $\QQ_0,\ \QQ_1$ conditions are satisfied. So, we only concentrate on $\QQ_0,\ \QQ_1$ conditions.
\begin{screen}
  Under the proper boundary conditions, $W_\OO^P(P_4,P_1|Q)$ has nontrivial configuration when there are nonnegative integers $N_L$ and $N_R$ such that
  \begin{align}
    \QQ_0&\equiv\frac{\EE_2-\EE_3}{2}=-\frac{n_L+n_R+n'_L+n'_R}{2}+N_L+N_R-\Delta_\phi+\Delta_\OO=\frac{\EE_4-\EE_1}{2}+N_L+N_R+\Delta_\OO\\
    \QQ_1&\equiv\frac{\PP_2-\PP_3}{2}=-\frac{n_L-n_R+n'_L-n'_R}{2}+N_L-N_R-s_\phi+s_\OO=\frac{\PP_4-\PP_1}{2}+N_L-N_R+s_\OO.
  \end{align}
\end{screen}
How about $W_\OO^P(P_4,P_1|-Q)$?
\begin{screen}
  Under the proper boundary conditions, $W_\OO^P(P_4,P_1|-Q)$ has nontrivial configuration when there are nonnegative integers $N'_L$ and $N'_R$ such that
  \begin{align}
    -\QQ_0&\equiv\frac{\EE_3-\EE_2}{2}=-\frac{n_L+n_R+n'_L+n'_R}{2}+N'_L+N'_R-\Delta_\phi+\Delta_\OO=\frac{\EE_4-\EE_1}{2}+N'_L+N'_R+\Delta_\OO\\
    -\QQ_1&\equiv\frac{\PP_3-\PP_2}{2}=-\frac{n_L-n_R+n'_L-n'_R}{2}+N'_L-N'_R-s_\phi+s_\OO=\frac{\PP_4-\PP_1}{2}+N'_L-N'_R+s_\OO.
  \end{align}
\end{screen}
So, the contribution from the $\OO$ primary conformal block in the equation is
\begin{multline}
  \sum_{N_L,N_R\geq0}[e^{-ia(\frac{\PP_4-\PP_1}{2}+N_L-N_R+s_\OO)}-e^{ia(\frac{\PP_4-\PP_1}{2}+N_L-N_R+s_\OO)}]W_\OO^P(P_4,P_1|N_L,N_R)\\
  =-2i\sum_{N_L,N_R\geq0}\sin\left[a\left(\frac{\PP_4-\PP_1}{2}+N_L-N_R+s_\OO\right)\right]W_\OO^P(P_4,P_1|N_L,N_R).\label{BE for primaries}
\end{multline}
As $Q$ is characterized by nonnegative integers $N_L$ and $N_R$, we write $W_\OO^P(P_4,P_1|N_L,N_R)$ instead of $W_\OO^P(P_4,P_1|Q)$.\par
It does not vanish in general, meaning the bootstrap equation is nontrivial. To summarize, the whole bootstrap equation is, for any $a\ (0<a<2\pi)$,
\begin{equation}
\sum_{Q_0,Q_1}e^{-iaQ_1}[W_\OO(P_4,P_1|Q)-W_\OO(P_4,P_1|-Q)]=0.
\end{equation}
\subsubsection{Comments on contributions from Virasoro descendants}
Next, consider contributions from intermediate states other than the $\OO$ primary conformal block. First, the three-point function of two chiral operators $\phi(z_3),\ \phi(z_2)$ with conformal weight $h_2=h_3=h_\phi$ and $L_{-n}\OO(z_1)$ with conformal weight $h_1=h_\OO+n$ is
\begin{equation}
  \langle\phi(z_3)\phi(z_2)(L_{-n}\OO)(z_1)\rangle=\sum_{j=2,3}\left[\frac{(n-1)h_\phi}{(z_j-z_1)^n}-\frac{1}{(z_j-z_1)^{n-1}}\pdv{z_j}\right]\langle\phi(z_3)\phi(z_2)\OO(z_1)\rangle.
\end{equation}
After acting the differential operator on the three-point function $\langle\phi(z_3)\phi(z_2)\OO(z_1)\rangle$, we perform integration by $z_1,z_2,z_3$. Each term in the differential operators acts as if they increase the conformal weights of the three operators. We can calculate the three-point function with one Virasoro descendant by considering the three-point function of primaries with changed conformal weights. For example,
\begin{align}
  \frac{(n-1)h_\phi}{(z_3-z_1)^n}\langle\phi(z_3)\phi(z_2)\OO(z_1)\rangle=\frac{(n-1)h_\phi}{(z_1-z_2)^{h_1+h_2-h_3}(z_2-z_3)^{h_2+h_3-h_1}(z_3-z_1)^{h_3+h_1-h_2+n}}.
\end{align}
So, it changes the conformal weight as follows.
\begin{align}
  h_1&=h_\OO\rightarrow h'_1=h_1+\frac{n}{2}=h_\OO+\frac{n}{2}\\
  h_2&=h_\phi\rightarrow h'_2=h_2=h_\phi\\
  h_3&=h_\phi\rightarrow h'_3=h_3+\frac{n}{2}=h_\phi+\frac{n}{2}
\end{align}
The differential operator $-(z_3-z_1)^{1-n}\partial_{z_3}$ acts on $(z_2-z_3)^{-h_2-h_3+h_1}$ and $(z_3-z_1)^{-h_3-h_1+h_2}$. In the former,
\begin{align}
  h_1&=h_\OO\rightarrow h'_1=h_1+\frac{n-1}{2}=h_\OO+\frac{n-1}{2}\\
  h_2&=h_\phi\rightarrow h'_2=h_2+\frac{1}{2}=h_\phi+\frac{1}{2}\\
  h_3&=h_\phi\rightarrow h'_3=h_3+\frac{n}{2}=h_\phi+\frac{n}{2}.
\end{align}
In the latter,
\begin{align}
  h_1&=h_\OO\rightarrow h'_1=h_1+\frac{n}{2}=h_\OO+\frac{n}{2}\\
  h_2&=h_\phi\rightarrow h'_2=h_2=h_\phi\\
  h_3&=h_\phi\rightarrow h'_3=h_3+\frac{n}{2}=h_\phi+\frac{n}{2}.
\end{align}
So, there are four ways to change the conformal weights of operators.
\begin{equation}
  (h_1,h_2,h_3)\rightarrow(h'_1,h'_2,h'_3)=
  \begin{cases}
    &(h_1+\frac{n}{2},h_2+\frac{n}{2},h_3)\\
    &(h_1+\frac{n}{2},h_2,h_3+\frac{n}{2})\\
    &(h_1+\frac{n-1}{2},h_2+\frac{n}{2},h_3+\frac{1}{2})\\
    &(h_1+\frac{n-1}{2},h_2+\frac{1}{2},h_3+\frac{n}{2})
  \end{cases}
\end{equation}
The three-point function vanishes for a fixed $J_3$ if $h_3$ becomes larger than $-J_3$. We can use those facts to calculate contributions from intermediate states other than the $\OO$ primary conformal block.
\subsubsection{Formalism of bootstrap equation without Virasoro descendants}
In this section, we summarize the formalism of the bootstrap equation without considering the contribution of Virasoro descendants. It tells us some information about CFT data.\par
Let us consider the four-point function of identical operators $\Phi$ without spin $(h,\bar{h})=(h_L,h_R)$. Of course, there are many contributions from various intermediate states, but here, we only consider the contributions from primary scalar states and their descendants. We neglect Virasoro descendants.\par
The intermediate state can be described by inserting primary operator $\Psi$ with conformal weight $(h,\bar{h})=(H_L,H_R)$. So, the contribution from $\Psi$ is
\begin{multline}
  \frac{\llangle\Phi\fft{\PP_4}{\EE_4}\Phi\fft{\PP_3}{\EE_3}\Psi\fft{\PP_5}{\EE_5}\rrangle\llangle\Psi\fft{-\PP_5}{-\EE_5}\Phi\fft{\PP_2}{\EE_2}\Phi\fft{\PP_1}{\EE_1}\rrangle}{\llangle\Psi\fft{-\PP_5}{-\EE_5}\Psi\fft{\PP_5}{\EE_5}\rrangle}\\
  =\frac{\llangle\phi\fft{J^{(+)}_4}{J^{(+)}_4}\phi\fft{J^{(+)}_3}{J^{(+)}_3}\psi\fft{J^{(+)}_5}{J^{(+)}_5}\rrangle\llangle\psi\fft{-J^{(+)}_5}{-J^{(+)}_5}\phi\fft{J^{(+)}_2}{J^{(+)}_2}\phi\fft{J^{(+)}_1}{J^{(+)}_1}\rrangle}{\llangle\psi\fft{-J^{(+)}_5}{-J^{(+)}_5}\psi\fft{J^{(+)}_5}{J^{(+)}_5}\rrangle}\times(+\rightarrow-)\\
  =\frac{\KK^{(h_L,h_L,H_L)}_{\mathrm{red}}(N_L,n'_L)\KK^{(H_L,h_L,h_L)}_{\mathrm{red}}(n_L,N_L)}{\frac{\Gamma(N_L+2H_L)}{\Gamma(N_L+1)\Gamma(2H_L)}}\times(L\rightarrow R).
\end{multline}
We defined the factorization as $\Phi(z,\bar{z})\equiv\phi(z)\bar{\phi}(\bar{z})$ and  $\Psi(z,\bar{z})\equiv\psi(z)\bar{\psi}(\bar{z})$. Remember that we can calculate the three-point function in the numerator as follows.
\begin{multline}
  \KK^{(h_3,h_2,h_1)}_{\mathrm{red}}(n,n')\\
  \equiv\tfrac{1}{\Gamma(b_1)\Gamma(b_2)\Gamma(b_3)}\sum_{q=0}^{\text{Min}\{n,n'\}}\tfrac{\Gamma(b_3+q+\text{Max}\{0,n-n'\})\Gamma(b_2-q+\text{Min}\{n,n'\})\Gamma(b_1+q+\text{Max}\{n'-n,0\})}{\Gamma(1+q+\text{Max}\{0,n-n'\})\Gamma(1-q+\text{Min}\{n,n'\})\Gamma(1+q+\text{Max}\{n'-n,0\})}
\end{multline}
where $b_i\equiv h_1+h_2+h_3-2h_i$.\par
By determining the boundary condition $(n_L,n_R,n'_L,n'_R)$, we get bootstrap equation. \\
(1)$\ (n_L,n_R,n'_L,n'_R)=(0,0,0,0)$\par
It does not give any nontrivial bootstrap equation because the phase factor in (\ref{BE for primaries}) is antisymmetric for $N_L$ and $N_R$ though $W^P_{\Psi}(P_4, P_1|N_L, N_R)$ is symmetric.\\
(2)$\ (n_L,n_R,n'_L,n'_R)=(1,0,0,0)$\par
\begin{figure}[tb]
  \begin{center}
  \includegraphics[width=13cm]{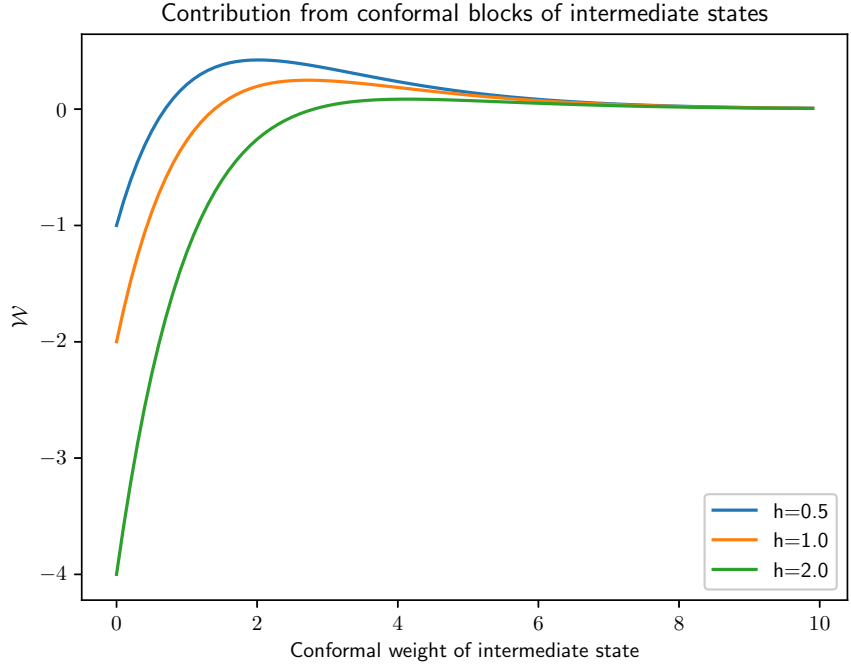}
  \caption{The horizontal axis represents the conformal weight of the intermediate state. We calculated them for the four-point function of identical scalar operators with $h=0.5,1.0,2.0$. We can see that they give negative contributions for the intermediate state with a small conformal weight, and on the other hand, they provide positive contributions above some conformal weight.}
  \label{ConformalBlock}
\end{center}
\end{figure}

It gives us the first nontrivial bootstrap equation.
\begin{equation}
  0=-2h_L(\lambda^{\Phi\Phi}_{\id})^2\sin\tfrac{a}{2}+\sum_{\Psi}(\lambda^{\Phi\Phi}_{\Psi})^2\sum_{N_L,N_R\geq0}\sin[a(-\tfrac{1}{2}+N_L-N_R)]W_\Psi(N_L,N_R)
\end{equation}
where
\begin{multline}
  W_\Psi(N_L,N_R)=\frac{[H_L(2h_L-H_L+1)+2h_L(N_L-1)]\Gamma(2H_L)\Gamma(H_L+N_L)\Gamma(H_L+N_L-1)}{(\Gamma(H_L))^2\Gamma(N_L+2H_L)\Gamma(N_L+1)}\\
  \frac{\Gamma(2H_R)(\Gamma(H_R+N_R))^2}{(\Gamma(H_R))^2\Gamma(N_R+2H_R)\Gamma(N_R+1)}.
\end{multline}
For example, when $a=\pi$,
\begin{align}
  0=-2h_L(\lambda^{\Phi\Phi}_{\id})^2-\sum_{\Psi}(\lambda^{\Phi\Phi}_{\Psi})^2\sum_{N_L,N_R\geq0}(-1)^{N_L+N_R}W_\Psi(N_L,N_R).
\end{align}
We can also write it by using the hypergeometric function.
\begin{align}
  0=(\lambda^{\Phi\Phi}_{\id})^2\mathcal{W}_{\id}+\sum_{\Psi}(\lambda^{\Phi\Phi}_{\Psi})^2\mathcal{W}_\Psi
\end{align}
where $\mathcal{W}_{\id}=-2h_L$ and
\begin{multline}
\mathcal{W}_\Psi={}_2F_1(H_R,H_R,2H_R,-1)[h_L{}_2F_1(H_L,H_L+1,2H_L+1,-1)\\
+(H_L-2h_L){}_2F_1(H_L-1,H_L,2H_L,-1)].
\end{multline}
We can calculate the conformal block for each intermediate state (Fig. \ref{ConformalBlock}). It includes all contributions from each conformal family created by the primary scalar operator and the action of $L_{-1}$, but here we neglect the contributions from Virasoro descendants.

\section{Conclusions and Discussion}
\paragraph{Summary}
In this paper, we have done the following:
\begin{itemize}
  \item We formulated one of the conformal bootstrap equations in momentum space representation at finite volume. We dealt with two-dimensional and three-dimensional CFT, but this approach applies to CFTs with $d>3$.
  \item In two-dimensional CFT, factorization helps calculate momentum space's two- and three-point functions. We explained three methods to calculate them, direct integral calculation, algebraic calculation, and WI method, though they are equivalent in principle. The most valuable and fascinating approach is the WI method since this applies to the Wightman function of operators with general conformal weight and the higher-dimensional CFTs.
  \item On the other hand, three-dimensional CFT is a little tricky to handle. One of the reasons is that the structure of the product of complete orthogonal basis, spherical harmonics $\{Y_{l,m}\}$, in three-dimensional CFT ($S^2\times\mathbb{R}$) is more complicated than that, $\{e^{-in\theta}\}$, in two-dimensional CFT ($S^1\times\mathbb{R}$). Because of the reason, it was too difficult to find a general solution for WIs. We have to solve the differential equations from (\ref{ODE1}) to (\ref{ODE6}) term by term with the computer.
  \item Taking large volume limit, we proved that the two-point function and the three-point function we got at finite volume are consistent with the previous result \cite{Gillioz:2018mto,Gillioz:2019iye,Gillioz:2019lgs,Gillioz:2021sce} at infinite volume.
  \item The main ingredient of this paper is that we explicitly constructed conformal blocks in two-dimensional CFT and showed how we could introduce bootstrap equations in momentum space. We applied the bootstrap equation \cite{Gillioz:2019iye} obtained from microcausality condition in infinite volume to finite volume case. One of the exciting properties of our formula is that we only have to take a discrete sum because of the quantization of energy and momentum. It helps us calculate conformal blocks analytically and numerically. We proved that our bootstrap equation is not satisfied trivially, which means it could give us a constraint for CFT data.
\end{itemize}
\paragraph{Remaining challenges}
We have some remaining challenges to be solved in the future.
\begin{itemize}
  \item In two-dimensional CFT, the problem is how, to sum up the contributions from Virasoro descendants. There is no exact result for summing them, so we have to calculate them by computer.
  \item In three-dimensional CFT, we have technical problems. The most challenging issue to be solved is we do not know the exact form of the three-point functions. We can calculate them term by term with recursion relations obtained from WIs, but it needs calculation by computer. If we could find the exact form, it would make it easier to calculate conformal blocks in three-dimensional CFT.
  \item In three-dimensional CFT, in this paper, we only dealt with scalar primaries because it is easy to handle. We have to consider various primaries with spin, but we can get them similarly as we did for scalar primaries. The generalization of our result in this paper is one of the remaining challenges, though it is not so hard.
\end{itemize}
\paragraph{Future research direction}
We formulated the basis for the conformal bootstrap equation in momentum space at finite volume. We end our paper with a discussion of future research directions.
\begin{itemize}
\item One of the future approaches is to consider better test functions. In this paper, we chose the delta function and step-like function as test functions because they are easy to derive bootstrap equations. In that case, we had to consider many contributions from the intermediate state, whose energy ranges from the lowest one to infinity. If we set a smoother function as a test function, we might get better bootstrap equations in which the contributions from an intermediate state with high energy are suppressed. Considering many types of test functions is one of our future directions.

\item This paper dealt with conformal bootstrap obtained from the microcausality condition. In the construction, we ignored the explicit form of the commutator by multiplying the test function that has support only at spacelike region. On the other hand, finding a concrete expression for the commutator is needed to find a valid, closed bootstrap equation for the Wightman function by comparing (43)(21), (42)(31), and (32)(41) channels. However, the commutator of the operators is not well-defined for a non-free CFT, so we need to consider the commutator of the smeared operators for time and space directions. This formulation has been constructed \cite{Nagano:2021tbu,Besken:2020snx} but has yet to apply it to the conformal bootstrap equations fully. It is a desirable research direction.

\end{itemize}

\section*{Acknowledgement}
The author thanks his supervisor Simeon Hellerman for his scientific advice. His insight into conformal bootstrap methods helped the author a lot. And the author also thanks their dear schoolmates for providing helpful guidance on research activities.\par
This project is supported by the WINGS-FMSP (World-leading Innovative Graduate Study for Frontiers of Mathematical Sciences and Physics) project.

\appendix

\section{Supplement to two-dimensional CFT}
\subsection{Solution for the Ward Identities}
\setcounter{equation}{0}
Let us solve ODEs (\ref{Ward Id2-1})-(\ref{Ward Id2-4}) obtained from WIs in two-dimensional CFT.
Remember that we used $L_0$ and $\tilde{L}_0$ WIs to determine the reduced form for the two-point function. First, shift $n$ by $n\rightarrow n-s_1$, and define $G_n\eq F_{n-s_1}$. We get
\begin{align}
L_{-1}&:\left(y\dv{y}-\Delta_1+n-s_1+1\right)G_{n+1}(y)+y\left(-y\dv{y}-\Delta_2-n-s_2\right)G_{n}(y)=0\label{Ward Id2-5}\\
\tilde{L}_{-1}&:y\left(-y\dv{y}-\Delta_2+n+1+s_2\right)G_{n+1}(y)+\left(y\dv{y}-\Delta_1-n+s_1\right)G_{n}(y)=0\label{Ward Id2-6}\\
L_1&:\left(-y\dv{y}+\Delta_2-n-1+s_2\right)G_{n+1}(y)+y\left(y\dv{y}+\Delta_1+s_1+n\right)G_{n}(y)=0\label{Ward Id2-7}\\
\tilde{L}_1&:y\left(y\dv{y}+\Delta_1-s_1-n-1\right)G_{n+1}(y)+\left(-y\dv{y}+\Delta_2+n-s_2\right)G_{n}(y)=0.\label{Ward Id2-8}
\end{align}
Adding equations (\ref{Ward Id2-5}) and (\ref{Ward Id2-7}) gives
\begin{equation}
0=2(h_2-h_1)[G_{n+1}(y)-yG_n(y)].
\end{equation}
And adding equations (\ref{Ward Id2-6}) and (\ref{Ward Id2-8}) gives
\begin{equation}
  0=2(\tilde{h}_1-\tilde{h}_2)[yG_{n+1}(y)-G_n(y)].
\end{equation}
There are four possibilities.
\begin{align}
  (1)&\ \ \ h_1\neq h_2,\  \tilde{h}_1\neq\tilde{h}_2 \ \ \text{supported only at }y=1\notag\\
  (2)&\ \ \ h_1=h_2,\ G_n(y)=y^{-n}G_0(y)\notag\\
  (3)&\ \ \ \tilde{h}_1=\tilde{h}_2,\ G_n(y)=y^nG_0(y)\notag\\
  (4)&\ \ \ h_1=h_2,\ \tilde{h}_1=\tilde{h}_2 \notag
\end{align}
The first solution is different from what we are looking for.
Substituting the second solution for (\ref{Ward Id2-5}) gives
\begin{equation}
  \dv{y}G_0(y)=2h\frac{1+y^2}{y(1-y^2)}G_0(y).
\end{equation}
Solving them, we get
\begin{equation}
  G_0(y)\propto\left(\frac{y}{|1-y|(1+y)}\right)^{2h}.
\end{equation}
So, the Wightman two-point function is
\begin{equation}
  C(n_1,n_2,r_1,r_2)\propto r_1^{-\Delta_1}r_2^{-\Delta_2}\delta(n_1+n_2+s_1+s_2)y^{-n_1-s_1}\left(\frac{y}{|1-y|(1+y)}\right)^{2h}.
\end{equation}
The second solution is wrong because it diverges when we take $y$ to $0$ for fixed $r_2$. In the same way, we can say that the third solution is wrong. So, we can conclude that $h_1=h_2\eq h$ and $\tilde{h}_1=\tilde{h}_2\eq\tilde{h}$. Then we only have two independent equations.
\begin{align}
  &\left(y\dv{y}-2h+n+1\right)G_{n+1}(y)+y\left(-y\dv{y}-2h-n\right)G_n(y)=0 \\
  &y\left(-y\dv{y}-2\tilde{h}+n+1\right)G_{n+1}(y)+\left(y\dv{y}-2\tilde{h}-n\right)G_n(y)=0
\end{align}
From them, we get
\begin{align}
  G_{n+1}&=\frac{1}{2h+2\tilde{h}-2n-2}[(y^{-1}-y)\hat{E}-2hy-2\tilde{h}y^{-1}-(y+y^{-1})n]G_n\\
  G_n&=\frac{1}{2h+2\tilde{h}+2n}[(y^{-1}-y)\hat{E}-2hy^{-1}-2\tilde{h}y+(y+y^{-1})(n+1)]G_{n+1}.
\end{align}
Let us fix $n$ to solve it. The most reasonable choice is $n=s=h-\tilde{h}$.
\begin{align}
  G_{s+1}&=\frac{1}{2\Delta-2s-2}[(y^{-1}-y)(\hat{E}+s)-(y^{-1}+y)(\Delta+s)]G_s\\
  G_s&=\frac{1}{2\Delta+2s}[(y^{-1}-y)(\hat{E}-s)+(y+y^{-1})(s+1-\Delta)]G_{s+1}
\end{align}
Combining them gives
\begin{equation}
  G''_s[y]+\frac{(2\Delta+1)y^2+(2\Delta-1)}{y^3-y}G'_s[y]+\left(\frac{\Delta^2}{y^2}+\frac{4s^2}{1-y^2}\right)G_s[y]=0.\label{ODE in 2D}
\end{equation}
From unitarity, we can assume that $G_s$ must be a discrete sum of the powers of $y$. And small-$y$ behavior of the solution are $G_s(y)\propto y^{\Delta}$. From these facts, we get the answer by series expansion.
\begin{align}
  G_s[y]&=K_sy^{\Delta}\sum_{l=0}^\infty c_ly^{2l}\\
  c_l&=\frac{\Gamma(\Delta+s+l)\Gamma(\Delta-s+l)}{\Gamma(\Delta+s)\Gamma(\Delta-s)\Gamma^2(l+1)}
\end{align}
$K_s$ is an undetermined constant depending on the normalization of operators.\par
Now we get a solution for the primary state with spin $s$. Next, we derive a solution for the excited (descendant) state. The recursion relation is
\begin{equation}
G_{n+s+1}=\frac{1}{2\Delta-2s-2n-2}[(y^{-1}-y)(\hat{E}+s)-(y^{-1}+y)(\Delta+s+n)]G_{n+s}.
\end{equation}
Assume that the solution has the following form.
\begin{align}
  &G_{n+s}=K_{n+s}y^{\Delta+n}\sum_{l\geq0}b_{n+s|l}y^{2l}=K_sy^{\Delta+n}\sum_{l\geq0}c_{n+s|l}y^{2l}\\
  &K_{n+s}b_{n+s|l}=K_sc_{n+s|l}\\
  &b_{s|l}=c_{s|l}=c_l=\frac{\Gamma(\Delta+s+l)\Gamma(\Delta-s+l)}{\Gamma(\Delta+s)\Gamma(\Delta-s)\Gamma^2(l+1)}\\
  &b_{n+s|l=0}=1
\end{align}
We use mathematical induction to get the solution.\\
(1) Calculation of $G_{s+1}$
\begin{align}
  G_{1+s}&=K_{1+s}y^{\Delta+1}\sum_{l\geq0}b_{1+s|l}y^{2l}\\
  &=\frac{1}{2\Delta-2s-2}\left[(1-y^2)\dv{y}-y^{-1}\Delta-y(2s+\Delta)\right]K_sy^{\Delta}\sum_{l\geq0}b_{s|l}y^{2l}
\end{align}
Picking up the coefficient of $y^{\Delta+1}$ gives
\begin{equation}
  K_{1+s}=(\Delta+s)K_s.
\end{equation}
Picking up the coefficient of $y^{\Delta+2l+1}$ gives
\begin{equation}
  K_{1+s}b_{1+s|l}=\frac{K_s}{l+1}\frac{\Gamma(\Delta+s+l+1)\Gamma(\Delta-s+l)}{\Gamma(\Delta+s)\Gamma(\Delta-s)\Gamma^2(l+1)}.
\end{equation}
So, we get
\begin{align}
  b_{1+s|l}&=\frac{\Gamma(\Delta+s+l+1)\Gamma(\Delta-s+l)}{\Gamma(\Delta+s+1)\Gamma(\Delta-s)\Gamma(l+2)\Gamma(l+1)}\\
  c_{1+s|l}&=\frac{K_{1+s}}{K_s}b_{1+s|l}=\frac{\Gamma(\Delta+s+l+1)\Gamma(\Delta-s+l)}{\Gamma(\Delta+s)\Gamma(\Delta-s)\Gamma(l+2)\Gamma(l+1)}.
\end{align}
(2) Calculation of $G_{s+2}$
\begin{align}
  G_{2+s}&=K_{2+s}y^{\Delta+2}\sum_{l\geq0}b_{2+s|l}y^{2l}\\
  &=\frac{1}{2\Delta-2s-4}\left[(1-y^2)\dv{y}-y^{-1}(\Delta+1)-y(2s+\Delta+1)\right]K_{1+s}y^{\Delta+1}\sum_{l\geq0}b_{1+s|l}y^{2l}
\end{align}
Picking up the coefficient of $y^{\Delta+2}$ gives
\begin{equation}
  K_{2+s}=\frac{(\Delta+s)(\Delta+s+1)}{2}K_s.
\end{equation}
Picking up the coefficient of $y^{\Delta+2+2l}$ gives
\begin{align}
  b_{2+s|l}&=2\frac{\Gamma(\Delta+s+l+2)\Gamma(\Delta-s+l)}{\Gamma(\Delta+s+2)\Gamma(\Delta-s)\Gamma(l+3)\Gamma(l+1)}\\
  c_{2+s|l}&=\frac{K_{2+s}}{K_s}b_{2+s|l}=\frac{\Gamma(\Delta+s+l+2)\Gamma(\Delta-s+l)}{\Gamma(\Delta+s)\Gamma(\Delta-s)\Gamma(l+3)\Gamma(l+1)}.
\end{align}
(3) General $G_{n+s}$\par
From (1) and (2), we can guess the form of general $G_{n+s}$.
\begin{align}
  K_{n+s}&=\frac{\Gamma(\Delta+s+n)}{\Gamma(n+1)\Gamma(\Delta+s)}K_s\label{guess for solution1}\\
  b_{n+s|l}&=\frac{\Gamma(1+n)\Gamma(\Delta+s+l+n)\Gamma(\Delta-s+l)}{\Gamma(\Delta+s+n)\Gamma(\Delta-s)\Gamma(l+n+1)\Gamma(l+1)}\label{guess for solution2}\\
  c_{n+s|l}&=\frac{\Gamma(\Delta+s+l+n)\Gamma(\Delta-s+l)}{\Gamma(\Delta+s)\Gamma(\Delta-s)\Gamma(l+n+1)\Gamma(l+1)}\label{guess for solution3}
\end{align}
We show it by mathematical induction. Picking up the coefficient of $y^{\Delta+2+2l}$ gives
\begin{align}
K_{n+s+1}b_{n+s+1|l}=\frac{K_{n+s}}{\Delta-s-n-1}[(l+1)b_{n+s|l+1}-(l+s+\Delta+n)b_{n+s|l}].
\end{align}
From this, we obtain recursion relations for $K_{n+s}$, $b_{n+s|l}$ and $c_{n+s|l}$.
\begin{align}
  K_{n+s+1}&=K_{n+s}\frac{b_{n+s|l=1}-(\Delta+s+n)}{\Delta-s-n-1}\\
  b_{n+s+1|l}&=\frac{(l+1)b_{n+s|l+1}-(l+s+\Delta+n)b_{n+s|l}}{b_{n+s|l=1}-(\Delta+s+n)}\\
  c_{n+s+1}&=\frac{K_{n+s+1}}{K_s}b_{n+s+1}
\end{align}
When the configuration (\ref{guess for solution1}), (\ref{guess for solution2}) and (\ref{guess for solution3}) are valid for $n$,
\begin{align}
&K_{n+s+1}\notag\\
&=\frac{\Gamma(\Delta+s+n)}{\Gamma(n+1)\Gamma(\Delta+s)}K_s\frac{1}{\Delta-s-n-1}\left[\frac{\Gamma(1+n)\Gamma(\Delta+s+1+n)\Gamma(\Delta-s+1)}{\Gamma(\Delta+s+n)\Gamma(\Delta-s)\Gamma(n+2)\Gamma(2)}-(\Delta+s+n)\right]\notag\\
&=\frac{\Gamma(\Delta+s+n+1)}{\Gamma(n+2)\Gamma(\Delta+s)}K_s
\end{align}
\begin{align}
  b_{n+s+1|l}&=\left[(l+1)\frac{\Gamma(1+n)\Gamma(\Delta+s+l+1+n)\Gamma(\Delta-s+l+1)}{\Gamma(\Delta+s+n)\Gamma(\Delta-s)\Gamma(l+n+2)\Gamma(l+2)}-\right.\notag\\
  &\left.\hspace{3cm}(l+s+\Delta+n)\frac{\Gamma(1+n)\Gamma(\Delta+s+l+n)\Gamma(\Delta-s+l)}{\Gamma(\Delta+s+n)\Gamma(\Delta-s)\Gamma(l+n+1)\Gamma(l+1)}\right]\notag\\
  &\hspace{5mm}\left[\frac{\Gamma(1+n)\Gamma(\Delta+s+1+n)\Gamma(\Delta-s+1)}{\Gamma(\Delta+s+n)\Gamma(\Delta-s)\Gamma(n+2)\Gamma(2)}-(\Delta+s+n)\right]^{-1}\notag\\
  &=\frac{\Gamma(2+n)\Gamma(\Delta+s+l+n)\Gamma(\Delta-s+l)}{\Gamma(\Delta+s+n)\Gamma(\Delta-s)\Gamma(l+n+2)\Gamma(l+2)}\frac{(l+1)(\Delta+s+l+n)(\Delta-s-n-1)}{(\Delta+s+n)(\Delta-s-n-1)}\notag\\
  &=\frac{\Gamma(2+n)\Gamma(\Delta+s+l+n+1)\Gamma(\Delta-s+l)}{\Gamma(\Delta+s+n+1)\Gamma(\Delta-s)\Gamma(l+n+2)\Gamma(l+1)}\\
  c_{n+s+1|l}&=\frac{\Gamma(\Delta+s+n+1)}{\Gamma(n+2)\Gamma(\Delta+s)}\frac{\Gamma(2+n)\Gamma(\Delta+s+l+n+1)\Gamma(\Delta-s+l)}{\Gamma(\Delta+s+n+1)\Gamma(\Delta-s)\Gamma(l+n+2)\Gamma(l+1)}\notag\\
  &=\frac{\Gamma(\Delta+s+l+n+1)\Gamma(\Delta-s+l)}{\Gamma(\Delta+s)\Gamma(\Delta-s)\Gamma(l+n+2)\Gamma(l+1)}.
\end{align}
So, the configurations (\ref{guess for solution1}), (\ref{guess for solution2}) and (\ref{guess for solution3}) are also valid for $n+1$. By mathematical induction, we get (\ref{guess for solution1}), (\ref{guess for solution2}) and (\ref{guess for solution3}) for positive integer $n$. In the same way, we can also get the solution for negative $n$.
\subsection{Direct integral calculation of three-point function}
Let us perform the Fourier transform directly for the three-point function.
\begin{equation}
  \langle\OO_{n_3}(r_3)\OO_{n_2}(r_2)\OO_{n_1}(r_1)\rangle=\frac{1}{(2\pi i)^3}\oint\frac{dz_3}{z_3^{1+n_3}}\oint\frac{dz_2}{z_2^{1+n_2}}\oint\frac{dz_1}{z_1^{1+n_1}}\lambda_{321}z_{21}^{-b_3}z_{32}^{-b_1}z_{31}^{-b_2}\label{integral0}
\end{equation}
First, integrate it with respect to $z_1$.
\begin{equation}
\frac{\lambda_{321}}{2\pi i}\oint\frac{dz_1}{z_1^{1+n_1}}z_{21}^{-b_3}z_{32}^{-b_1}z_{31}^{-b_2}=\frac{\lambda_{321}}{\Gamma(b_3)\Gamma(b_2)}\sum_{k=0}^{n_1}\frac{\Gamma(b_3+k)\Gamma(b_2+n_1-k)}{\Gamma(1+k)\Gamma(1+n_1-k)}z_2^{-b_3-k}z_3^{-b_2-n_1+k}z_{32}^{-b_1}\label{integral1}
\end{equation}
Next, define $\omega_3=1/z_3$ and integrate it with respect to $\omega_3$. Under this transformation, the contour integral around $z=\infty$ becomes contour integral around $\omega_3=0$.
\begin{flalign}
&\frac{1}{2\pi i}\oint\frac{dz_3}{z_3^{1+n_3}}\text{(\ref{integral1})}=\frac{1}{2\pi i}\oint\frac{d\omega_3}{\omega_3^{1+m_3}}\frac{\lambda_{321}}{\Gamma(b_3)\Gamma(b_2)}\sum_{k=0}^{n_1}\frac{\Gamma(b_3+k)\Gamma(b_2+n_1-k)}{\Gamma(1+k)\Gamma(1+n_1-k)}\frac{z_2^{-b_3-k}\omega_3^{n_1-k}}{(1-z_2\omega_3)^{b_1}}&\label{integral2}
\end{flalign}
Here, we defined $m_3\eq-n_3-2h_3$. There are two cases. $m_3\geq n_1$ or $m_3\leq n_1$.\\
(1) $m_3\leq n_1$
\begin{equation}
  \text{(\ref{integral2})}=\frac{\lambda_{123}}{\Gamma(b_3)\Gamma(b_2)}\sum_{k=n_1-m_3}^{n_1}\frac{\Gamma(b_3+k)\Gamma(b_2+n_1-k)}{\Gamma(1+k)\Gamma(1+n_1-k)}\frac{\Gamma(b_1+m_3-n_1+k)}{\Gamma(m_3-n_1+k+1)\Gamma(b_1)}z_2^{-b_3+m_3-n_1}
\end{equation}
So, the three-point function is
\begin{equation}
  \text{(\ref{integral0})}=\frac{\delta\left(\sum_{i=1}^3(n_i+h_i)\right)\lambda_{123}}{\Gamma(b_3)\Gamma(b_2)\Gamma(b_1)}\sum_{k=n_1-m_3}^{n_1}\frac{\Gamma(b_3+k)\Gamma(b_2+n_1-k)}{\Gamma(1+k)\Gamma(1+n_1-k)}\frac{\Gamma(b_1+m_3-n_1+k)}{\Gamma(1+m_3-n_1+k)}.
\end{equation}
(2) $m_3\geq n_1$
\begin{equation}
  \text{(\ref{integral2})}=\frac{\lambda_{123}}{\Gamma(b_3)\Gamma(b_2)}\sum_{k=0}^{n_1}\frac{\Gamma(b_3+k)\Gamma(b_2+n_1-k)}{\Gamma(1+k)\Gamma(1+n_1-k)}\frac{\Gamma(b_1+m_3-n_1+k)}{\Gamma(m_3-n_1+k+1)\Gamma(b_1)}z_2^{-b_3+m_3-n_1}
\end{equation}
So, the three-point function is
\begin{equation}
  \text{(\ref{integral0})}=\frac{\delta\left(\sum_{i=1}^3(n_i+h_i)\right)\lambda_{123}}{\Gamma(b_3)\Gamma(b_2)\Gamma(b_1)}\sum_{k=0}^{n_1}\frac{\Gamma(b_3+k)\Gamma(b_2+n_1-k)}{\Gamma(1+k)\Gamma(1+n_1-k)}\frac{\Gamma(b_1+m_3-n_1+k)}{\Gamma(1+m_3-n_1+k)}.
\end{equation}
In the end, we get
\begin{equation}
  \text{(\ref{integral0})}=\frac{\delta\left(\sum_{i=1}^3(n_i+h_i)\right)\lambda_{123}}{\Gamma(b_3)\Gamma(b_2)\Gamma(b_1)}\sum_{k=\rm{Max}\{0,n_1-m_3\}}^{n_1}\frac{\Gamma(b_3+k)\Gamma(b_2+n_1-k)}{\Gamma(1+k)\Gamma(1+n_1-k)}\frac{\Gamma(b_1+m_3-n_1+k)}{\Gamma(1+m_3-n_1+k)}.
\end{equation}
Define a new variable $q\eq k-\mathrm{Max}\{0,n_1-m_3\}$.
Then, we get
\begin{align}
&\langle\OO_{n_3}(r_3)\OO_{n_2}(r_2)\OO_{n_1}(r_1)\rangle=\delta(n_1+n_2+n_3+h_1+h_2+h_3)\frac{\lambda_{321}}{\Gamma(b_1)\Gamma(b_2)\Gamma(b_3)}\\
&\sum_{q=0}^{\mathrm{Min}\{n_1,m_3\}}\frac{\Gamma[b_3+q+\mathrm{Max}\{{0,n_1-m_3}\}]\Gamma[b_2-q+\mathrm{Min}\{{n_1,m_3}\}]\Gamma[b_1+q+\mathrm{Max}\{{m_3-n_1,0}\}]}{\Gamma[1+q+\mathrm{Max}\{{0,n_1-m_3}\}]\Gamma[1-q+\mathrm{Min}\{{n_1,m_3}\}]\Gamma[1+q+\mathrm{Max}\{{m_3-n_1,0}\}]}\notag
\end{align}
It is valid for all integer $n_1$ and $m_3$, and invariant under the time-reversal transformation ($n_1\Leftrightarrow m_3$, $b_1\Leftrightarrow b_3$).
\subsection{Consistency check for three-point function}
\subsubsection{Ward Identities for three-point function}
Define the complete Wightman three-point function as
\begin{equation}
  C(n_1,n_2,n_3,r_1,r_2,r_3)\eq\langle\mathcal{\tilde{O}}^{(3)}_{n_3}(r_3)\mathcal{\tilde{O}}^{(2)}_{n_2}(r_2)\mathcal{\tilde{O}}^{(1)}_{n_1}(r_1)\rangle.
\end{equation}
We get a reduced three-point function using $L_0\pm\tilde{L}_0$ WIs.
\begin{equation}
  C(n_1,n_2,n_3)=r_1^{-\Delta_1}r_2^{-\Delta_2}r_3^{-\Delta_3}\delta(n_1+n_2+n_3+s_1+s_2+s_3)F(n_1,n_3,y_1,y_3)
\end{equation}
with $y_1\eq\frac{r_1}{r_2}$ and $y_3\eq\frac{r_3}{r_2}$. Next, consider $P_z,\ P_{\bar{z}},\ K^z $ and $K^{\bar{z}}$ WIs.
\begin{align}
  &P_z\cdot C(n_1,n_2,n_3)=0\notag\\
  &=\frac{1}{2r_3}\left(r_3\dv{r_3}+n_3+1\right)C(n_1,n_2,n_3+1)+\frac{1}{2r_2}\left(r_2\dv{r_2}+n_2+1\right)C(n_1,n_2+1,n_3)\notag\\
  &\ \ \ \ \ \ \ \ \ \ \ \ +\frac{1}{2r_1}\left(r_1\dv{r_1}+n_1+1\right)C(n_1+1,n_2,n_3)\\
  &\notag\\
  &K^z\cdot C(n_1,n_2,n_3)=0\notag\\
  &=\frac{r_3}{2}\left(r_3\dv{r_3}+n_3+4h_3-1\right)C(n_1,n_2,n_3-1)+\frac{r_2}{2}\left(r_2\dv{r_2}+n_2+4h_2-1\right)C(n_1,n_2-1,n_3)\notag\\
  &\ \ \ \ \ \ \ \ \ \ \ \ +\frac{r_1}{2}\left(r_1\dv{r_1}+n_1+4h_1-1\right)C(n_1-1,n_2,n_3)\\
  &\notag\\
  &P_{\bar{z}}\cdot C(n_1,n_2,n_3)=0\notag\\
  &=\frac{1}{2r_3}\left(r_3\dv{r_3}-n_3+1\right)C(n_1,n_2,n_3-1)+\frac{1}{2r_2}\left(r_2\dv{r_2}-n_2+1\right)C(n_1,n_2-1,n_3)\notag\\
  &\ \ \ \ \ \ \ \ \ \ \ \ +\frac{1}{2r_1}\left(r_1\dv{r_1}-n_1+1\right)C(n_1-1,n_2,n_3)\\
  &\notag\\
  &K^{\bar{z}}\cdot C(n_1,n_2,n_3)=0\notag\\
  &=\frac{r_3}{2}\left(r_3\dv{r_3}-n_3+4\tilde{h}_3-1\right)C(n_1,n_2,n_3+1)+\frac{r_2}{2}\left(r_2\dv{r_2}-n_2+4\tilde{h}_2-1\right)C(n_1,n_2+1,n_3)\notag\\
  &\ \ \ \ \ \ \ \ \ \ \ \ +\frac{r_1}{2}\left(r_1\dv{r_1}-n_1+4\tilde{h}_1-1\right)C(n_1+1,n_2,n_3)
\end{align}
Substitute the reduced form for WIs. We get
\begin{align}
  \mathrm{P_z\ Ward\ Identity}:\ \ \ \ \ 0=&\frac{1}{y_3}\left(y_3\dv{y_3}+n_3-\Delta_3+1\right)F(n_1,n_3+1,y_1,y_3)\notag\\
  &+\left(-y_1\dv{y_1}-y_3\dv{y_3}-n_1-n_3-s_1-s_3-2h_2\right)F(n_1,n_3,y_1,y_3)\notag\\
  &+\frac{1}{y_1}\left(y_1\dv{y_1}+n_1-\Delta_1+1\right)F(n_1+1,n_3,y_1,y_3)\\
  \mathrm{P_{\bar{z}}\ Ward\ Identity}:\ \ \ \ \ 0=&\frac{1}{y_3}\left(y_3\dv{y_3}-n_3-\Delta_3+1\right)F(n_1,n_3-1,y_1,y_3)\notag\\
  &+\left(-y_1\dv{y_1}-y_3\dv{y_3}+n_1+n_3+s_1+s_3-2\tilde{h}_2\right)F(n_1,n_3,y_1,y_3)\notag\\
  &+\frac{1}{y_1}\left(y_1\dv{y_1}-n_1-\Delta_1+1\right)F(n_1-1,n_3,y_1,y_3)\\
  \mathrm{K^z\ Ward\ Identity}:\ \ \ \ \ 0=&y_3\left(y_3\dv{y_3}+n_3+4h_3-\Delta_3-1\right)F(n_1,n_3-1,y_1,y_3)\notag\\
  &+\left(-y_1\dv{y_1}-y_3\dv{y_3}-n_1-n_3-s_1-s_3+2h_2\right)F(n_1,n_3,y_1,y_3)\notag\\
  &+y_1\left(y_1\dv{y_1}+n_1+4h_1-\Delta_1-1\right)F(n_1-1,n_3,y_1,y_3)\\
  \mathrm{K^{\bar{z}}\ Ward\ Identity}:\ \ \ \ \ 0=&y_3\left(y_3\dv{y_3}-n_3+4\tilde{h}_3-\Delta_3-1\right)F(n_1,n_3+1,y_1,y_3)\notag\\
  &+\left(-y_1\dv{y_1}-y_3\dv{y_3}+n_1+n_3+s_1+s_3+2\tilde{h}_2\right)F(n_1,n_3,y_1,y_3)\notag\\
  &+y_1\left(y_1\dv{y_1}-n_1+4\tilde{h}_1-\Delta_1-1\right)F(n_1+1,n_3,y_1,y_3).
\end{align}
\subsubsection{Reduction of Ward Identities for holomorphic three-point function}
The relation between $\KK_{\mathrm{red}}(n_1,m_3)$ and $F(n_1,n_3,y_1,y_3)$ is
\begin{equation}
  \KK_{\mathrm{red}}(n_1,m_3)r_1^{n_1}r_2^{n_2}r_3^{n_3}=r_1^{-\Delta_1}r_2^{-\Delta_2}r_3^{-\Delta_3}F(n_1,n_3,y_1,y_3).
\end{equation}
Remember that $m_3$ is defined as $m_3=-(n_3+2h_3)$. Then,
\begin{equation}
  F(n_1,n_3,y_1,y_3)=y_1^{n_1+h_1}y_3^{n_3+h_3}\KK_{\mathrm{red}}(n_1,m_3).\label{reduced form}
\end{equation}
For holomorphic operators, WIs for $F(n_1,n_3,y_1,y_3)$ are
\begin{align}
  \mathrm{P_z\ Ward\ Identity}:\ \ \ \ \ 0=&\frac{1}{y_3}\left(y_3\dv{y_3}+n_3-h_3+1\right)F(n_1,n_3+1,y_1,y_3)\notag\\
  &+\left(-y_1\dv{y_1}-y_3\dv{y_3}-n_1-n_3-h_1-h_3-2h_2\right)F(n_1,n_3,y_1,y_3)\notag\\
  &+\frac{1}{y_1}\left(y_1\dv{y_1}+n_1-h_1+1\right)F(n_1+1,n_3,y_1,y_3)\\
  \mathrm{P_{\bar{z}}\ Ward\ Identity}:\ \ \ \ \ 0=&\frac{1}{y_3}\left(y_3\dv{y_3}-n_3-h_3+1\right)F(n_1,n_3-1,y_1,y_3)\notag\\
  &+\left(-y_1\dv{y_1}-y_3\dv{y_3}+n_1+n_3+h_1+h_3\right)F(n_1,n_3,y_1,y_3)\notag\\
  &+\frac{1}{y_1}\left(y_1\dv{y_1}-n_1-h_1+1\right)F(n_1-1,n_3,y_1,y_3)\\
  \mathrm{K^z\ Ward\ Identity}:\ \ \ \ \ 0=&y_3\left(y_3\dv{y_3}+n_3+3h_3-1\right)F(n_1,n_3-1,y_1,y_3)\notag\\
  &+\left(-y_1\dv{y_1}-y_3\dv{y_3}-n_1-n_3-h_1-h_3+2h_2\right)F(n_1,n_3,y_1,y_3)\notag\\
  &+y_1\left(y_1\dv{y_1}+n_1+3h_1-1\right)F(n_1-1,n_3,y_1,y_3)\\
  \mathrm{K^{\bar{z}}\ Ward\ Identity}:\ \ \ \ \ 0=&y_3\left(y_3\dv{y_3}-n_3-h_3-1\right)F(n_1,n_3+1,y_1,y_3)\notag\\
  &+\left(-y_1\dv{y_1}-y_3\dv{y_3}+n_1+n_3+h_1+h_3\right)F(n_1,n_3,y_1,y_3)\notag\\
  &+y_1\left(y_1\dv{y_1}-n_1-h_1-1\right)F(n_1+1,n_3,y_1,y_3).
\end{align}
Substituting (\ref{reduced form}) for them gives the following relations.
\begin{align}
  P_z&:0=(n_3+1)\KK_{\mathrm{red}}(n_1,n_3+1)-(h_1+h_2+h_3+n_1+n_3)\KK_{\mathrm{red}}(n_1,n_3)\notag\\
  &\hspace{1cm}+(n_1+1)\KK_{\mathrm{red}}(n_1+1,n_3)\\
  P_{\bar{z}}&:\mathrm{satisfied\  trivially}\notag\\
  K^z&:0=(n_3+2h_3-1)\KK_{\mathrm{red}}(n_1,n_3-1)-(n_1+n_3+h_1+h_3-h_2)\KK_{\mathrm{red}}(n_1,n_3)\notag\\
  &\hspace{1cm}+(n_1+2h_1-1)\KK_{\mathrm{red}}(n_1-1,n_3)\\
  K^{\bar{z}}&:\mathrm{satisfied\  trivially}\notag
\end{align}
They are the reduced WIs for the holomorphic three-point function. All we have to do is to check whether this $P_z$ WI and $K^z$ WI are satisfied.
\subsubsection{Consistency check}
We have to prove the following equations. For reduced $P_z$ WI, we get
\begin{align}
  0&=(n_3+1)\sum_{k=\mathrm{MAX}\{0,n_1-m_3+1\}}^{n_1}\frac{\Gamma(b_3+k)\Gamma(b_2+n_1-k)\Gamma(b_1+m_3-n_1+k-1)}{\Gamma(1+k)\Gamma(1+n_1-k)\Gamma(1+m_3-n_1+k-1)}\notag\\
  &\ -(n_1+n_3+h_1+h_2+h_3)\sum_{k=\mathrm{MAX}\{0,n_1-m_3\}}^{n_1}\frac{\Gamma(b_3+k)\Gamma(b_2+n_1-k)\Gamma(b_1+m_3-n_1+k)}{\Gamma(1+k)\Gamma(1+n_1-k)\Gamma(1+m_3-n_1+k)}\notag\\
  &\ +(n_1+1)\sum_{k=\mathrm{MAX}\{0,n_1-m_3+1\}}^{n_1+1}\frac{\Gamma(b_3+k)\Gamma(b_2+n_1+1-k)\Gamma(b_1+m_3-n_1+k-1)}{\Gamma(1+k)\Gamma(1+n_1+1-k)\Gamma(1+m_3-n_1+k-1)}.\label{Pz Ward Id}
\end{align}
For reduced $K^z$ WI, we get
\begin{align}
  0&=(2h_3+n_3-1)\sum_{k=\mathrm{MAX}\{0,n_1-1-m_3\}}^{n_1}\frac{\Gamma(b_3+k)\Gamma(b_2+n_1-k)\Gamma(b_1+m_3-n_1+k+1)}{\Gamma(1+k)\Gamma(1+n_1-k)\Gamma(1+m_3-n_1+k+1)}\notag\\
  &-(h_1-h_2+h_3+n_1+n_3)\sum_{k=\mathrm{MAX}\{0,n_1-m_3\}}^{n_1}\frac{\Gamma(b_3+k)\Gamma(b_2+n_1-k)\Gamma(b_1+m_3-n_1+k)}{\Gamma(1+k)\Gamma(1+n_1-k)\Gamma(1+m_3-n_1+k)}\notag\\
  &+(2h_1+n_1-1)\sum_{k=\mathrm{MAX}\{0,n_1-1-m_3\}}^{n_1-1}\frac{\Gamma(b_3+k)\Gamma(b_2+n_1-1-k)\Gamma(b_1+m_3-n_1+k+1)}{\Gamma(1+k)\Gamma(1+n_1-1-k)\Gamma(1+m_3-n_1+k+1)}.\label{Kz Ward Id}
\end{align}
(A-1) $P_z$ WI for $n_1+1\leq m_3$\par
When $n_1+1\leq m_3$, RHS of (\ref{Pz Ward Id}) is
\begin{align}
  \mathrm{(RHS)}&=(-m_3-b_1-b_2+1)\sum_{k=0}^{n_1}\frac{\Gamma(b_3+k)\Gamma(b_2+n_1-k)\Gamma(b_1+m_3-n_1+k-1)}{\Gamma(1+k)\Gamma(1+n_1-k)\Gamma(1+m_3-n_1+k-1)}\notag\\
  &\ -(n_1-m_3+b_3)\sum_{k=0}^{n_1}\frac{\Gamma(b_3+k)\Gamma(b_2+n_1-k)\Gamma(b_1+m_3-n_1+k)}{\Gamma(1+k)\Gamma(1+n_1-k)\Gamma(1+m_3-n_1+k)}\notag\\
  &\ +(n_1+1)\sum_{k=0}^{n_1+1}\frac{\Gamma(b_3+k)\Gamma(b_2+n_1+1-k)\Gamma(b_1+m_3-n_1+k-1)}{\Gamma(1+k)\Gamma(1+n_1+1-k)\Gamma(1+m_3-n_1+k-1)}.
\end{align}
We call the first term (A), the second term (B), and the third term (C).
\begin{align}
  \mathrm{(A)}&=(-m_3-b_1-b_2+1)\sum_{k=0}^{n_1}\frac{\Gamma(b_3+k)\Gamma(b_2+n_1-k)\Gamma(b_1+m_3-n_1+k-1)}{\Gamma(1+k)\Gamma(1+n_1-k)\Gamma(1+m_3-n_1+k-1)}\notag\\
  \mathrm{(B)}&=-(n_1-m_3+b_3)\sum_{k=0}^{n_1}\frac{\Gamma(b_3+k)\Gamma(b_2+n_1-k)\Gamma(b_1+m_3-n_1+k)}{\Gamma(1+k)\Gamma(1+n_1-k)\Gamma(1+m_3-n_1+k)}\notag\\
  \mathrm{(C)}&=(n_1+1)\sum_{k=0}^{n_1+1}\frac{\Gamma(b_3+k)\Gamma(b_2+n_1+1-k)\Gamma(b_1+m_3-n_1+k-1)}{\Gamma(1+k)\Gamma(1+n_1+1-k)\Gamma(1+m_3-n_1+k-1)}
\end{align}
Then, we get
\begin{align}
  \mathrm{(A)}&=-\sum_{k=0}^{n_1}(b_1+b_2+m_3-1)\frac{\Gamma(b_3+k)\Gamma(b_2+n_1-k)\Gamma(b_1+m_3-n_1+k-1)}{\Gamma(1+k)\Gamma(1+n_1-k)\Gamma(1+m_3-n_1+k-1)}\notag\\
  &=-\sum_{k=0}^{n_1}(m_3-n_1+k)\frac{\Gamma(b_3+k)\Gamma(b_2+n_1-k)\Gamma(b_1+m_3-n_1+k)}{\Gamma(1+k)\Gamma(1+n_1-k)\Gamma(m_3-n_1+k+1)}\notag\\
  &\ \ \ \ \ -\sum_{k=0}^{n_1}(1+n_1-k)\frac{\Gamma(b_3+k)\Gamma(b_2+n_1-k+1)\Gamma(b_1+m_3-n_1+k-1)}{\Gamma(1+k)\Gamma(2+n_1-k)\Gamma(m_3-n_1+k)}.\notag\\
\end{align}
We call the first term (A-1) and the second (A-2). Then, we get
\begin{align}
&\mathrm{(RHS)}=\mathrm{((A-1)}+\mathrm{(B))}+\mathrm{((A-2)}+\mathrm{(C))}\notag\\
&=\frac{\Gamma(b_3+n_1+1)\Gamma(b_2)\Gamma(b_1+m_3)}{\Gamma(1+n_1)\Gamma(1)\Gamma(1+m_3)}+\sum_{k=0}^{n_1}\frac{\Gamma(b_3+k)\Gamma(b_2+n_1-k+1)\Gamma(b_1+m_3-n_1+k-1)}{\Gamma(k)\Gamma(2+n_1-k)\Gamma(m_3-n_1+k)}\notag\\
&\ \ \ \ \ -\sum_{k=0}^{n_1}(b_3+k)\frac{\Gamma(b_3+k)\Gamma(b_2+n_1-k)\Gamma(b_1+m_3-n_1+k)}{\Gamma(1+k)\Gamma(1+n_1-k)\Gamma(m_3-n_1+k+1)}\notag\\
&=\frac{\Gamma(b_3+n_1+1)\Gamma(b_2)\Gamma(b_1+m_3)}{\Gamma(1+n_1)\Gamma(1)\Gamma(1+m_3)}+\sum_{k=1}^{n_1}\frac{\Gamma(b_3+k)\Gamma(b_2+n_1-k+1)\Gamma(b_1+m_3-n_1+k-1)}{\Gamma(k)\Gamma(2+n_1-k)\Gamma(m_3-n_1+k)}\notag\\
&\ \ \ \ \ -\sum_{k=1}^{n_1+1}\frac{\Gamma(b_3+k)\Gamma(b_2+n_1-k+1)\Gamma(b_1+m_3-n_1+k-1)}{\Gamma(k)\Gamma(2+n_1-k)\Gamma(m_3-n_1+k)}\notag\\
&=\frac{\Gamma(b_3+n_1+1)\Gamma(b_2)\Gamma(b_1+m_3)}{\Gamma(1+n_1)\Gamma(1)\Gamma(1+m_3)}-\frac{\Gamma(b_3+n_1+1)\Gamma(b_2)\Gamma(b_1+m_3)}{\Gamma(1+n_1)\Gamma(1)\Gamma(1+m_3)}\notag\\
&=0.
\end{align}
The above calculation shows that $P_z$ WI is satisfied for $n_1+1\leq m_3$.\\
\\
(A-2) $P_z$ WI for $n_1\geq m_3$\\
When $n_1\geq m_3$, RHS of (\ref{Pz Ward Id}) is
\begin{align}
  \mathrm{(RHS)}&=(-m_3-b_1-b_2+1)\sum_{k=n_1-m_3+1}^{n_1}\frac{\Gamma(b_3+k)\Gamma(b_2+n_1-k)\Gamma(b_1+m_3-n_1+k-1)}{\Gamma(1+k)\Gamma(1+n_1-k)\Gamma(1+m_3-n_1+k-1)}\notag\\
  &\ -(n_1-m_3+b_3)\sum_{k=n_1-m_3}^{n_1}\frac{\Gamma(b_3+k)\Gamma(b_2+n_1-k)\Gamma(b_1+m_3-n_1+k)}{\Gamma(1+k)\Gamma(1+n_1-k)\Gamma(1+m_3-n_1+k)}\notag\\
  &\ +(n_1+1)\sum_{k=n_1-m_3+1}^{n_1+1}\frac{\Gamma(b_3+k)\Gamma(b_2+n_1+1-k)\Gamma(b_1+m_3-n_1+k-1)}{\Gamma(1+k)\Gamma(1+n_1+1-k)\Gamma(1+m_3-n_1+k-1)}.
\end{align}
We call the first term (D), the second term (E), and the third term (F).
\begin{align}
  \mathrm{(D)}&=(-m_3-b_1-b_2+1)\sum_{k=n_1-m_3+1}^{n_1}\frac{\Gamma(b_3+k)\Gamma(b_2+n_1-k)\Gamma(b_1+m_3-n_1+k-1)}{\Gamma(1+k)\Gamma(1+n_1-k)\Gamma(1+m_3-n_1+k-1)}\notag\\
  \mathrm{(E)}&=-(n_1-m_3+b_3)\sum_{k=n_1-m_3}^{n_1}\frac{\Gamma(b_3+k)\Gamma(b_2+n_1-k)\Gamma(b_1+m_3-n_1+k)}{\Gamma(1+k)\Gamma(1+n_1-k)\Gamma(1+m_3-n_1+k)}\notag\\
  \mathrm{(F)}&=(n_1+1)\sum_{k=n_1-m_3+1}^{n_1+1}\frac{\Gamma(b_3+k)\Gamma(b_2+n_1+1-k)\Gamma(b_1+m_3-n_1+k-1)}{\Gamma(1+k)\Gamma(1+n_1+1-k)\Gamma(1+m_3-n_1+k-1)}
\end{align}
Then, we get
\begin{align}
  \mathrm{(D)}&=-\sum_{k=n_1-m_3+1}^{n_1}(m_3+b_1+b_2-1)\frac{\Gamma(b_3+k)\Gamma(b_2+n_1-k)\Gamma(b_1+m_3-n_1+k-1)}{\Gamma(1+k)\Gamma(1+n_1-k)\Gamma(1+m_3-n_1+k-1)}\notag\\
  &=-\sum_{k=n_1-m_3+1}^{n_1}(m_3-n_1+k)\frac{\Gamma(b_3+k)\Gamma(b_2+n_1-k)\Gamma(b_1+m_3-n_1+k)}{\Gamma(1+k)\Gamma(1+n_1-k)\Gamma(1+m_3-n_1+k)}\notag\\
  &\hspace{1cm}-\sum_{k=n_1-m_3+1}^{n_1}(1+n_1-k)\frac{\Gamma(b_3+k)\Gamma(b_2+n_1-k+1)\Gamma(b_1+m_3-n_1+k-1)}{\Gamma(1+k)\Gamma(1+n_1-k+1)\Gamma(1+m_3-n_1+k-1)}.
\end{align}
We call the first term (D-1) and the second (D-2). Then, we get
\begin{align}
  &\mathrm{(RHS)}=\mathrm{((D-1)}+\mathrm{(E))}+\mathrm{((D-2)}+\mathrm{(F))}\notag\\
  &=-(n_1-m_3+b_3)\frac{\Gamma(b_3+n_1-m_3)\Gamma(b_2+m_3)\Gamma(b_1)}{\Gamma(1+n_1-m_3)\Gamma(1+m_3)\Gamma(1)}\notag\\
  &\hspace{1cm}-\sum_{k=n_1-m_3+1}^{n_1}(b_3+k)\frac{\Gamma(b_3+k)\Gamma(b_2+n_1-k)\Gamma(b_1+m_3-n_1+k)}{\Gamma(1+k)\Gamma(1+n_1-k)\Gamma(1+m_3-n_1+k)}\notag\\
  &+(n_1+1)\frac{\Gamma(b_3+n_1+1)\Gamma(b_2)\Gamma(b_1+m_3)}{\Gamma(1+n_1+1)\Gamma(1)\Gamma(1+m_3)}\notag\\
  &\hspace{1cm}+\sum_{k=n_1-m_3}^{n_1-1}(b_3+k)\frac{\Gamma(b_3+k)\Gamma(b_2+n_1-k)\Gamma(b_1+m_3-n_1+k)}{\Gamma(1+k)\Gamma(1+n_1-k)\Gamma(1+m_3-n_1+k)}\notag\\
  &=-\frac{\Gamma(b_3+n_1-m_3+1)\Gamma(b_2+m_3)\Gamma(b_1)}{\Gamma(1+n_1-m_3)\Gamma(1+m_3)\Gamma(1)}+\frac{\Gamma(b_3+n_1-m_3+1)\Gamma(b_2-m_3)\Gamma(b_1)}{\Gamma(1+n_1-m_3)\Gamma(1+m_3)\Gamma(1)}\notag\\
  &\hspace{1cm}+(n_1+1)\frac{\Gamma(b_3+n_1+1)\Gamma(b_2)\Gamma(b_1+m_3)}{\Gamma(1+n_1+1)\Gamma(1)\Gamma(1+m_3)}-(b_3+n_1)\frac{\Gamma(b_3+n_1)\Gamma(b_2)\Gamma(b_1+m_3)}{\Gamma(1+n_1)\Gamma(1)\Gamma(1+m_3)}\notag\\
  &=0.
\end{align}
The above calculation shows that $P_z$ WI is satisfied for $n_1\geq m_3$.\\
\\
(B-1) $K^z$ WI for $n_1\leq m_3$\\
When $n_1\leq m_3$, RHS of (\ref{Kz Ward Id}) is
\begin{align}
  \mathrm{(RHS)}&=-(m_3+1)\sum_{k=0}^{n_1}\frac{\Gamma(b_3+k)\Gamma(b_2+n_1-k)\Gamma(b_1+m_3-n_1+k+1)}{\Gamma(1+k)\Gamma(1+n_1-k)\Gamma(1+m_3-n_1+k+1)}\notag\\
  &\hspace{1cm}+(m_3-n_1+b_1)\sum_{k=0}^{n_1}\frac{\Gamma(b_3+k)\Gamma(b_2+n_1-k)\Gamma(b_1+m_3-n_1+k)}{\Gamma(1+k)\Gamma(1+n_1-k)\Gamma(1+m_3-n_1+k)}\notag\\
  &\hspace{1cm}+(b_2+b_3+n_1-1)\sum_{k=0}^{n_1-1}\frac{\Gamma(b_3+k)\Gamma(b_2+n_1-1-k)\Gamma(b_1+m_3-n_1+k+1)}{\Gamma(1+k)\Gamma(1+n_1-1-k)\Gamma(1+m_3-n_1+k+1)}.
\end{align}
We call the first term (G), the second term (H), and the third term (I).
\begin{align}
  \mathrm{(I)}&=\sum_{k=0}^{n_1-1}(b_2+n_1-1-k)\frac{\Gamma(b_3+k)\Gamma(b_2+n_1-1-k)\Gamma(b_1+m_3-n_1+k+1)}{\Gamma(1+k)\Gamma(1+n_1-1-k)\Gamma(1+m_3-n_1+k+1)}\notag\\
  &\hspace{1cm}+\sum_{k=0}^{n_1-1}(b_3+k)\frac{\Gamma(b_3+k)\Gamma(b_2+n_1-1-k)\Gamma(b_1+m_3-n_1+k+1)}{\Gamma(1+k)\Gamma(1+n_1-1-k)\Gamma(1+m_3-n_1+k+1)}\notag\\
  &=\sum_{k=0}^{n_1-1}(n_1-k)\frac{\Gamma(b_3+k)\Gamma(b_2+n_1-k)\Gamma(b_1+m_3-n_1+k+1)}{\Gamma(1+k)\Gamma(1+n_1-k)\Gamma(1+m_3-n_1+k+1)}\notag\\
  &\hspace{1cm}+\sum_{k=0}^{n_1-1}(1+k)\frac{\Gamma(b_3+k+1)\Gamma(b_2+n_1-1-k)\Gamma(b_1+m_3-n_1+k+1)}{\Gamma(1+k+1)\Gamma(1+n_1-1-k)\Gamma(1+m_3-n_1+k+1)}\notag\\
  &=\sum_{k=0}^{n_1-1}(n_1-k)\frac{\Gamma(b_3+k)\Gamma(b_2+n_1-k)\Gamma(b_1+m_3-n_1+k+1)}{\Gamma(1+k)\Gamma(1+n_1-k)\Gamma(1+m_3-n_1+k+1)}\notag\\
  &\hspace{1cm}+\sum_{k=1}^{n_1}k\frac{\Gamma(b_3+k)\Gamma(b_2+n_1-k)\Gamma(b_1+m_3-n_1+k)}{\Gamma(1+k)\Gamma(1+n_1-k)\Gamma(1+m_3-n_1+k)}
\end{align}
We call the first term (I-1) and the second (I-2). Then, we get
\begin{align}
  \mathrm{(RHS)}&=\mathrm{((I-1)+(G))}+\mathrm{((I-2)}+\mathrm{(H))}\notag\\
  &=-\sum_{k=0}^{n_1-1}(1+m_3-n_1+k)\frac{\Gamma(b_3+k)\Gamma(b_2+n_1-k)\Gamma(b_1+m_3-n_1+k+1)}{\Gamma(1+k)\Gamma(1+n_1-k)\Gamma(1+m_3-n_1+k+1)}\notag\\
  &\hspace{1cm}-(m_3+1)\frac{\Gamma(b_3+n_1)\Gamma(b_2)\Gamma(b_1+m_3+1)}{\Gamma(1+n_1)\Gamma(1)\Gamma(1+m_3+1)}\notag\\
  &\hspace{1cm}+\sum_{k=1}^{n_1}(m_3-n_1+b_1+k)\frac{\Gamma(b_3+k)\Gamma(b_2+n_1-k)\Gamma(b_1+m_3-n_1+k)}{\Gamma(1+k)\Gamma(1+n_1-k)\Gamma(1+m_3-n_1+k)}\notag\\
  &\hspace{1cm}+(m_3-n_1+b_1)\frac{\Gamma(b_3)\Gamma(b_2+n_1)\Gamma(b_1+m_3-n_1)}{\Gamma(1)\Gamma(1+n_1)\Gamma(1+m_3-n_1)}\notag\\
  &=-(1+m_3-n_1)\frac{\Gamma(b_3)\Gamma(b_2+n_1)\Gamma(b_1+m_3-n_1+1)}{\Gamma(1)\Gamma(1+n_1)\Gamma(2+m_3-n_1)}\notag\\
  &\hspace{1cm}+(m_3-n_1+b_1)\frac{\Gamma(b_3)\Gamma(b_2+n_1)\Gamma(b_1+m_3-n_1)}{\Gamma(1)\Gamma(1+n_1)\Gamma(1+m_3-n_1)}\notag\\
  &\hspace{1cm}(m_3+b_1)\frac{\Gamma(b_3+n_1)\Gamma(b_2)\Gamma(b_1+m_3)}{\Gamma(1+n_1)\Gamma(1)\Gamma(1+m_3)}\notag\\
  &\hspace{1cm}-(m_3+1)\frac{\Gamma(b_3+n_1)\Gamma(b_2)\Gamma(b_1+m_3+1)}{\Gamma(1+n_1)\Gamma(1)\Gamma(1+m_3+1)}\notag\\
  &=0.
\end{align}
The above calculation shows that $K_z$ WI is satisfied for $n_1\leq m_3$.\\
\\
(B-2) $K^z$ WI for $n_1\geq m_3+1$\\
When $n_1\geq m_3+1$, RHS of (\ref{Kz Ward Id}) is
\begin{align}
  0&=-(m_3+1)\sum_{k=n_1-1-m_3}^{n_1}\frac{\Gamma(b_3+k)\Gamma(b_2+n_1-k)\Gamma(b_1+m_3-n_1+k+1)}{\Gamma(1+k)\Gamma(1+n_1-k)\Gamma(1+m_3-n_1+k+1)}\notag\\
  &\hspace{1cm}+(m_3-n_1+b_1)\sum_{k=n_1-m_3}^{n_1}\frac{\Gamma(b_3+k)\Gamma(b_2+n_1-k)\Gamma(b_1+m_3-n_1+k)}{\Gamma(1+k)\Gamma(1+n_1-k)\Gamma(1+m_3-n_1+k)}\notag\\
  &\hspace{1cm}+(b_2+b_3+n_1-1)\sum_{k=n_1-1-m_3}^{n_1-1}\frac{\Gamma(b_3+k)\Gamma(b_2+n_1-1-k)\Gamma(b_1+m_3-n_1+k+1)}{\Gamma(1+k)\Gamma(1+n_1-1-k)\Gamma(1+m_3-n_1+k+1)}.
\end{align}
We call the first term (J), the second term (K), and the third term (L).
\begin{align}
  \mathrm{(L)}&=\sum_{k=n_1-m_3-1}^{n_1-1}(n_1-k)\frac{\Gamma(b_3+k)\Gamma(b_2+n_1-k)\Gamma(b_1+m_3-n_1+k+1)}{\Gamma(1+k)\Gamma(1+n_1-k)\Gamma(1+m_3-n_1+k+1)}\notag\\
  &\hspace{1cm}+\sum_{k=n_1-m_3}^{n_1}k\frac{\Gamma(b_3+k)\Gamma(b_2+n_1-k)\Gamma(b_1+m_3-n_1+k)}{\Gamma(1+k)\Gamma(1+n_1-k)\Gamma(1+m_3-n_1+k)}
\end{align}
We call the first term of this (L-1) and the second term (L-2). Then, we get
\begin{align}
  &\mathrm{(RHS)}=\mathrm{((L-1)+(J))}+\mathrm{((L-2)}+\mathrm{(K))}\notag\\
  &=-\sum_{k=n_1-m_3}^{n_1-1}(1+m_3-n_1+k)\frac{\Gamma(b_3+k)\Gamma(b_2+n_1-k)\Gamma(b_1+m_3-n_1+k+1)}{\Gamma(1+k)\Gamma(1+n_1-k)\Gamma(1+m_3-n_1+k+1)}\notag\\
  &\hspace{1cm}-(m_3+1)\frac{\Gamma(b_3+n_1)\Gamma(b_2)\Gamma(b_1+m_3+1)}{\Gamma(1+n_1)\Gamma(1)\Gamma(1+m_3+1)}\notag\\
  &\hspace{1cm}+\sum_{k=n_1-m_3}^{n_1}(b_1+m_3-n_1+k)\frac{\Gamma(b_3+k)\Gamma(b_2+n_1-k)\Gamma(b_1+m_3-n_1+k)}{\Gamma(1+k)\Gamma(1+n_1-k)\Gamma(1+m_3-n_1+k)}
  &\hspace{1cm}\notag\\
  &=-(m_3+1)\frac{\Gamma(b_3+n_1)\Gamma(b_2)\Gamma(b_1+m_3+1)}{\Gamma(1+n_1)\Gamma(1)\Gamma(1+m_3+1)}+(b_1+m_3)\frac{\Gamma(b_3+n_1)\Gamma(b_2)\Gamma(b_1+m_3)}{\Gamma(1+n_1)\Gamma(1)\Gamma(1+m_3)}\notag\\
  &=0.
\end{align}
The above calculation shows that $K_z$ WI is satisfied for $n_1\geq m_3+1$.\par
Proof completed. Our formula for the three-point function satisfies $P_z$ and $K_z$ WIs. So, our procedure is consistent with WIs.
\subsection{Factorization method}
This section summarizes the factorization method for the three-point function in two-dimensional CFT. The same argument can be made for a two-point function.
\subsubsection{Definitions of new modes}
Let us define
\begin{equation}
\OO\fft{J/R}{L/R}\eq \text{Fourier component of $\OO$ carrying momentum $\tfrac{J}{R}$ and energy $\tfrac{L}{R}$}.\label{def of new mode}
\end{equation}
For a product $C(z,\bar{z}) \equiv A(z)\tilde{B}(\bar{z})$ of a holomorphic local operator
$A(z)$ and an antiholomorphic
local operator $\tilde{B}(\bar{z})$, we have
\begin{equation}
C\fft{J}{L}=A\fft{\frac{J+L}{2}}{\frac{J+L}{2}}\tilde{B}\fft{\frac{J-L}{2}}{-\frac{J-L}{2}}.
\end{equation}
Define a local $\omega$-frame operator.
\begin{equation}
\OO_{\{\omega\}}(\omega,\bar{\omega})\eq\left(\dv{z}{\omega}\right)^h\left(\dv{\bar{z}}{\bar{\omega}}\right)^{\bar{h}}\OO(z,\bar{z})
\end{equation}
Here, $z=Re^{i\omega/R},\ \bar{z}=Re^{-i\bar{\omega}/R}$ and $\sigma^1\equiv\sigma\equiv\text{Re}[\omega],\ \sigma^2\equiv-\tau\equiv\text{Im}[\omega]$. The Jacobian is
\begin{equation}
  \dv{z}{\omega}=i\frac{z}{R}=ie^{i\omega/R},\hspace{1cm}\dv{\bar{z}}{\bar{\omega}}=-i\frac{\bar{z}}{R}=-ie^{-i\bar{\omega}/R}.
\end{equation}
So,
\begin{align}
  C_{\{\omega\}}=i^sR^{-\Delta}e^{is\sigma_1/R}e^{-\Delta\sigma_2/R}C_{\{z\}}.
\end{align}
Now we would like to express our $\omega$ frame
quantities as discrete sums of Fourier modes.
\begin{equation}
C_{\{\omega\}}(\sigma,t)=\sum_{P,E}e^{-iP\sigma+iEt}C\fft{P}{E}=\sum_{J,\mathcal{E}}e^{-iJ\sigma/R+i\mathcal{E} t/R}C\fft{J}{\mathcal{E}}
\end{equation}
We always have $P=J/R$. We will also
call $\mathcal{E}\equiv ER$ so $E=\mathcal{E}/R$.
The quantity $\mathcal{E}$ is dimensionless and
denotes the amount by which the operator
raises or lowers the eigenvalue of the dilatation
generator $\hat{\Delta}\equiv RH$.  It is not
quantized in integer units however, except for
modes of a holomorphic or an antiholomorphic
operator.
\subsubsection{Correlators of spacetime fourier modes}
Let us find correlators of the $C\fft{\PP}{\EE}$. We have
\begin{align}
  &\bra{0}C\fft{\PP_3}{\EE_3}C\fft{\PP_2}{\EE_2}C\fft{\PP_1}{\EE_1}\ket{0}\notag\\
  &=\bra{0}A\fft{J^{(+)}_3}{J^{(+)}_3}A\fft{J^{(+)}_2}{J^{(+)}_2}A\fft{J^{(+)}_1}{J^{(+)}_1}\ket{0}|_{J^{(+)}_I\eq\frac{1}{2}(\PP_I+\EE_I)}\bra{0}B\fft{J^{(-)}_3}{-J^{(-)}_3}B\fft{J^{(-)}_2}{-J^{(-)}_2}B\fft{J^{(-)}_1}{-J^{(-)}_1}\ket{0}|_{J^{(-)}_I\eq\frac{1}{2}(\PP_I-\EE_I)}\notag\\
  &=\delta(\PP_1+\PP_2+\PP_3)\delta(\EE_1+\EE_2+\EE_3)\KK_{\mathrm{red}}(n_L,n'_L)\KK_{\mathrm{red}}(n_R,n'_R),
\end{align}
where
\begin{align}
  n_L&=\frac{1}{2}(\PP_1+\EE_1)-h_1, & n_R&=-\frac{1}{2}(\PP_1-\EE_1)-\tilde{h}_1\notag\\
  n'_L&=-\frac{1}{2}(\PP_3+\EE_3)-h_3, & n'_R&=\frac{1}{2}(\PP_3-\EE_3)-\tilde{h}_3.
\end{align}
All the $n_L,n_R,n'_L,n'_R$ are nonnegative. So the minimum energies of the in-state and out-state are when $n_L=n_R=n'_L=n'_R=0$. Now define
 \begin{equation}
 l\eq\text{min}\{n_L,n_R\},\hspace{1cm}
 l'\eq\text{min}\{n'_L,n'_R\}.
 \end{equation}
 Then, we get
 \begin{equation}
   \EE_1=\Delta_1+|\PP_1-s_1|+2l,\hspace{1cm}\EE_3=\Delta_3+|\PP_3-s_3|+2l'.
 \end{equation}
 The interpretation of these equations is as follows.
 The energy in a conformal family is
 minimized by the primary state.  The primary state is the one for which $\PP_1=s_1$ (for the
 in-state) or $\PP_3=s_3$ (for the out-state),
 and for which the excitation energy within
 a given momentum sector of the conformal family is as small as possible -- namely zero.  The
 excitation energy (in units of $1/R$
 as always, of course) within
 a given momentum sector of the conformal family is the thing we're calling $2l$ (for
 the in-state) or $2l'$ (for the out-state).
 Then, for a given momentum sector within a conformal family, the minimum energy is
$\Delta_1+|\PP_1-s_1|$ (for the in-state), or $\Delta_3+|\PP_3-s_3|$ (for the out-state).
 Then, for a given momentum-sector within a conformal family, the minimum energy is
$\Delta_1+|\PP_1-s_1|$ (for the in-state), or $\Delta_3+|\PP_3-s_3|$ (for the out-state).
\subsection{Analytic continuation for conformal weight}
\subsubsection{New notation}
To extend the domain of conformal weight, we need a new notation. Remember that $\tilde{\OO}_n(r)$ is defined as
\begin{equation}
  \tilde{\OO}_n(r)\eq\int\frac{\dd\theta}{2\pi}\OO_n(r,\theta)e^{-in\theta}.
\end{equation}
Let us rescale this mode so that it becomes dimensionless.
\begin{equation}
  \tilde{\OO}_{\{n\}}\eq r^{-\Delta}\tilde{\OO}_n(r)=r^{-h-\tilde{h}}\tilde{\OO}_n(r)
\end{equation}
Using this new mode, we can write the actions of the conformal operators on the operator as follows.
\begin{align}
  L_{-1}\cdot\tilde{\OO}_{\{n\}}&=\frac{1}{2r}(\hat{E}-h-\tilde{h}+n+1)\tilde{\OO}_{\{n+1\}}\\
  \tilde{L}_{-1}\cdot\tilde{\OO}_{\{n\}}&=\frac{1}{2r}(\hat{E}-h-\tilde{h}-n+1)\tilde{\OO}_{\{n-1\}}\\
  L_0\cdot\tilde{\OO}_{\{n\}}&=\frac{1}{2}(\hat{E}+h-\tilde{h}+n)\tilde{\OO}_{\{n\}}\\
  \tilde{L}_0\cdot\tilde{\OO}_{\{n\}}&=\frac{1}{2}(\hat{E}-h+\tilde{h}-n)\tilde{\OO}_{\{n\}}\\
  L_1\cdot\tilde{\OO}_{\{n\}}&=\frac{r}{2}(\hat{E}+3h-\tilde{h}+n-1)\tilde{\OO}_{\{n-1\}}\\
  \tilde{L}_1\cdot\tilde{\OO}_{\{n\}}&=\frac{r}{2}(\hat{E}-h+3\tilde{h}-n-1)\tilde{\OO}_{\{n+1\}}
\end{align}
where $\hat{E}$ is the differential operator $\hat{E}\eq r\partial_r$.
Next, define
\begin{equation}
  \tilde{\OO}_{\langle J\rangle}\eq\tilde{\OO}_{\{J-h+\tilde{h}\}}.
\end{equation}
Then, we get
\begin{align}
  L_{-1}\cdot\tilde{\OO}_{\langle J\rangle}&=\frac{1}{2r}(\hat{E}-2h+J+1)\tilde{\OO}_{\langle J+1\rangle}\\
  \tilde{L}_{-1}\cdot\tilde{\OO}_{\langle J\rangle}&=\frac{1}{2r}(\hat{E}-2\tilde{h}-J+1)\tilde{\OO}_{\langle J-1\rangle}\\
  L_0\cdot\tilde{\OO}_{\langle J\rangle}&=\frac{1}{2}(\hat{E}+J)\tilde{\OO}_{\langle J\rangle}\\
  \tilde{L}_0\cdot\tilde{\OO}_{\langle J\rangle}&=\frac{1}{2}(\hat{E}-J)\tilde{\OO}_{\langle J\rangle}\\
  L_1\cdot\tilde{\OO}_{\langle J\rangle}&=\frac{r}{2}(\hat{E}+2h+J-1)\tilde{\OO}_{\langle J-1\rangle}\\
  \tilde{L}_1\cdot\tilde{\OO}_{\langle J\rangle}&=\frac{r}{2}(\hat{E}+2\tilde{h}-J-1)\tilde{\OO}_{\langle J+1\rangle}.
\end{align}
Next, expand this by $r/R$.
\begin{equation}
  \tilde{\OO}_{\langle J\rangle}(r)=\sum_{\epsilon}\tilde{\OO}_{\langle J,\varepsilon\rangle}\left(\frac{r}{R}\right)^{\varepsilon}
\end{equation}
For $\tilde{\OO}_{\langle J,\varepsilon\rangle}$, the actions of conformal generators are described as
\begin{align}
  L_{-1}\cdot\tilde{\OO}_{\langle J,\varepsilon\rangle}&=\frac{1}{2R}(\varepsilon-2h+J+2)\tilde{\OO}_{\langle J+1,\varepsilon+1\rangle}\\
  \tilde{L}_{-1}\cdot\tilde{\OO}_{\langle J,\varepsilon\rangle}&=\frac{1}{2R}(\varepsilon-2\tilde{h}-J+2)\tilde{\OO}_{\langle J-1,\varepsilon+1\rangle}\\
  L_{0}\cdot\tilde{\OO}_{\langle J,\varepsilon\rangle}&=\frac{1}{2R}(\varepsilon+J)\tilde{\OO}_{\langle J,\varepsilon\rangle}\\
  \tilde{L}_{-1}\cdot\tilde{\OO}_{\langle J,\varepsilon\rangle}&=\frac{1}{2R}(\varepsilon-J)\tilde{\OO}_{\langle J,\varepsilon\rangle}\\
  L_{1}\cdot\tilde{\OO}_{\langle J,\varepsilon\rangle}&=\frac{R}{2}(\varepsilon+2h+J-2)\tilde{\OO}_{\langle J-1,\varepsilon-1\rangle}\\
  \tilde{L}_{1}\cdot\tilde{\OO}_{\langle J,\varepsilon\rangle}&=\frac{R}{2}(\varepsilon+2\tilde{h}-J-2)\tilde{\OO}_{\langle J+1,\varepsilon-1\rangle}.
\end{align}
Define the variables $\hat{\varepsilon}^{\pm}\eq\frac{1}{2}(\varepsilon\pm J)$, and rewrite the above equations in terms of them. We define new mode as
\begin{equation}
\tilde{\OO}_{\langle J,\varepsilon\rangle}=\tilde{\OO}_{[\frac{\varepsilon+J}{2},\frac{\varepsilon-J}{2}]}.
\end{equation}
Then, we get
\begin{align}
  L_{-1}\cdot\tilde{\OO}_{[\hat{\varepsilon}^+,\hat{\varepsilon}^-]}&=\frac{1}{R}(\hat{\varepsilon}^++1-h)\tilde{\OO}_{[\hat{\varepsilon}^++1,\hat{\varepsilon}^-]}\\
  \tilde{L}_{-1}\cdot\tilde{\OO}_{[\hat{\varepsilon}^+,\hat{\varepsilon}^-]}&=\frac{1}{R}(\hat{\varepsilon}^-+1-\tilde{h})\tilde{\OO}_{[\hat{\varepsilon}^+,\hat{\varepsilon}^-+1]}\\
  L_{0}\cdot\tilde{\OO}_{[\hat{\varepsilon}^+,\hat{\varepsilon}^-]}&=\hat{\varepsilon}^+\tilde{\OO}_{[\hat{\varepsilon}^+,\hat{\varepsilon}^-]}\\
  \tilde{L}_{0}\cdot\tilde{\OO}_{[\hat{\varepsilon}^+,\hat{\varepsilon}^-]}&=\hat{\varepsilon}^-\tilde{\OO}_{[\hat{\varepsilon}^+,\hat{\varepsilon}^-]}\\
  L_{1}\cdot\tilde{\OO}_{[\hat{\varepsilon}^+,\hat{\varepsilon}^-]}&=R(\hat{\varepsilon}^+-1+h)\tilde{\OO}_{[\hat{\varepsilon}^+-1,\hat{\varepsilon}^-]}\\
  \tilde{L}_{1}\cdot\tilde{\OO}_{[\hat{\varepsilon}^+,\hat{\varepsilon}^-]}&=R(\hat{\varepsilon}^--1+\tilde{h})\tilde{\OO}_{[\hat{\varepsilon}^+,\hat{\varepsilon}^--1]}.
\end{align}
Absorbing powers of $R$ into conformal generator $L_m\rightarrow R^{-m}L_m$ gives
\begin{align}
  L_{-1}\cdot\tilde{\OO}_{[\hat{\varepsilon}^+,\hat{\varepsilon}^-]}&=(\hat{\varepsilon}^++1-h)\tilde{\OO}_{[\hat{\varepsilon}^++1,\hat{\varepsilon}^-]}\\
  \tilde{L}_{-1}\cdot\tilde{\OO}_{[\hat{\varepsilon}^+,\hat{\varepsilon}^-]}&=(\hat{\varepsilon}^-+1-\tilde{h})\tilde{\OO}_{[\hat{\varepsilon}^+,\hat{\varepsilon}^-+1]}\\
  L_{0}\cdot\tilde{\OO}_{[\hat{\varepsilon}^+,\hat{\varepsilon}^-]}&=\hat{\varepsilon}^+\tilde{\OO}_{[\hat{\varepsilon}^+,\hat{\varepsilon}^-]}\\
  \tilde{L}_{0}\cdot\tilde{\OO}_{[\hat{\varepsilon}^+,\hat{\varepsilon}^-]}&=\hat{\varepsilon}^-\tilde{\OO}_{[\hat{\varepsilon}^+,\hat{\varepsilon}^-]}\\
  L_{1}\cdot\tilde{\OO}_{[\hat{\varepsilon}^+,\hat{\varepsilon}^-]}&=(\hat{\varepsilon}^+-1+h)\tilde{\OO}_{[\hat{\varepsilon}^+-1,\hat{\varepsilon}^-]}\\
  \tilde{L}_{1}\cdot\tilde{\OO}_{[\hat{\varepsilon}^+,\hat{\varepsilon}^-]}&=(\hat{\varepsilon}^--1+\tilde{h})\tilde{\OO}_{[\hat{\varepsilon}^+,\hat{\varepsilon}^--1]}.
\end{align}
They can be summarized as
\begin{align}
  L_m\cdot\tilde{\OO}_{[\hat{\varepsilon}^+,\hat{\varepsilon}^-]}&=[(h-1)m+\hat{\varepsilon}^+]\tilde{\OO}_{[\hat{\varepsilon}^+-m,\hat{\varepsilon}^-]}\\
  \tilde{L}_m\cdot\tilde{\OO}_{[\hat{\varepsilon}^+,\hat{\varepsilon}^-]}&=[(\tilde{h}-1)m+\hat{\varepsilon}^+]\tilde{\OO}_{[\hat{\varepsilon}^+,\hat{\varepsilon}^--m]}.
\end{align}
\subsubsection{Two-point function}
Let us analyze the two-point function again. First, define
\begin{equation}
  \YY2[h,\tilde{h}|\hat{\varepsilon}^+,\hat{\varepsilon}^-]\eq\bra{0}\tilde{\OO}_{[-\hat{\varepsilon}^+,-\hat{\varepsilon}^-]}\tilde{\OO}_{[\hat{\varepsilon}^+,\hat{\varepsilon}^-]}\ket{0}.
\end{equation}
Let us calculate the commutator of $L_{-1}$ and $\tilde{\OO}_{[-\hat{\varepsilon}^+,-\hat{\varepsilon}^-]}\tilde{\OO}_{[\hat{\varepsilon}^+-1,\hat{\varepsilon}^-]}$.
\begin{align}
  0&=\bra{0}\left[L_{-1},\tilde{\OO}_{[-\hat{\varepsilon}^+,-\hat{\varepsilon}^-]}\tilde{\OO}_{[\hat{\varepsilon}^+-1,\hat{\varepsilon}^-]}\right]\ket{0}\notag\\
  &=[1-h-\hat{\varepsilon}^+]\bra{0}\tilde{\OO}_{[1-\hat{\varepsilon}^+,-\hat{\varepsilon}^-]}\tilde{\OO}_{[\hat{\varepsilon}^+-1,\hat{\varepsilon}^-]}\ket{0}+[\hat{\varepsilon}^+-h]\bra{0}\tilde{\OO}_{[-\hat{\varepsilon}^+,-\hat{\varepsilon}^-]}\tilde{\OO}_{[\hat{\varepsilon}^+,\hat{\varepsilon}^-]}\ket{0}\notag\\
  &=[1-h-\hat{\varepsilon}^+]\YY2[h,\tilde{h}|\hat{\varepsilon}^+-1,\hat{\varepsilon}^-]+[\hat{\varepsilon}^+-h]\YY2[h,\tilde{h}|\hat{\varepsilon}^+,\hat{\varepsilon}^-]
\end{align}
So, we have
\begin{equation}
  \YY2[h,\tilde{h}|\hat{\varepsilon}^+,\hat{\varepsilon}^-]=\frac{\hat{\varepsilon}^++h-1}{\hat{\varepsilon}^+-h}\YY2[h,\tilde{h}|\hat{\varepsilon}^+-1,\hat{\varepsilon}^-].
\end{equation}
Shifitng $\hat{\varepsilon}^+$ by 1 gives
\begin{equation}
  \YY2[h,\tilde{h}|\hat{\varepsilon}^++1,\hat{\varepsilon}^-]=\frac{\hat{\varepsilon}^++h}{\hat{\varepsilon}^+-h+1}\YY2[h,\tilde{h}|\hat{\varepsilon}^+,\hat{\varepsilon}^-].
\end{equation}
Similarly, the $\tilde{L}_{-1}$ identity gives
\begin{equation}
  \YY2[h,\tilde{h}|\hat{\varepsilon}^+,\hat{\varepsilon}^-+1]=\frac{\hat{\varepsilon}^-+\tilde{h}}{\hat{\varepsilon}^--\hat{h}+1}\YY2[h,\tilde{h}|\hat{\varepsilon}^+,\hat{\varepsilon}^-].
\end{equation}
In terms of $n_L$ and $n_R$,
\begin{align}
  \YY2[h,\tilde{h}|n_L+1,n_R]&=\frac{n_L+2h}{n_L+1}\YY2[h,\tilde{h}|n_L,n_R]\\
  \YY2[h,\tilde{h}|n_L,n_R+1]&=\frac{n_R+2\tilde{h}}{n_R+1}\YY2[h,\tilde{h}|n_L,n_R].
\end{align}
To write this recursion relation simpler, we define
\begin{equation}
  \RRR\YY2[h,\tilde{h},n_L,n_R]=\Gamma(n_L+1)\Gamma(n_R+1)\YY2[h,\tilde{h}|n_L,n_R].
\end{equation}
And in terms of it, we can write the recursion relations as
\begin{align}
  \RRR\YY2[h,\tilde{h},n_L+1,n_R]&=(n_L+2h)\RRR\YY2[h,\tilde{h},n_L,n_R]\\
  \RRR\YY2[h,\tilde{h},n_L,n_R+1]&=(n_R+2\tilde{h})\RRR\YY2[h,\tilde{h},n_L,n_R].
\end{align}
From these recursion relations, we get
\begin{equation}
  \RRR\YY2[h,\tilde{h},n_L,n_R]=\frac{\Gamma(n_R+2h)}{\Gamma(2h)}\frac{\Gamma(n_L+2\tilde{h})}{\Gamma(2\tilde{h})}\RRR\YY2[h,\tilde{h},0,0].
\end{equation}
So,
\begin{equation}
\YY2[h,\tilde{h},n_L,n_R]=\frac{\Gamma(n_R+2h)}{\Gamma(n_L+1)\Gamma(2h)}\frac{\Gamma(n_L+2\tilde{h})}{\Gamma(n_R+1)\Gamma(2\tilde{h})}\RRR\YY2[h,\tilde{h},0,0].
\end{equation}
We can calculate $\RRR\YY2[h,\tilde{h},0,0]$ with ease. And this has the same form as our result, for example (\ref{two-point-func in two-dim CFT}), in the main text. It is important to note that this result can be applied to all pairs of real numbers except negative integers, $(h,\tilde{h})$.
\subsubsection{Three-point function}
In the same way, we can extend the domain of $(h_1,\tilde{h}_1,h_2,\tilde{h}_2,h_3,\tilde{h}_3)$ in our formula for the three-point function. First, define the three-point function as
\begin{equation}
  \YY 3[h_{3,2,1},\tilde{h}_{3,2,1}|\hat{\varepsilon}^+_3,\hat{\varepsilon}^+_1,\hat{\varepsilon}^-_3,\hat{\varepsilon}^-_1]\eq\bra{0}\tilde{\OO}^{(3)}_{[-\hat{\varepsilon}^+_3,-\hat{\varepsilon}^-_3]}\tilde{\OO}^{(2)}_{[\hat{\varepsilon}^+_3-\hat{\varepsilon}^+_1,\hat{\varepsilon}^-_3-\hat{\varepsilon}^-_1]}\tilde{\OO}^{(1)}_{[\hat{\varepsilon}^+_1,\hat{\varepsilon}^-_1]}\ket{0}.
\end{equation}
Commuting $L_{-1}$ with $\tilde{\OO}^{(3)}_{[-\hat{\varepsilon}^+_3,-\hat{\varepsilon}^-_3]}\tilde{\OO}^{(2)}_{[\hat{\varepsilon}^+_3-\hat{\varepsilon}^+_1,\hat{\varepsilon}^-_3-\hat{\varepsilon}^-_1]}\tilde{\OO}^{(1)}_{[\hat{\varepsilon}^+_1-1,\hat{\varepsilon}^-_1]}$ and taking the vacuum expectation value give,
\begin{multline}
  0=[\hat{\varepsilon}^+_1-h_1]\YY 3[h_{3,2,1},\tilde{h}_{3,2,1}|\hat{\varepsilon}^+_3,\hat{\varepsilon}^+_1,\hat{\varepsilon}^-_3,\hat{\varepsilon}^-_1]\\
  +[\hat{\varepsilon}^+_3-\hat{\varepsilon}^+_1+1-h_2]\YY 3[h_{3,2,1},\tilde{h}_{3,2,1}|\hat{\varepsilon}^+_3,\hat{\varepsilon}^+_1-1,\hat{\varepsilon}^-_3,\hat{\varepsilon}^-_1]\\
  +[-\hat{\varepsilon}_3+1-h_3]\YY 3[h_{3,2,1},\tilde{h}_{3,2,1}|\hat{\varepsilon}^+_3-1,\hat{\varepsilon}^+_1-1,\hat{\varepsilon}^-_3,\hat{\varepsilon}^-_1].
\end{multline}
Next, commuting $L_1$ with $\tilde{\OO}^{(3)}_{[-\hat{\varepsilon}^+_3,-\hat{\varepsilon}^-_3]}\tilde{\OO}^{(2)}_{[\hat{\varepsilon}^+_3-\hat{\varepsilon}^+_1,\hat{\varepsilon}^-_3-\hat{\varepsilon}^-_1]}\tilde{\OO}^{(1)}_{[\hat{\varepsilon}^+_1+1,\hat{\varepsilon}^-_1]}$ and taking the vacuum expectation value give,
\begin{multline}
  0=[\hat{\varepsilon}^+_1+h_1]\YY 3[h_{3,2,1},\tilde{h}_{3,2,1}|\hat{\varepsilon}^+_3,\hat{\varepsilon}^+_1,\hat{\varepsilon}^-_3,\hat{\varepsilon}^-_1]\\
  +[\hat{\varepsilon}^+_3-\hat{\varepsilon}^+_1-1+h_2]\YY 3[h_{3,2,1},\tilde{h}_{3,2,1}|\hat{\varepsilon}^+_3,\hat{\varepsilon}^+_1+1,\hat{\varepsilon}^-_3,\hat{\varepsilon}^-_1]\\
  +[-\hat{\varepsilon}_3-1+h_3]\YY 3[h_{3,2,1},\tilde{h}_{3,2,1}|\hat{\varepsilon}^+_3+1,\hat{\varepsilon}^+_1+1,\hat{\varepsilon}^-_3,\hat{\varepsilon}^-_1].
\end{multline}
In this case, each $\hat{\varepsilon}$ can be written as
\begin{align}
  \hat{\varepsilon}^+_1&=n_{1,L}+h_1, \hspace{1cm} \hat{\varepsilon}^-_1=n_{1,R}+\tilde{h}_1\notag\\
  \hat{\varepsilon}^+_3&=n_{3,L}+h_3, \hspace{1cm} \hat{\varepsilon}^-_3=n_{3,R}+\tilde{h}_3.
\end{align}
Then, we get
\begin{align}
  &\YY 3[h_{3,2,1},\tilde{h}_{3,2,1}|n_{3,L}+1,n_{1,L}+1,n_{3,R},n_{1,R}]\notag\\
  &=\frac{n_{1,L}-n_{3,L}-h_3+h_1+h_2-1}{n_{1,L}+1}\YY 3[h_{3,2,1},\tilde{h}_{3,2,1}|n_{3,L}+1,n_{1,L},n_{3,R},n_{1,R}]\notag\\
  &\hspace{1cm}+\frac{n_{3,L}+2h_3}{n_{1,L}+1}\YY 3[h_{3,2,1},\tilde{h}_{3,2,1}|n_{3,L},n_{1,L},n_{3,R},n_{1,R}].\label{recursion1}
\end{align}
And
\begin{align}
  &\YY 3[h_{3,2,1},\tilde{h}_{3,2,1}|n_{3,L}+1,n_{1,L}+1,n_{3,R},n_{1,R}]\notag\\
  &=\frac{n_{3,L}-n_{1,L}-h_1+h_3+h_2-1}{n_{3,L}+1}\YY 3[h_{3,2,1},\tilde{h}_{3,2,1}|n_{3,L},n_{1,L}+1,n_{3,R},n_{1,R}]\notag\\
  &\hspace{1cm}+\frac{n_{1,L}+2h_1}{n_{3,L}+1}\YY 3[h_{3,2,1},\tilde{h}_{3,2,1}|n_{3,L},n_{1,L},n_{3,R},n_{1,R}].\label{recursion2}
\end{align}
In the same way, we can get the following two equations from WI for $\tilde{L}_{-1}$ and $\tilde{L}_1$.
\begin{align}
  &\YY 3[h_{3,2,1},\tilde{h}_{3,2,1}|n_{3,L},n_{1,L},n_{3,R}+1,n_{1,R}+1]\notag\\
  &=\frac{n_{1,R}-n_{3,R}-\tilde{h}_3+\tilde{h}_1+\tilde{h}_2-1}{n_{1,R}+1}\YY 3[h_{3,2,1},\tilde{h}_{3,2,1}|n_{3,L},n_{1,L},n_{3,R}+1,n_{1,R}]\notag\\
  &\hspace{1cm}+\frac{n_{3,R}+2\tilde{h}_3}{n_{1,R}+1}\YY 3[h_{3,2,1},\tilde{h}_{3,2,1}|n_{3,L},n_{1,L},n_{3,R},n_{1,R}]\label{recursion3}
\end{align}
\begin{align}
  &\YY 3[h_{3,2,1},\tilde{h}_{3,2,1}|n_{3,L},n_{1,L},n_{3,R}+1,n_{1,R}+1]\notag\\
  &=\frac{n_{3,R}-n_{1,R}-\tilde{h}_1+\tilde{h}_3+\tilde{h}_2-1}{n_{3,R}+1}\YY 3[h_{3,2,1},\tilde{h}_{3,2,1}|n_{3,L},n_{1,L},n_{3,R},n_{1,R}+1]\notag\\
  &\hspace{1cm}+\frac{n_{1,R}+2\tilde{h}_1}{n_{3,R}+1}\YY 3[h_{3,2,1},\tilde{h}_{3,2,1}|n_{3,L},n_{1,L},n_{3,R},n_{1,R}]\label{recursion4}
\end{align}
And from unitarity, we have
\begin{equation}
  \YY 3[h_{3,2,1},\tilde{h}_{3,2,1}|n_{3,L},n_{1,L},n_{3,R},n_{1,R}]=0\ \ \ \mathrm{if}\ n_{3,L}<0\ \mathrm{or}\ n_{1,L}<0\ \mathrm{or}\ n_{3,R}<0\ \mathrm{or}\ n_{1,R}<0. \label{unitary}
\end{equation}
From the recursion relations and the unitarity condition, we can get the three-point functions for semi-positive integers $(n_{3,L},n_{1,L},n_{3,R},n_{3,L})$. From here, we summarize how we can extend the domain of $(n_{3,L},n_{1,L},n_{3,R},n_{3,L})$.\par
First, fix $n_{3,R}=n_{1,R}=0$ and derive all $\YY 3[h_{3,2,1},\tilde{h}_{3,2,1}|n_{3,L},n_{1,L},0,0]$ values using recursion relations (\ref{recursion1}) and (\ref{recursion2}). From (\ref{recursion1}) with $n_{1,R}=n_{3,R}=0$ and $n_{3,L}=-1$, we get
\begin{equation}
    \YY 3[h_{3,2,1},\tilde{h}_{3,2,1}|0,n_{1,L}+1,0,0]=\frac{n_{1,L}-h_3+h_1+h_2}{n_{1,L}+1}\YY 3[h_{3,2,1},\tilde{h}_{3,2,1}|0,n_{1,L},0,0].
\end{equation}
From this relation, we can see that $\YY 3[h_{3,2,1},\tilde{h}_{3,2,1}|0,n_{1,L},0,0]$ is a $n_{1,L}$-th degree polynomial in $h_{3,2,1}$.
And from (\ref{recursion2}) with $n_{1,R}=n_{3,R}=0$ and $n_{1,L}=-1$, we get
\begin{equation}
    \YY 3[h_{3,2,1},\tilde{h}_{3,2,1}|n_{3,L}+1,0,0,0]=\frac{n_{3,L}-h_1+h_3+h_2}{n_{3,L}+1}\YY 3[h_{3,2,1},\tilde{h}_{3,2,1}|n_{3,L},0,0,0].
\end{equation}
From this relation, we can see that $\YY 3[h_{3,2,1},\tilde{h}_{3,2,1}|n_{3,L},0,0,0]$ is a $n_{3,L}$-th degree polynomial in $h_{3,2,1}$.\par
Next, from (\ref{recursion1}) with $n_{1,R}=n_{3,R}=0$ and $n_{3,L}=0$, we get
\begin{align}
    \YY 3[h_{3,2,1},\tilde{h}_{3,2,1}|1,n_{1,L}+1,0,0]&=\frac{n_{1,L}-h_3+h_1+h_2-1}{n_{1,L}+1}\YY 3[h_{3,2,1},\tilde{h}_{3,2,1}|1,n_{1,L},0,0]\notag\\
    &\hspace{1cm}+\frac{2h_3}{n_{1,L}+1}\YY 3[h_{3,2,1},\tilde{h}_{3,2,1}|0,n_{1,L},0,0]
\end{align}
When $n_L=0$, the first term on the RHS is a quadratic polynomial in $h_{1,2,3}$, and the second term is a linear polynomial in $h_{1,2,3}$, so the LHS is a quadratic polynomial in $h_{1,2,3}$.\par
When $n_L=1$, the first term on the RHS is a cubic polynomial in $h_{1,2,3}$, and the second term is a quadratic polynomial in $h_{1,2,3}$, so the LHS is a cubic polynomial in $h_{1,2,3}$.\par
In this way, we can determine all of $ \YY 3[h_{3,2,1},\tilde{h}_{3,2,1}|1,n_{1,L},0,0]$ for $n_{1,L}\geq0$. And $ \YY 3[h_{3,2,1},\tilde{h}_{3,2,1}|1,n_{1,L},0,0]$ is a $n_{1,L}+1$-th polynomial in $h_{1,2,3}$. Similarly, we can determine all of $ \YY 3[h_{3,2,1},\tilde{h}_{3,2,1}|n_{3,L},1,0,0]$ for $n_{3,L}\geq0$, and it is a polynomial in $h_{1,2,3}$.\par
Mathematical induction on $\mathrm{Min}\{n_{3,L},n_{1,L}\}$ shows that $\YY 3[h_{3,2,1},\tilde{h}_{3,2,1}|n_{3,L},n_{1,L},0,0]$ is a $n_{1,L}+n_{3,L}$-th polynomial in $h_{3,2,1}$. In the antiholomorphic part, fixing the value of $(n_{3,L},n_{1,L})$, and using (\ref{recursion3}) and (\ref{recursion4}), we can perform mathematical induction similarly.\par
The point is that all of  $\YY 3[h_{3,2,1},\tilde{h}_{3,2,1}|n_{3,L},n_{1,L},n_{3,R},n_{1,R}]$ are determined by the followings.\par
\begin{itemize}
  \item The unitarity condition (\ref{unitary})
  \item The arbitrarily normalized initial value $\YY 3[h_{3,2,1},\tilde{h}_{3,2,1}|0,0,0,0]$
  \item The left moving recursion relations (\ref{recursion1}) and (\ref{recursion2}), and the right moving recursion relations (\ref{recursion3}) and (\ref{recursion4}). They are derived from the $L_{\pm1}$ and the $\tilde{L}_{\pm1}$ WIs respectively.
\end{itemize}
And another important thing is that the three-point function is a polynomial in $h_{1,2,3}$, and its degree is not greater than $n_{3,L}+n_{1,L}$, and that the three-point function is a polynomial in $\tilde{h}_{1,2,3}$, and its degree is not greater than $n_{3,R}+n_{1,R}$.\\
\begin{figure}[ht]
  \begin{minipage}{0.47\columnwidth}
    \centering
    \includegraphics[width=\columnwidth]{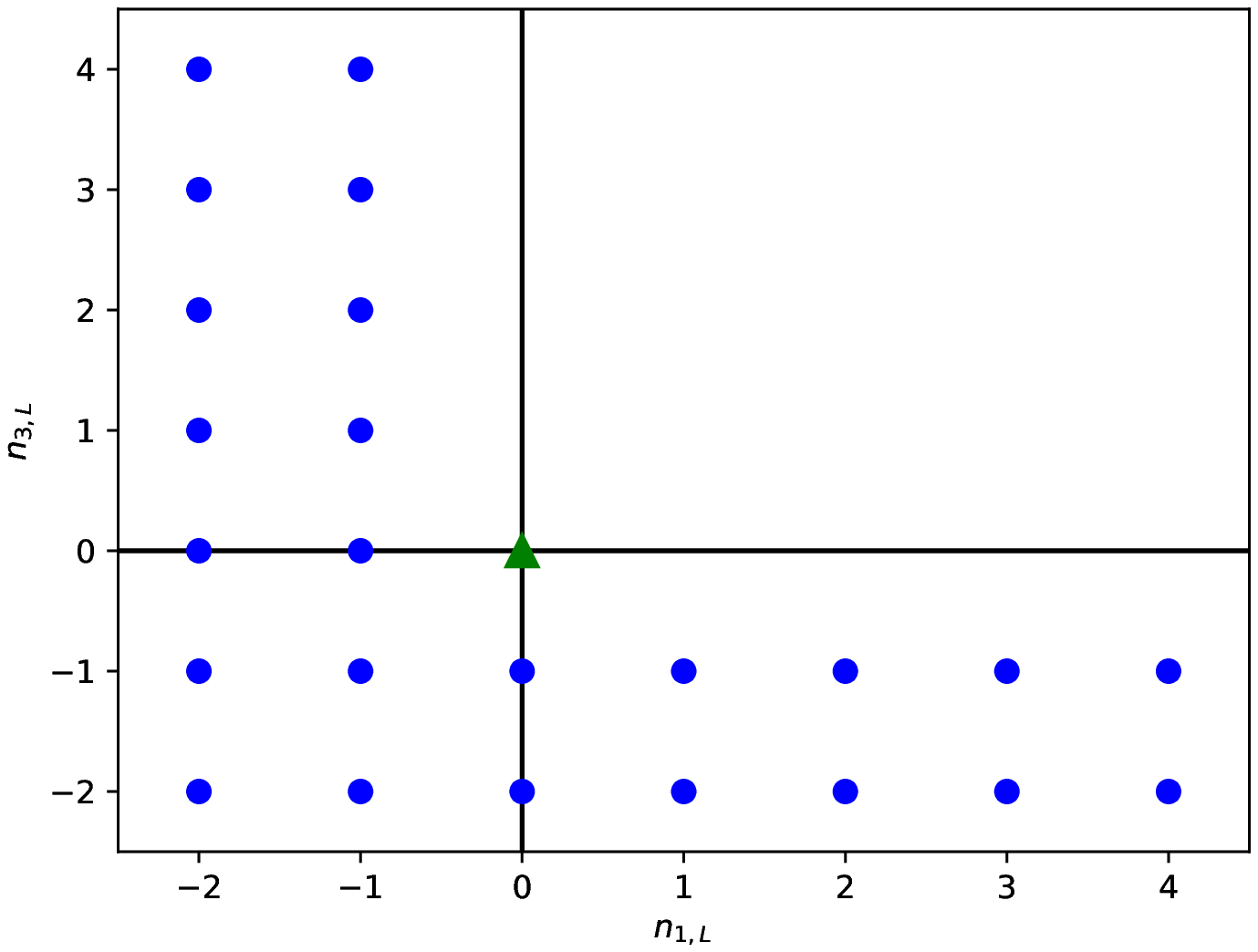}
    \caption{Prepare the arbitrarily normalized initial value (green triangle) and unitary condition (blue circle).}
    \label{analytic_conti1}
  \end{minipage}
  \hspace{0.04\columnwidth}
  \begin{minipage}{0.47\columnwidth}
    \centering
    \includegraphics[width=\columnwidth]{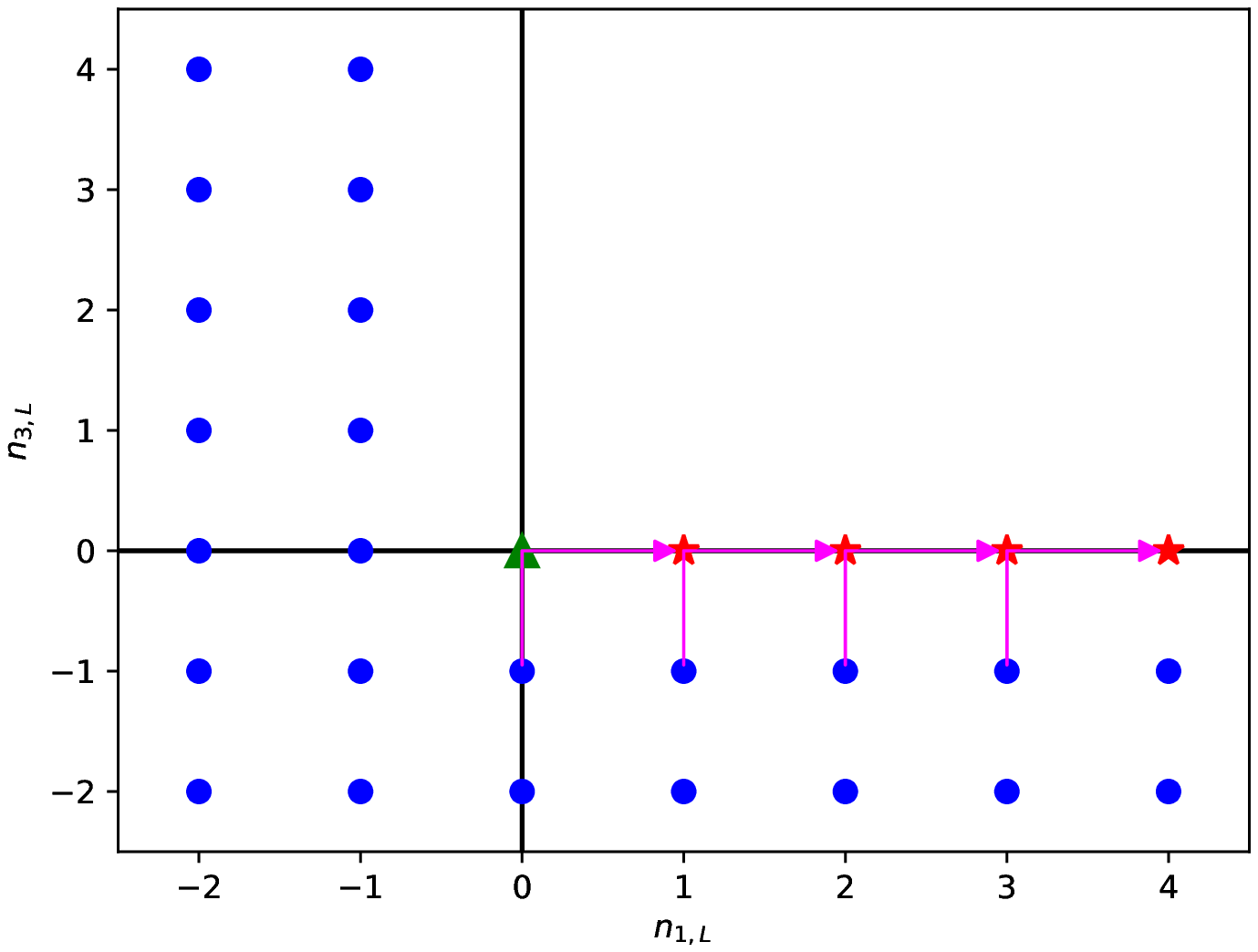}
    \caption{The recursion relation (\ref{recursion1}) with $n_{1,R}=n_{3,R}=0$ and $n_{3,L}=-1$ gives three-point function at horizontal axis.}
    \label{analytic_conti2}
  \end{minipage}
\end{figure}
\begin{figure}[ht]
  \begin{minipage}{0.47\columnwidth}
    \centering
    \includegraphics[width=\columnwidth]{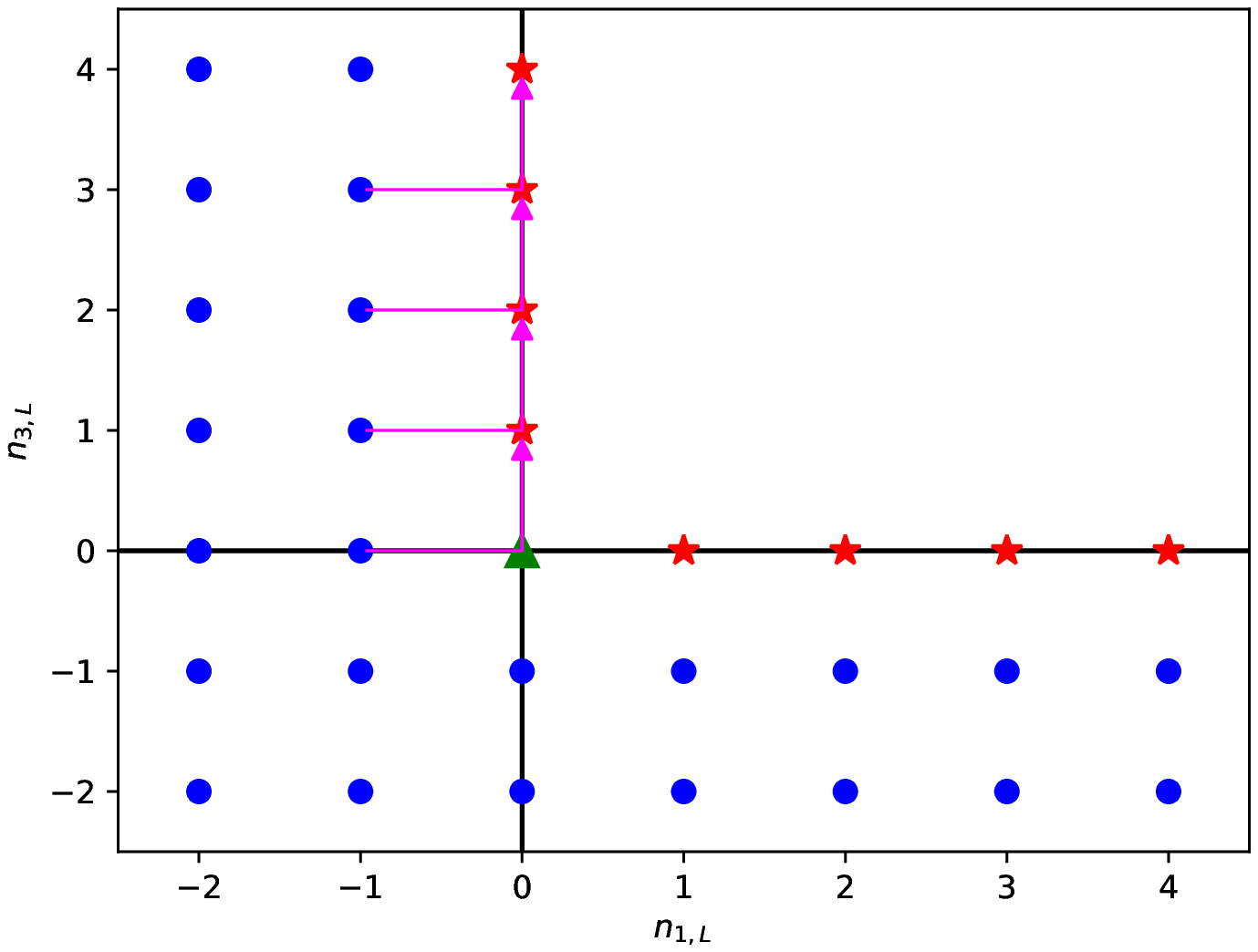}
    \caption{The recursion relation (\ref{recursion2}) with $n_{1,R}=n_{3,R}=0$ and $n_{1,L}=-1$ gives three-point functions at vertical axis.}
    \label{analytic_conti3}
  \end{minipage}
  \hspace{0.04\columnwidth}
  \begin{minipage}{0.47\columnwidth}
    \centering
    \includegraphics[width=\columnwidth]{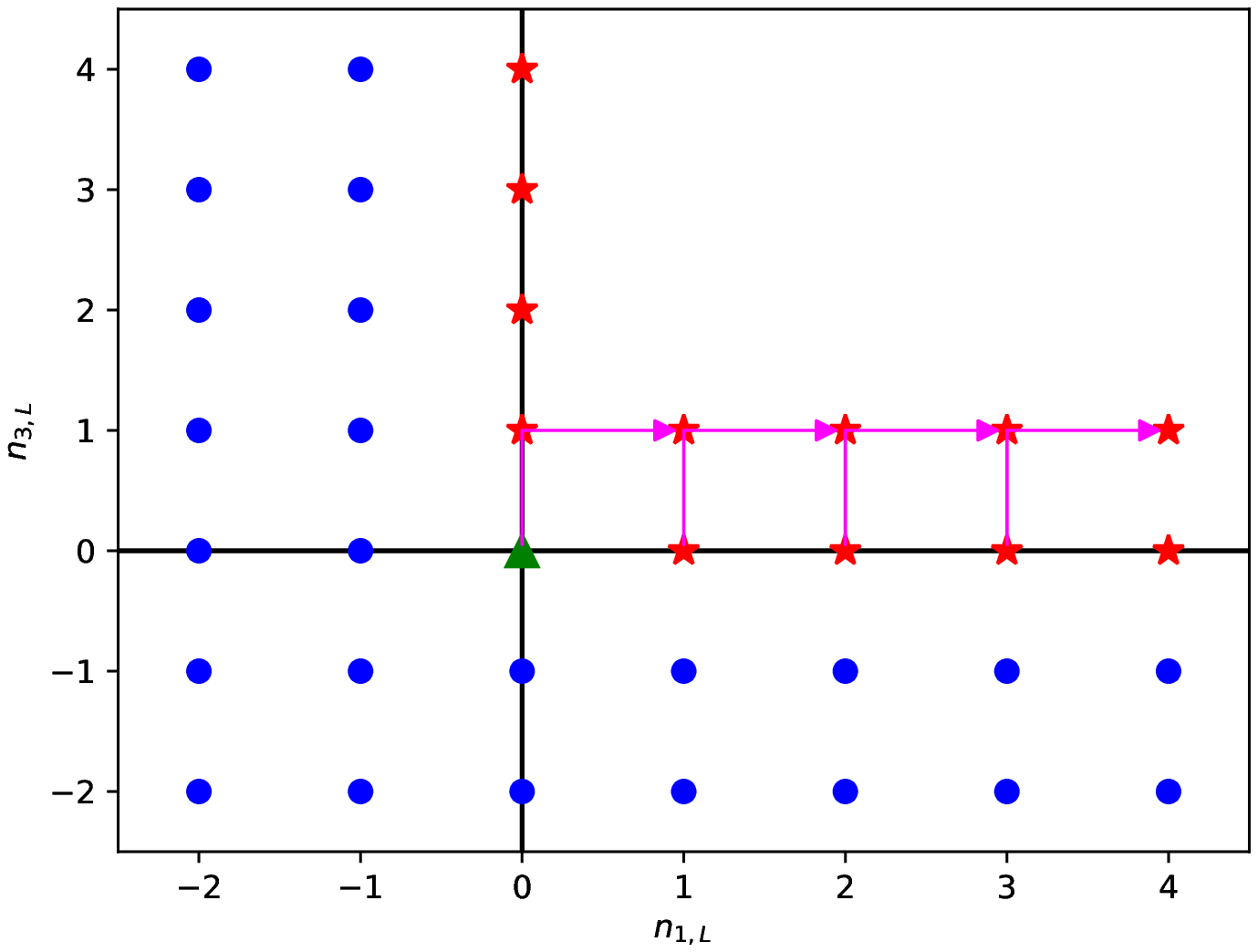}
    \caption{The recursion ralation (\ref{recursion1}) with $n_{1,R}=n_{3,R}=0$ and $n_{3,L}=0$ gives three-point functions at $n_{3,L}=1$}
    \label{analytic_conti4}
  \end{minipage}
\end{figure}
\begin{figure}[ht]
  \begin{minipage}{0.47\columnwidth}
    \centering
    \includegraphics[width=\columnwidth]{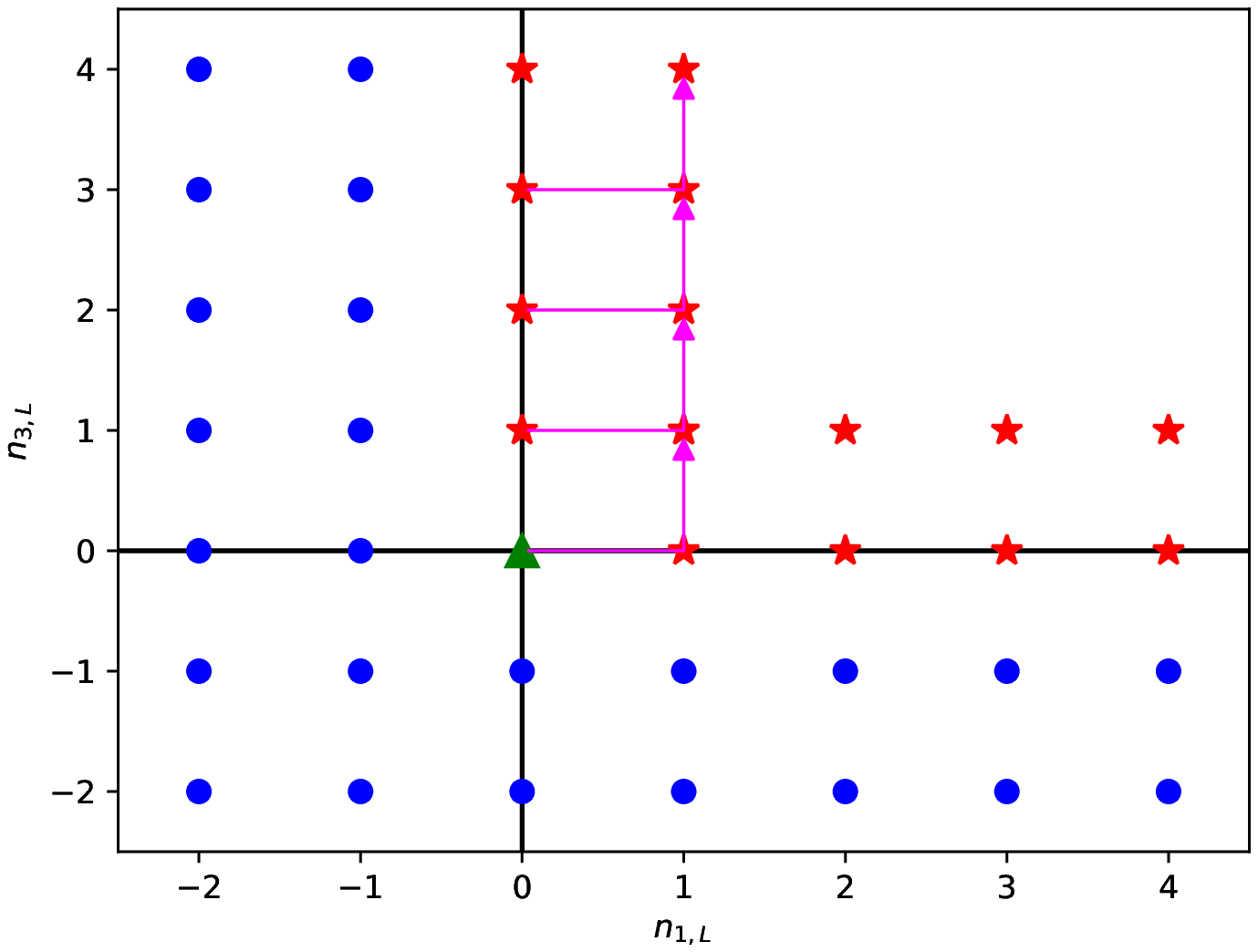}
    \caption{The recursion ralation (\ref{recursion2}) with $n_{1,R}=n_{3,R}=0$ and $n_{1,L}=0$ gives three-point functions at $n_{1,L}=1$}
    \label{analytic_conti5}
  \end{minipage}
  \hspace{0.04\columnwidth}
  \begin{minipage}{0.47\columnwidth}
    \centering
    \includegraphics[width=\columnwidth]{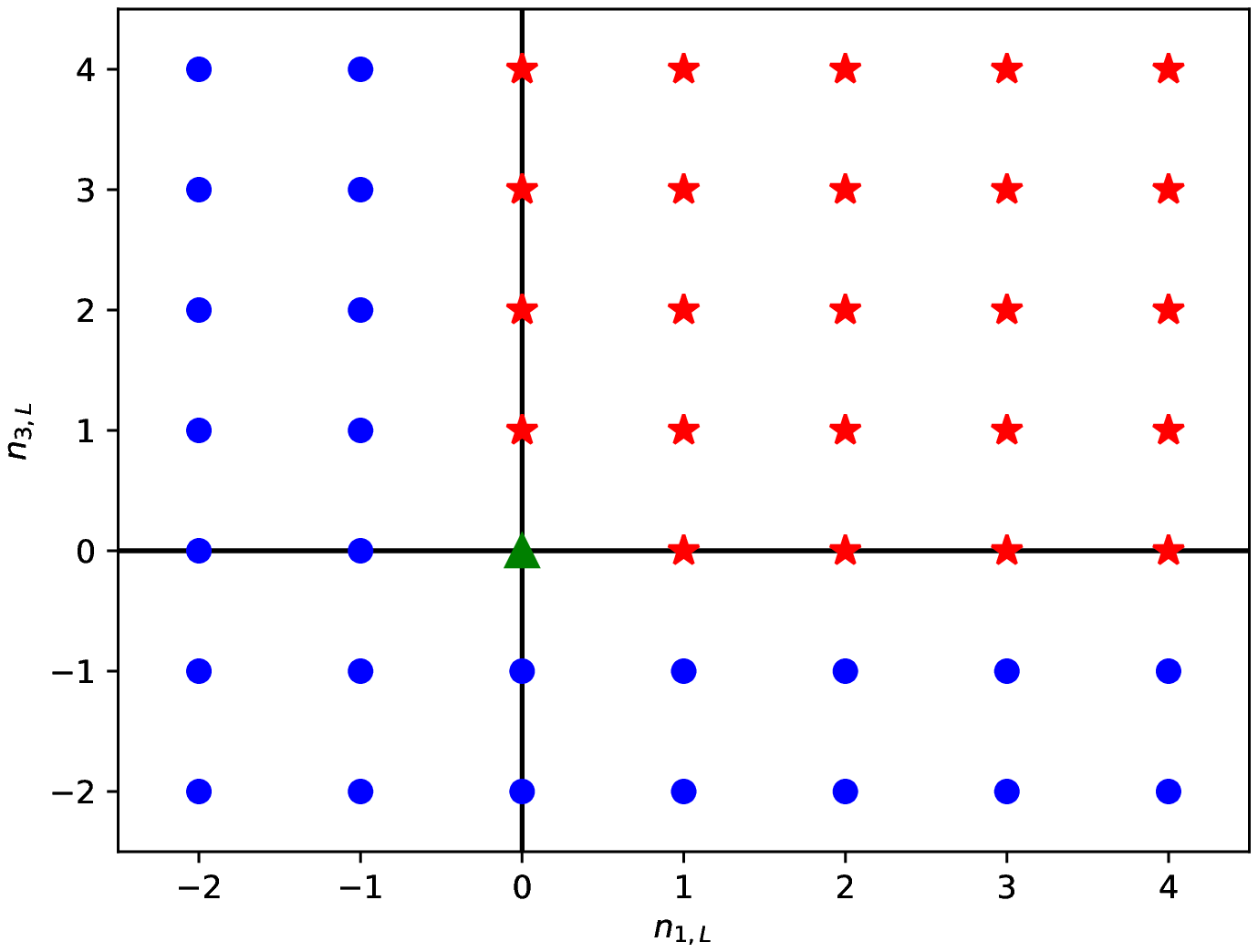}
    \caption{We can get three-point functions at all points.}
    \label{analytic_conti6}
  \end{minipage}
\end{figure}
The derivation from the recursion relations of all three-point functions is summarized in the images. The horizontal axis represents $n_{1,L}$ and the vertical axis represents $n_{3,L}$.\par
First, prepare the arbitrarily normalized initial value $\YY 3[h_{3,2,1},\tilde{h}_{3,2,1}|0,0,0,0]$, which is green triangle in the Fig. \ref{analytic_conti1}. From the unitarity condition, three-point functions at blue points vanish. At blue points, $n_{1,L}<0$ or $n_{3_L}<0$ is satisfied.\par
Next, from the recursion relation (\ref{recursion1}) with $n_{1,R}=n_{3,R}=0$ and $n_{3,L}=-1$, we can get three-point functions at horizontal axis (Fig. \ref{analytic_conti2}). And in the same way, three-point functions at the vertical axis can be calculated from the recursion relation (\ref{recursion2}) with $n_{1,R}=n_{3,R}=0$ and $n_{1,L}=-1$ (Fig. \ref{analytic_conti3}).\par
Recursion relation (\ref{recursion1}) with $n_{1,R}=n_{3,R}=0$ and $n_{3,L}=0$ gives three-point functions at $n_{3,L}=1$ (Fig. \ref{analytic_conti4}). And three-point functions at $n_{1,L}=1$ can be gotten in the same way (Fig. \ref{analytic_conti5}).\par
Finally, we can get three-point functions for all integer pairs of $(n_{1,L},n_{3,L},n_{1,R},n_{3,R})$ (Fig. \ref{analytic_conti6}).\par
Next, consider an explicit solution for the three-point functions.\par
First, start from the pair of nonnegative integer conformal weight $(h_{1,2,3},\tilde{h}_{1,2,3})$. In these particular cases, the three-point function factorizes globally rather than just locally into a product of a holomorphic and antiholomorphic three-point function. Fourier transform of three-point functions also factorizes and can be calculated individually by contour integration. We performed the contour integration in the previous appendix, and the result is
\begin{align}
  &\YY 3[h_{3,2,1},\tilde{h}_{3,2,1}|n_{3,L},n_{1,L},n_{3,R},n_{1,R}]\notag\\
  &=K\YY 3^{(\mathrm{chiral})}[h_{3,2,1}|n_{3,L},n_{1,L}]\YY 3^{(\mathrm{antichiral})}[\tilde{h}_{3,2,1}|n_{3,R},n_{1,R}],\notag
\end{align}
where $K$ is an unfixed constant related to the OPE coefficient. The chiral part for $n_{1,L},n_{3,L}\geq0$ is
\begin{align}
  &\YY 3^{(\mathrm{chiral})}[h_{3,2,1}|n_{3,L},n_{1,L}]=\frac{1}{\Gamma(b_1)\Gamma(b_2)\Gamma(b_3)}\notag\\
  &\sum_{q=0}^{\mathrm{Min}\{n_{1,L},n_{3,L}\}}\tfrac{\Gamma[b_3+q+\mathrm{Max}\{{0,n_{1,L}-n_{3,L}}\}]\Gamma[b_2-q+\mathrm{Min}\{{n_{1,L},n_{3,L}}\}]\Gamma[b_1+q+\mathrm{Max}\{{n_{3,L}-n_{1,L},0}\}]}{\Gamma[1+q+\mathrm{Max}\{{0,n_{1,L}-n_{3,L}}\}]\Gamma[1-q+\mathrm{Min}\{{n_{1,L},n_{3,L}}\}]\Gamma[1+q+\mathrm{Max}\{{n_{3,L}-n_{1,L},0}\}]},\label{Direct Integral for holo}
\end{align}
and $0$ for otherwise. And the antichiral part for $n_{1,R},n_{3,R}\geq0$ is
\begin{align}
  &\YY 3^{(\mathrm{antichiral})}[\tilde{h}_{3,2,1}|n_{3,R},n_{1,R}]=\frac{1}{\Gamma(\tilde{b}_1)\Gamma(\tilde{b}_2)\Gamma(\tilde{b}_3)}\notag\\
  &\sum_{q=0}^{\mathrm{Min}\{n_{1,R},n_{3,R}\}}\tfrac{\Gamma[\tilde{b}_3+q+\mathrm{Max}\{{0,n_{1,R}-n_{3,R}}\}]\Gamma[\tilde{b}_2-q+\mathrm{Min}\{{n_{1,R},n_{3,R}}\}]\Gamma[\tilde{b}_1+q+\mathrm{Max}\{{n_{3,R}-n_{1,R},0}\}]}{\Gamma[1+q+\mathrm{Max}\{{0,n_{1,R}-n_{3,R}}\}]\Gamma[1-q+\mathrm{Min}\{{n_{1,R},n_{3,R}}\}]\Gamma[1+q+\mathrm{Max}\{{n_{3,R}-n_{1,R},0}\}]},\label{Direct Integral for antiholo}
\end{align}
and $0$ for otherwise. \par
The contour integral derivation of (\ref{Direct Integral for holo}) and (\ref{Direct Integral for antiholo}) are only valid for a pair of nonnegative integers $(h_{3,2,1},\tilde{h}_{3,2,1})$. \par
We proved that any solution to the $L_{\pm1}$ and $\tilde{L}_{\pm}$ WIs, which also satisfies the unitarity-derived condition, is uniquely determined by the primary-to-primary three-point coefficient $\YY 3[h_{3,2,1},\tilde{h}_{3,2,1}|0,0,0,0]$. So what we have to do is only to prove that the formula (\ref{Direct Integral for holo}) and (\ref{Direct Integral for antiholo}) actually satisfy the WIs for general conformal weights.\par
In the previous section, we proved that the formula satisfies WIs. We dealt with positive integer conformal weights there, but we can use the same proof to show that WIs are satisfied for general conformal weights because we didn't use the fact that conformal weights are positive integers there. \par
So the proof is completed. We can use our formula of the three-point function for operators with general conformal weights, except for negative integers.
\section{Supplement to three-dimensional CFT}
\subsection{Wigner \texorpdfstring{$3j$}{TEXT} symbol}
\setcounter{equation}{0}
The Wigner $3j$ symbol is a more symmetric form of Clebsch-Gordan coefficients.
\begin{equation}
\mqty(l_1&l_2&l_3\\m_1&m_2&m_3)\eq\frac{(-1)^{l_1}}{\sqrt{2l_3+1}}\bra{l_1,m_1,l_2,m_2}\ket{l_3,-m_3}
\end{equation}
It has the following properties.\\
(1) Symmetry\\
It is symmetric under an even permutation, but the phase factor appears under an odd permutation. For example,
\begin{align}
&\mqty(l_1&l_2&l_3\\m_1&m_2&m_3)=\mqty(l_3&l_1&l_2\\m_3&m_1&m_2)=\mqty(l_2&l_3&l_1\\m_2&m_3&m_1)\\
&\mqty(l_1&l_2&l_3\\m_1&m_2&m_3)=(-1)^{l_1+l_2+l_3}\mqty(l_2&l_1&l_3\\m_2&m_1&m_3)=(-1)^{l_1+l_2+l_3}\mqty(l_3&l_2&l_1\\m_3&m_2&m_1).
\end{align}
(2) Selection rule\\
The $3j$ symbol is 0 unless all of the following conditions are met.
\begin{align}
&\text{(A)}\ \ m_1+m_2+m_3=0\label{selection rule A}\\
&\text{(B)}\ \ l_1+l_2+l_3 \text{\ \ is an integer.} \notag\\
&\ \ \ \ \ \ \text{Moreover, an even integer if  } m_1=m_2=m_3=0\label{selection rule B}\\
&\text{(C)}\ \ |m_i|\leq l_i\label{selection rule C}\\
&\text{(D)}\ \ |l_1-l_2|\leq l_3\leq |l_1+l_2|\label{selection rule D}
\end{align}
(3) Relationship with spherical harmonics\\
The $3j$ symbol is deeply related to the product of spherical harmonics.
\begin{align}
&Y_{l_1,m_1}(\theta,\phi)Y_{l_2,m_2}(\theta,\phi)\notag\\
&=\sqrt{\frac{(2l_1+1)(2l_2+1)}{4\pi}}\sum_{l=0}^{\infty}\sum_{m=-l}^{l}(-1)^m\sqrt{2l+1}\mqty(l_1&l_2&l\\m_1&m_2&-m)\mqty(l_1&l_2&l_3\\0&0&0)Y_{l,m}(\theta,\phi)
\end{align}
The $3j$ symbols can be computed using the Racah formula.
\begin{align}
&\mqty(l_1&l_2&l_3\\m_1&m_2&m_3)=(-1)^{l_1-l_2-l_3}\sqrt{\frac{(l_1+l_2-l_3)!(l_1-l_2+l_3)!(-l_1+l_2+l_3)!}{(l_1+l_2+l_3+1)!}}\notag\\
&\sqrt{(l_1+m_1)!(l_1-m_1)!(l_2+m_2)!(l_2-m_2)!(l_3+m_3)!(l_3-m_3)!}\notag\\
&\sum_k\frac{(-1)^k}{k!(l_3-l_2+k+m_1)!(l_3-l_1+k-m_2)!(l_1+l_2-l_3-k)!(l_1-k-m_1)!(l_2-k+m_2)!}
\end{align}
Though the formula above is a little complicated, we often use the following specific simple case in this paper.
\begin{align}
&\mqty(l_1&l_2&l_1+l_2\\m_1&m_2&-m_1-m_2)\notag\\
&=(-1)^{l_1-l_2+m_1+m_2}\sqrt{\frac{(2l_1)!(2l_2)!}{(2l_1+2l_2+1)!}}\sqrt{\frac{(l_1+l_2+m_1+m_2)!(l_1+l_2-m_1-m_2)!}{(l_1+m_1)!(l_1-m_1)!(l_2+m_2)!(l_2-m_2)!}}
\end{align}
Using them, we can calculate conformal generators' action on spatially-integrated operators' modes in three-dimensional CFT. Here, we show the calculation for $P_x+iP_y$ as an example. We can get relations for other generators similarly.
\begin{equation}
(P_x+iP_y)\cdot\OO_{l,m}=\int \dd\Omega Y^{*}_{l,m}(\theta,\phi)\left[\sin\theta e^{i\phi}\pdv{r}+\frac{\cos\theta}{r}e^{i\phi}\pdv{\theta}+i\frac{e^{i\phi}}{r\sin\theta}\pdv{\phi}\right]\OO(r,\theta,\phi)\label{The action of P+}
\end{equation}
We calculate each term in (\ref{The action of P+}) separately.
\begin{flalign}
&(\text{First term})=\int \dd\Omega Y^*_{l,m}(\theta,\phi)\sin\theta e^{i\phi}\pdv{r}\OO(r,\theta,\phi)=-2\sqrt{\frac{2\pi}{3}}\pdv{r}\int\dd\Omega Y^*_{l,m}Y_{1,1}\OO&\\
&(\text{Second term})=\int\dd\Omega Y^*_{l,m}(\theta,\phi)\frac{\cos\theta}{r}e^{i\phi}\pdv{\theta}\OO(r,\theta,\phi)&\notag\\
&=\frac{1}{r}\int\dd\theta\dd\phi\left[-\cos^2\theta Y^*_{l,m}+\sin^2\theta Y^*_{l,m}-\sin\theta\cos\theta\pdv{\theta}Y^*_{l,m}\right]e^{i\phi}\OO&\notag\\
&=\frac{1}{r}\int\dd\theta\dd\phi\left[-1+2\sin^2\theta-\cos\theta\{-m\cos\theta-\sin\theta\sqrt{(l+m)(l-m+1)}e^{-i\phi}\}\right]Y^*_{l,m}e^{i\phi}\OO&\notag\\
&=\frac{m-1}{r}\int\dd\theta\dd\phi Y^*_{l,m}e^{i\phi}\OO+\frac{2(m-2)}{r}\sqrt{\frac{2\pi}{3}}\int\dd\Omega Y^*_{l,m}Y_{1,1}\OO\notag&\\
&\hspace{5cm}+\frac{2\sqrt{(l+m)(l-m+1)}}{r}\sqrt{\frac{\pi}{3}}\int\dd\Omega Y^*_{l,m-1}Y_{1,0}\OO&\\
&(\text{Third term})=\int\dd\Omega Y^*_{l,m}(\theta,\phi)\frac{i}{r\sin\theta}e^{i\phi}\pdv{\phi}\OO=-\frac{m-1}{r}\int\dd\theta\dd\phi Y^*_{l,m}e^{i\phi}\OO&
\end{flalign}
In total,
\begin{flalign}
&(P_x+iP_y)\cdot\OO_{l,m}&\notag\\
&=2\sqrt{\frac{2\pi}{3}}\left(\frac{m-2}{r}-\pdv{r}\right)\int\dd\Omega Y^*_{l,m}Y_{1,1}\OO+2\sqrt{\frac{\pi}{3}}\frac{\sqrt{(l+m)(l-m+1)}}{r}\int\dd\Omega Y^*_{l,m-1}Y_{1,0}\OO&
\end{flalign}
We would like to write the action of $P_x+iP_y$ by a linear combination of modes of operators, so let us apply the contraction rule to the product of spherical harmonics.
\begin{align}
Y^*_{l,m}Y_{1,0}=(-1)^m\sqrt{\frac{3(2l+1)}{4\pi}}\sum_{c=0}^{\infty}\sum_{\gamma=-c}^c(-1)^\gamma\sqrt{2c+1}\mqty(l&1&c\\-m&0&-\gamma)\mqty(l&1&c\\0&0&0)Y_{c,\gamma}
\end{align}
From (\ref{selection rule B}) and (\ref{selection rule D}), only $c=l\pm1$ survive when $l\geq1$, and only $c=1$ survives when $l=0$. And from (\ref{selection rule A}), only $\gamma=-m$ survives.
\begin{multline}
Y^*_{l,m}Y_{1,0}=\sqrt{\frac{3(2l+1)}{4\pi}}\left[\sqrt{2l+3}\mqty(l&1&l+1\\-m&0&m)\mqty(l&1&l+1\\0&0&0)Y_{l+1,-m}\right.\\
\left.+\sqrt{2l-1}\mqty(l&1&l-1\\-m&0&m)\mqty(l&1&l-1\\0&0&0)Y_{l-1,-m}\right]
\end{multline}
In the same way,
\begin{multline}
Y^*_{l,m}Y_{1,1}=-\sqrt{\frac{3(2l+1)}{4\pi}}\left[\sqrt{2l+3}\mqty(l&1&l+1\\-m&1&m-1)\mqty(l&1&l+1\\0&0&0)Y_{l+1,-m+1}\right.\\
\left.+\sqrt{2l-1}\mqty(l&1&l-1\\-m&1&m-1)\mqty(l&1&l-1\\0&0&0)Y_{l-1,-m+1}\right].
\end{multline}
By using the above contraction rule, we get the following relation.
\begin{align}
&(P_x+iP_y)\cdot\OO_{l,m}\notag\\
&=(-1)^{m-1}\sqrt{(2l+1)(2l+3)}\mqty(l&1&l+1\\0&0&0)\left[\sqrt{2}\mqty(l&1&l+1\\-m&1&m-1)\left(\pdv{r}-\frac{m-2}{r}\right)\right.\notag\\
&\hspace{5cm}\left.+\mqty(l&1&l+1\\-m+1&0&m-1)\frac{\sqrt{(l+m)(l-m+1)}}{r}\right]\OO_{l+1,m-1}\notag\\
&+(-1)^{m-1}\sqrt{(2l+1)(2l-1)}\mqty(l&1&l-1\\0&0&0)\left[\sqrt{2}\mqty(l&1&l-1\\-m&1&m-1)\left(\pdv{r}-\frac{m-2}{r}\right)\right.\notag\\
&\hspace{5cm}\left.+\mqty(l&1&l-1\\-m+1&0&m-1)\frac{\sqrt{(l+m)(l-m+1)}}{r}\right]\OO_{l-1,m-1}
\end{align}
We can get the actions of other conformal generators similarly.
\subsection{Exact two-point function}
We show the detail of the calculation of exact two-point function in three-dimensional CFT by solving ODEs obtained from WIs. First, consider the $P_z$ WI.
\begin{align}
&P_z\cdot\langle\OO_{l_2,m_2}(r_2)\OO_{l_1,m_1}(r_1)\rangle\notag\\
&=(-1)^{m_2}\sqrt{(2l_2+1)(2l_2+3)}\mqty(l_2&1&l_2+1\\0&0&0)\left[\mqty(l_2&1&l_2+1\\-m_2&0&m_2)\left(\pdv{r_2}-\frac{m_2-2}{r_2}\right)\right.\notag\\
&\left.\hspace{1cm}+\sqrt{2}\mqty(l_2&1&l_2+1\\-m_2+1&-1&m_2)\frac{\sqrt{(l_2+m_2)(l_2-m_2+1)}}{r_2}\right]\langle\OO_{l_2+1,m_2}\OO_{l_1,m_1}\rangle\notag\\
&+(-1)^{m_2}\sqrt{(2l_2+1)(2l_2-1)}\mqty(l_2&1&l_2-1\\0&0&0)\left[\mqty(l_2&1&l_2-1\\-m_2&0&m_2)\left(\pdv{r_2}-\frac{m_2-2}{r_2}\right)\right.\notag\\
&\left.\hspace{1cm}+\sqrt{2}\mqty(l_2&1&l_2-1\\-m_2+1&-1&m_2)\frac{\sqrt{(l_2+m_2)(l_2-m_2+1)}}{r_2}\right]\langle\OO_{l_2-1,m_2}\OO_{l_1,m_1}\rangle\notag\\
&+(-1)^{m_1}\sqrt{(2l_1+1)(2l_1+3)}\mqty(l_1&1&l_1+1\\0&0&0)\left[\mqty(l_1&1&l_1+1\\-m_1&0&m_1)\left(\pdv{r_1}-\frac{m_1-2}{r_1}\right)\right.\notag\\
&\left.\hspace{1cm}+\sqrt{2}\mqty(l_1&1&l_1+1\\-m_1+1&-1&m_1)\frac{\sqrt{(l_1+m_1)(l_1-m_1+1)}}{r_1}\right]\langle\OO_{l_2,m_2}\OO_{l_1+1,m_1}\rangle\notag\\
&+(-1)^{m_1}\sqrt{(2l_1+1)(2l_1-1)}\mqty(l_1&1&l_1-1\\0&0&0)\left[\mqty(l_1&1&l_1-1\\-m_1&0&m_1)\left(\pdv{r_1}-\frac{m_1-2}{r_1}\right)\right.\notag\\
&\left.\hspace{1cm}+\sqrt{2}\mqty(l_1&1&l_1-1\\-m_1+1&-1&m_1)\frac{\sqrt{(l_1+m_1)(l_1-m_1+1)}}{r_1}\right]\langle\OO_{l_2,m_2}\OO_{l_1-1,m_1}\rangle
\end{align}
The first and fourth terms have $\delta(l_1-l_2-1)$, and the second and third have $\delta(l_1-l_2+1)$. We have two independent recursion relations for them. From the latter, we get
\begin{multline}
0=y\left[\mqty(l_1&1&l_1+1\\0&0&0)\left(-y\dv{y}-\Delta_2+2\right)+\sqrt{2}\mqty(l_1&1&l_1+1\\0&-1&1)\sqrt{(l_1+1)(l_1+2)}\right]G^{l_1}\\
+\left[\mqty(l_1&1&l_1+1\\0&0&0)\left(y\dv{y}-\Delta_1+2\right)+\sqrt{2}\mqty(l_1&1&l_1+1\\1&-1&0)\sqrt{l_1(l_1+1)}\right]G^{l_1+1}.\label{Pz Ward Identity}
\end{multline}
And from the $K_z$ WI, we get
\begin{multline}
0=y\left[\mqty(l_1&1&l_1+1\\0&0&0)\left(y\dv{y}+\Delta_1-2\right)-\sqrt{2}\mqty(l_1&1&l_1+1\\0&-1&1)\sqrt{(l_1+1)(l_1+2)}\right]G^{l_1}\\
+\left[\mqty(l_1&1&l_1+1\\0&0&0)\left(-y\dv{y}+\Delta_2-2\right)-\sqrt{2}\mqty(l_1&1&l_1+1\\1&-1&0)\sqrt{l_1(l_1+1)}\right]G^{l_1+1}.\label{Kz Ward Identity}
\end{multline}
From (\ref{Pz Ward Identity}) and (\ref{Kz Ward Identity}), we get
\begin{equation}
(\Delta_1-\Delta_2)[yG^{l_1}-G^{l_1+1}]=0.
\end{equation}
First, assume that $G^{l_1+1}=yG^{l_1}$ is satisfied. Substituting it for (\ref{Pz Ward Identity}) yields
\begin{equation}
(\Delta_1+\Delta_2+1)yG^{l_1}=0.
\end{equation}
However, as $\Delta_1+\Delta_2+1>0$, $G^{l_1}$ must vanish. So $\Delta_1=\Delta_2$ must be satisfied for two-point function not to vanish. Next, consider $P_x\pm iP_y$ and $K_x+iK_y$ WIs. There are only two independent identities.
\begin{multline}
0=\left[\sqrt{2}\mqty(l&1&l+1\\-1&1&0)\left(y\dv{y}-\Delta+2\right)+\mqty(l&1&l+1\\-1&0&1)\sqrt{(l+2)(l+1)}\right]G^{l}\\
+y\left[\sqrt{2}\mqty(l&1&l+1\\-1&1&0)\left(-y\dv{y}-\Delta+1\right)+\mqty(l&1&l+1\\0&0&0)\sqrt{(l+1)l}\right]G^{l+1}
\end{multline}
\begin{multline}
0=y\left[\sqrt{2}\mqty(l&1&l+1\\-1&1&0)\left(y\dv{y}+\Delta-2\right)-\mqty(l&1&l+1\\-1&0&1)\sqrt{(l+2)(l+1)}\right]G^{l}\notag\\
+\left[\sqrt{2}\mqty(l&1&l+1\\-1&1&0)\left(-y\dv{y}+\Delta-1\right)-\mqty(l&1&l+1\\0&0&0)\sqrt{(l+1)l}\right]G^{l+1}
\end{multline}
Substituting the explicit form of the Wigner $3j$ symbol, we get
\begin{align}
&\left[y\dv{y}-\Delta+l+2\right]G^{l+1}+y\left[-y\dv{y}-\Delta-l\right]G^l=0\\
&y\left[-y\dv{y}-\Delta+l+2\right]G^{l+1}+\left[y\dv{y}-\Delta-l\right]G^l=0.
\end{align}
Combining them gives
\begin{equation}
\dv[2]{y}G^l+\frac{2\Delta y^2+2(\Delta-1)}{y^3-y}\dv{y}G^l+\frac{\Delta(\Delta-1)}{y^2}G^l=0.
\end{equation}
It is very similar to the ODE (\ref{ODE in 2D}) for the two-point function in two-dimensional CFT. We can solve it by series expansion as we did in two-dimensional CFT. The answer is shown in the main text.
\subsection{Reduction of the three-point function}
\subsubsection{Strategy}
In the calculation of the two-point functions in three-dimensional CFT, we found that the ODE (\ref{ODE for two pt func in 3 dim}) obtained from the WIs was independent of the subscripts $m_1$ and $m_2$ because of the linear relationship (\ref{Jpm Ward Identity result}) between $F_{m1}^{l_1,l_2}$ and $F_{m_1+1}^{l_1,l_2}$. Similarly, we can get a linear relationship between $F_{m_1,m_3}^{l_1,l_2,l_3}$ and $F_{m'_1,m'_3}^{l_1,l_2,l_3}$, which enables us to get the simpler ODEs. \par
\begin{figure}[tb]
  \begin{center}
  \includegraphics[width=8cm]{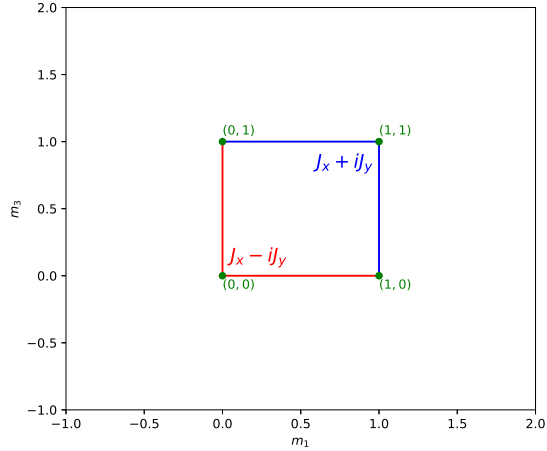}
  \caption{$J_x+iJ_y$ relates three points connected by the blue line, and $J_x-iJ_y$ relates three points connected by the red line.}
  \label{reduction5}
\end{center}
\end{figure}

First, remember that the $J_x+iJ_y$ WI relates $F_{m_1,m_3}^{l_1,l_2,l_3}$, $F_{m_1-1,m_3}^{l_1,l_2,l_3}$ and $F_{m_1,m_3-1}^{l_1,l_2,l_3}$, and that the $J_x-iJ_y$ WI relates $F_{m_1,m_3}^{l_1,l_2,l_3}$, $F_{m_1+1,m_3}^{l_1,l_2,l_3}$ and $F_{m_1,m_3+1}^{l_1,l_2,l_3}$. The situation can be explained graphically. The Fig. \ref{reduction5} shows examples of how $J_x\pm iJ_y$ relates $F_{m_1,m_3}^{l_1,l_2,l_3}$ at three points in the $m_1-m_3$ plane. In the figure, for example, if we know the value of $F_{1,0}^{l_1,l_2,l_3}$ and $F_{1,1}^{l_1,l_2,l_3}$, we can derive $F_{1,0}^{l_1,l_2,l_3}$ with the $J_x+iJ_y$ WI.\par
\begin{figure}[tb]
  \begin{center}
  \includegraphics[width=8cm]{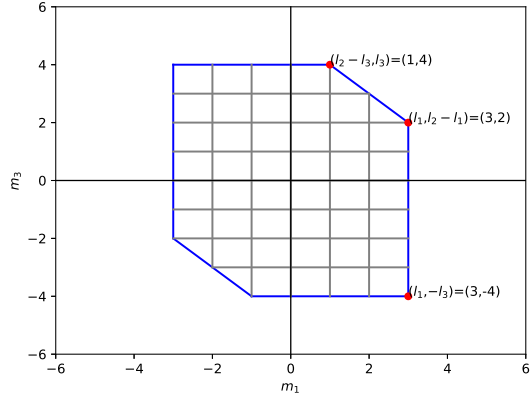}
  \caption{Example of selection rule $(l_1,l_2,l_3)=(3,4,5)$. $F_{m_1,m_3}^{l_1,l_2,l_3}$ vanish in areas outside the blue line.}
  \label{reduction1}
\end{center}
\end{figure}

Next, consider the selection rule for $F_{m_1,m_3}^{l_1,l_2,l_3}$. Prepare $l_1$, $l_2$ and $l_3$ which satisfy $|l_1-l_2|\leq l_3\leq l_1+l_2$ , $|l_2-l_3|\leq l_1\leq l_2+l_3$ and $|l_3-l_1|\leq l_2\leq l_3+l_1$. For this fixed pairs of $l$, the rule is as follows. $F_{m_1,m_3}^{l_1,l_2,l_3}$ vanishes unless the all following inequalities are satisfied.
\begin{equation}
 -l_1\leq m_1\leq l_1, \hspace{1cm} -l_3\leq m_3\leq l_3, \hspace{1cm} -l_2\leq m_1+m_3\leq l_2
\end{equation}
The last condition comes from $m_2=-m_1-m_3$ and $-l_2\leq m_2\leq l_2$. The area which satisfies the above conditions is diamond-shaped as Fig. \ref{reduction1}.
Our goal is to get all $F_{m_1,m_3}^{l_1,l_2,l_3}$s in this area using $F_{l_1,-l_3}^{l_1,l_2,l_3}$ and the recursion relations obtained from $J\pm iJ_y$ WIs.\par
\begin{figure}[tb]
  \begin{center}
  \includegraphics[width=8cm]{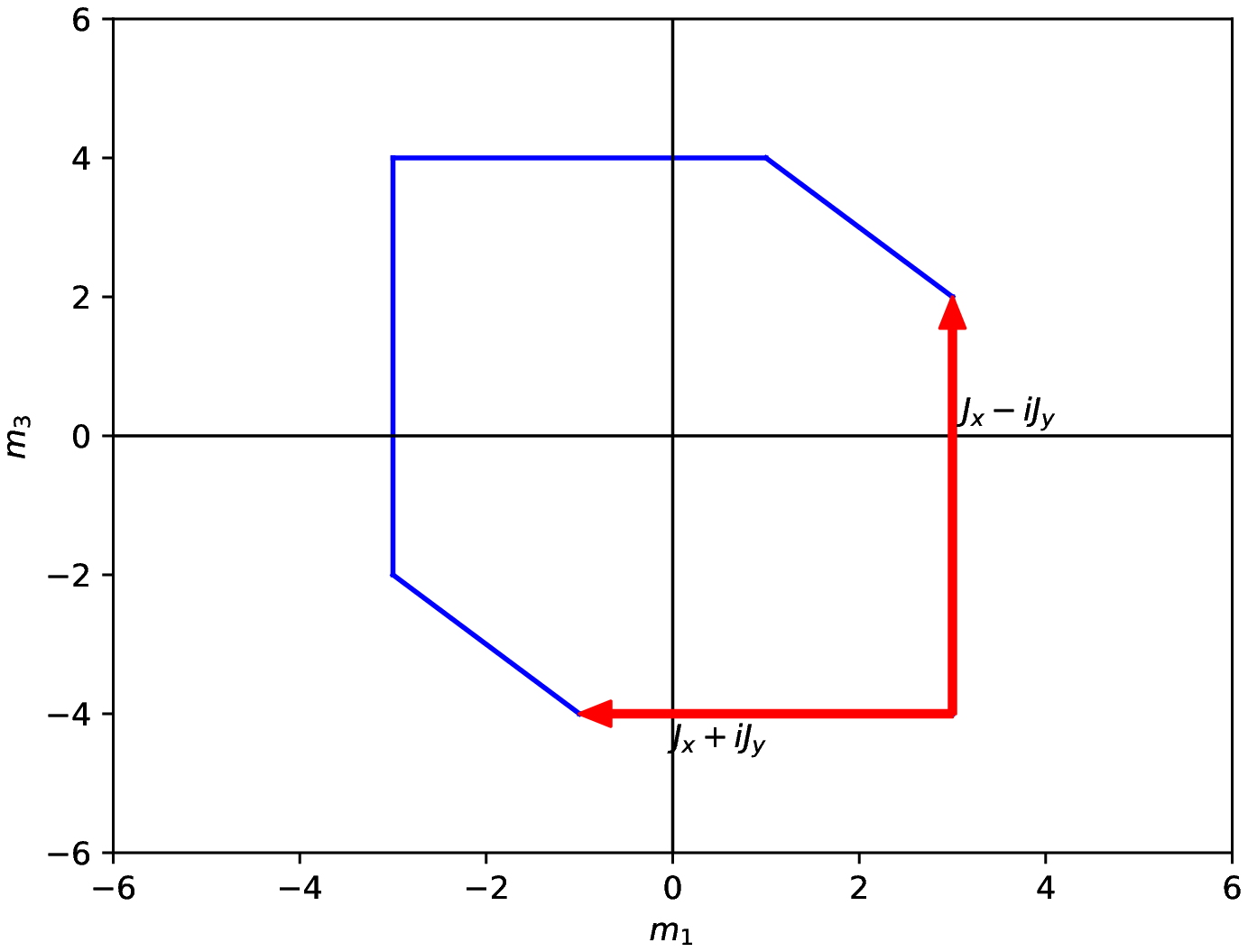}
  \caption{We can get the $F$s at the right boundary and at the bottom boundary by applying $J_x\pm iJ_y$ WIs respectively.}
  \label{reduction2}
\end{center}
\end{figure}

First, apply the $J_x-iJ_y$ recursion relation to $(m_1,m_3)=(l_1,-l_3+k)\ (0\leq k\leq l_2-l_1-l_3-1)$. $F_{m_1+1,m_3}^{l_1,l_2,l_3}$ vanishes for this $m_1$ and $m_3$, so the $J_x-iJ_y$ recursion relation relates two points. We can find all the values of $F$ at the right boundary. Similarly, we can find all the values of $F$ at the bottom boundary. We show the situation in the Fig. \ref{reduction2}.\par
\begin{figure}[ht]
  \begin{minipage}{0.47\columnwidth}
    \centering
    \includegraphics[width=\columnwidth]{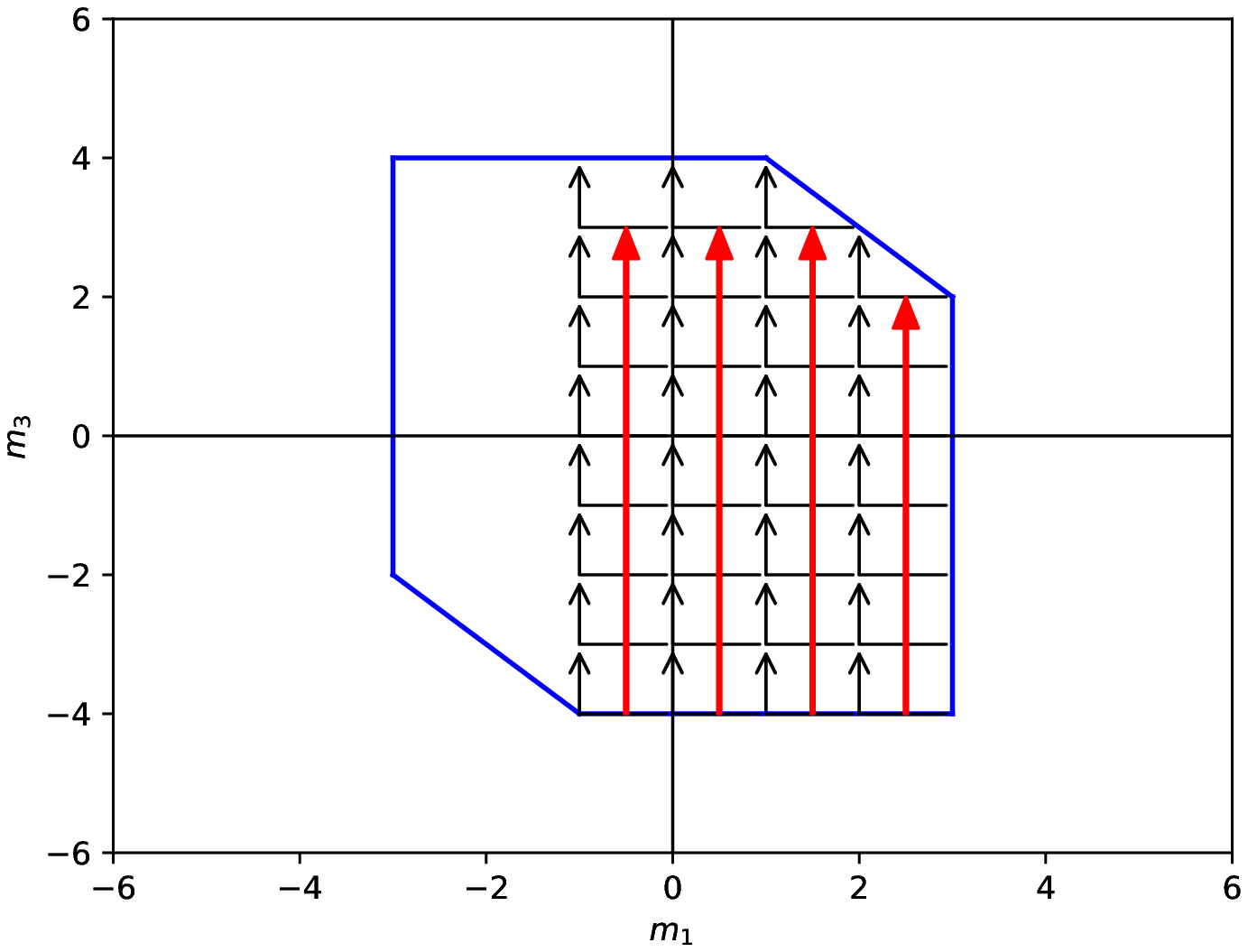}
    \caption{Apply the $J_x-iJ_y$ WI to the right part of the area}
    \label{reduction3}
  \end{minipage}
  \hspace{0.04\columnwidth}
  \begin{minipage}{0.47\columnwidth}
    \centering
    \includegraphics[width=\columnwidth]{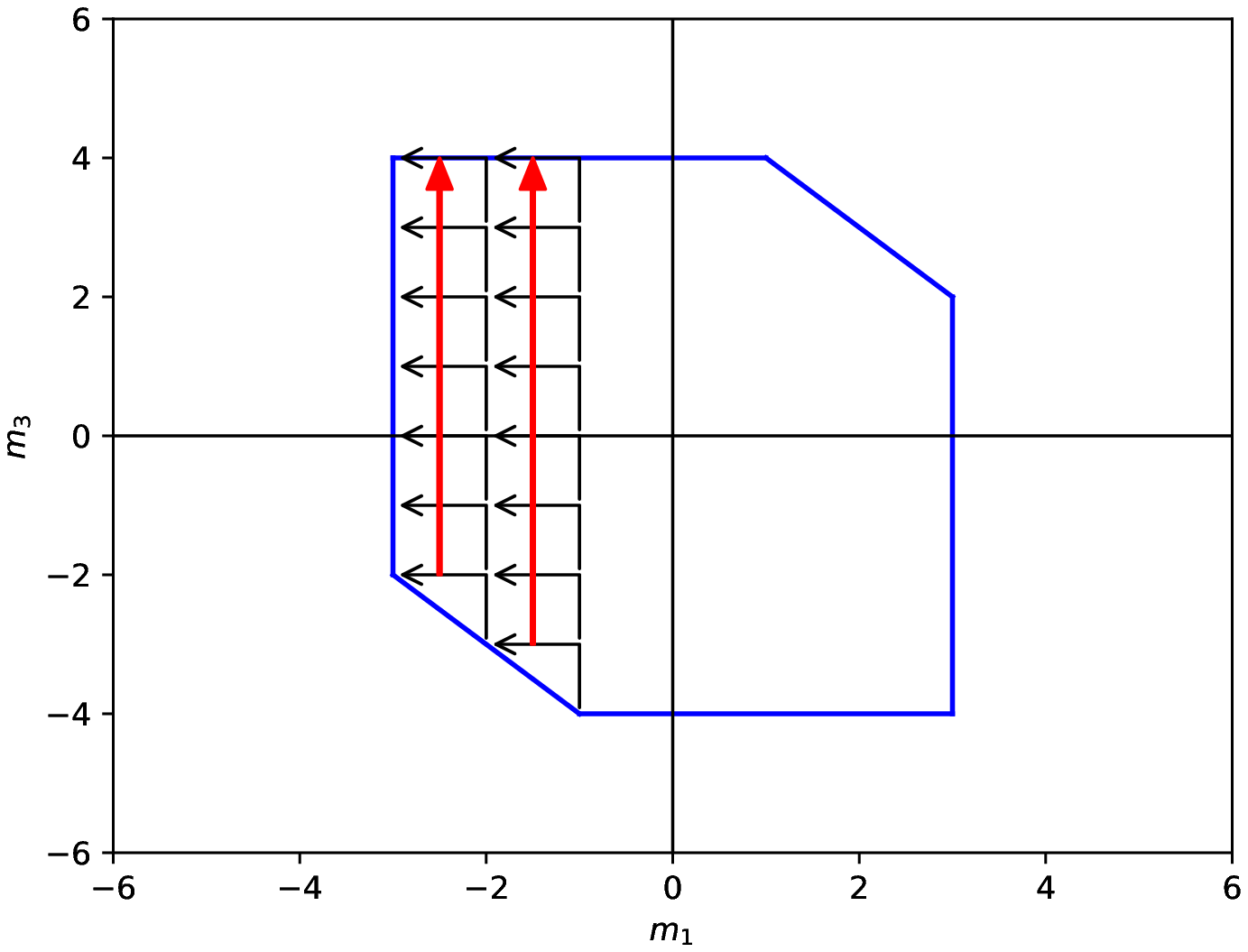}
    \caption{Apply the $J_x+iJ_y$ WI to the left part of the area}
    \label{reduction4}
  \end{minipage}
\end{figure}

Next, apply the $J_x-iJ_y$ WI to the right part of the area and apply the $J_x+iJ_y$ WI to the left part of the area Fig. \ref{reduction3} and Fig. \ref{reduction4}. Then we can find all the $F$s in the area.
\subsubsection{Solution}
Now that we know the strategy, all we have to do is solve the recursion relations. Applying the $J_x-iJ_y$ WI to the right boundary gives
\begin{equation}
F_{l_1,-l_3+k}^{l_1,l_2,l_3}=(-1)^k\sqrt{\frac{(l_2-l_1+l_3)!(l_2+l_1-l_3+k)!(2l_3-k)!}{(l_2-l_1+l_3-k)!(l_2+l_1-l_3)!(2l_3)!k!}}F_{l_1,-l_3}^{l_1,l_2,l_3}.
\end{equation}
And applying the $J_x+iJ_y$ WI to the bottom boundary gives
\begin{equation}
F_{l_1-n,-l_3}^{l_1,l_2,l_3}=(-1)^n\sqrt{\frac{(l_2+l_1-l_3)!(l_2-l_1+l_3+n)!(2l_1-n)!}{(l_2+l_1-l_3-n)!(l_2-l_1+l_3)!(2l_1)!n!}}F_{l_1,-l_3}^{l_1,l_2,l_3}.\label{f_{n,0}}
\end{equation}
Next, applying the $J_x+iJ_y$ WI to $m_1=l_1, m_3=-l_3+k$ gives
\begin{multline}
F_{l_1-1,-l_3+k}=\frac{(-1)^k}{\sqrt{2l_1}}\sqrt{\frac{(l_2-l_1+l_3)!(l_2+l_1-l_3+k-1)!(2l_3-k)!}{(l_2-l_1+l_3-k+1)!(l_2+l_1-l_3)!(2l_3)!k!}}\\
[2l_1k-(l_2+l_1-l_3)(l_2-l_1+l_3+1)]F_{l_1,-l_3}^{l_1,l_2,l_3}.
\end{multline}
And applying the $J_x+iJ_y$ WI to $m_1=l_1-1, m_3=-l_3+k$ gives
\begin{flalign}
&F_{l_1-2,-l_3+k}=\frac{(-1)^{k+1}}{2\sqrt{l_1(2l_1-1)}}\sqrt{\frac{(l_2-l_1+l_3)!(l_2+l_1-l_3+k-2)!(2l_3-k)!}{(l_2-l_1+l_3-k+2)!(l_2+l_1-l_3)!(2l_3)!k!}}[2l_1k(2l_3-k+1)-\notag\\
&\{2l_1k-(l_2+l_1-l_3)(l_2-l_1+l_3+1)\}\{2(l_1-1)k-(l_2+l_1-l_3-1)(l_2-l_1+l_3+2)\}]F_{l_1,-l_3}^{l_1,l_2,l_3}.
\end{flalign}
They induce us to write $F_{l_1-n,-l_3+k}^{l_1,l_2,l_3}$ as
\begin{equation}
F_{l_1-n,-l_3+k}^{l_1,l_2,l_3}=(-1)^k\sqrt{\frac{(2l_1-n)!(2l_3-k)!(l_2-l_1+l_3)!(l_2+l_1-l_3+k-n)!}{(2l_1)!n!(2l_3)!k!(l_2-l_1+l_3-k+n)!(l_2+l_1-l_3)!}}f_{n,k}F_{l_1,-l_3}^{l_1,l_2,l_3}.
\end{equation}
The $J_x\pm iJ_y$ recursion relations for this $f_{n,k}$ are
\begin{align}
f_{n+1,k}&=k(2l_3-k+1)f_{n,k-1}-(l_2+l_1-l_3-n+k)(l_2-l_1+l_3+n-k+1)f_{n,k}\\
f_{n,k+1}&=n(2l_1-n+1)f_{n-1,k}+f(n,k).
\end{align}
From its definition, $f_{0,k}=1$. The following results can be obtained from actual calculations using the recursion relations.
\begin{align}
f_{n,0}&=(-1)^n\frac{(l_2+l_1-l_3)!(l_2-l_1+l_3+n)!}{(l_2+l_1-l_3-n)!(l_2-l_1+l_3)!}\\
f_{n,1}&=(-1)^n\frac{(l_2+l_1-l_3+1)!(l_2-l_1+l_3+n-1)!}{(l_2+l_1-l_3-n+1)!(l_2-l_1+l_3-1)!}\notag\\
&\hspace{5mm}+(-1)^{n-1}\frac{(l_2+l_1-l_3)!(l_2-l_1+l_3+n-1)!}{(l_2+l_1-l_3-n+1)!(l_2-l_1+l_3)!}2nl_3\\
f_{n,2}&=(-1)^n\frac{(l_2+l_1-l_3+2)!(l_2-l_1+l_3+n-2)!}{(l_2+l_1-l_3-n+2)!(l_2-l_1+l_3-2)!}\notag\\
&\hspace{5mm}+(-1)^{n-1}\frac{(l_2+l_1-l_3+1)!(l_2-l_1+l_3+n-2)!}{(l_2+l_1-l_3-n+2)!(l_2-l_1+l_3-1)!}2n(2l_3-1)\notag\\
&\hspace{5mm}+(-1)^n\frac{(l_2+l_1-l_3)!(l_2-l_1+l_3+n-2)!}{(l_2+l_1-l_3-n+2)!(l_2-l_1+l_3)!}n(n-1)2l_3(2l_3-1)\\
f_{n,3}&=(-1)^n\frac{(l_2+l_1-l_3+3)!(l_2-l_1+l_3+n-3)!}{(l_2+l_1-l_3-n+3)!(l_2-l_1+l_3-3)!}\notag\\
&\hspace{5mm}+(-1)^{n-1}\frac{(l_2+l_1-l_3+2)!(l_2-l_1+l_3+n-3)!}{(l_2+l_1-l_3-n+3)!(l_2-l_1+l_3-2)!}3n(2l_3-2)\notag\\
&\hspace{5mm}+(-1)^n\frac{(l_2+l_1-l_3+1)!(l_2-l_1+l_3+n-3)!}{(l_2+l_1-l_3-n+3)!(l_2-l_1+l_3-1)!}3n(n-1)(2l_3-1)(2l_3-2)\notag\\
&\hspace{5mm}+(-1)^{n-1}\frac{(l_2+l_1-l_3)!(l_2-l_1+l_3+n-3)!}{(l_2+l_1-l_3-n+3)!(l_2-l_1+l_3)!}n(n-1)(n-2)2l_3(2l_3-1)(2l_3-2)\notag\\
\end{align}
From them, we can assume that $f_{n,k}$ has the following form.
\begin{align}
f_{n,k}&=\sum_{l=0}^kP_{n,k,l}(-1)^{n-l}\frac{(l_2+l_1-l_3+k-l)!(l_2-l_1+l_3+n-k)!}{(l_2+l_1-l_3-n+k)!(l_2-l_1+l_3-k+l)!}\label{assumption f_{n,k}}\\
P_{n,k,l}&={}_kC_l\frac{n!}{(n-l)!}\frac{(2l_3-k+l)!}{(2l_3-k)!}=\frac{k!n!(2l_3-k+l)!}{l!(k-l)!(n-l)!(2l_3-k)!}\label{assumption P_{n,k,l}}
\end{align}
We prove it by mathematical induction. First, the above configuration is consistent with the result for $(n,k)=(n,0)$ shown in (\ref{f_{n,0}}). The recursion relations of $P_{n,k,l}$ for $k$ are
\begin{align}
&P_{n,k+1,0}=P_{n,k,0}=1\\
&P_{n,k+1,1}=P_{n,k,1}-n(2l_1-2l_3-n+2k+1)P_{n,k,0}+n(2l_1-n+1)P_{n-1,k,0}\\
&P_{n,k+1,2}=P_{n,k,2}-(n-1)(2l_1-2l_3-n+2k)P_{n,k,1}+n(2l_1-n+1)P_{n-1,k,1}\\
&\cdots\notag\\
&P_{n,k+1,k}=P_{n,k,k}-(n-k)(2l_1-2l_3-n+k+2)P_{n,k,0}+n(2l_1-n+1)P_{n-1,k,k-1}\\
&P_{n,k+1,k+1}=-(n-k)(2l_1-2l_3-n+k+1)P_{n,k,k}+n(2l_1-n+1)P_{n-1,k,k}.
\end{align}
Assume that the above configuration (\ref{assumption f_{n,k}}) and (\ref{assumption P_{n,k,l}}) are correct for $k\geq0$. Then, from the above recursion relations, we get
\begin{align}
&P_{n,k+1,0}=1\\
&P_{n,k+1,m}=P_{n,k,m}-(n-m+1)(2l_1-2l_3-n+2k+2-m)P_{n,k,m-1}+n(2l_1-n+1)P_{n-1,k,m-1}\notag\\
&={}_kC_m\frac{n!}{(n-m)!}\frac{(2l_3-k+m)!}{(2l_3-k)!}+n{}_kC_{m-1}\frac{(n-1)!}{(n-m)!}\frac{(2l_3-k+m-1)!}{(2l_3-k)!}(2l_1-n+1)\notag\\
&\hspace{0.5cm}-(n-m+1){}_kC_{m-1}\frac{n!}{(n-m+1)!}\frac{(2l_3-k+m-1)!}{(2l_3-k)!}(2l_1-2l_3-n+2k+2-m)\notag\\
&={}_{k+1}C_m\frac{n!}{(n-m)!}\frac{(2l_3-k-1+m)!}{(2l_3-k-1)!}\hspace{1cm} (1\leq m\leq k)\\
&P_{n,k+1,k+1}\notag\\
&=-(n-k)(2l_1-2l_3-n+k+1)\frac{n!}{(n-k)!}\frac{(2l_3)!}{(2l_3-k)!}+n(2l_1-n+1)\frac{(n-1)!}{(n-k-1)!}\frac{(2l_3)!}{(2l_3-k)!}\notag\\
&=\frac{n!}{(n-k-1)!}\frac{(2l_3)!}{(2l_3-k-1)!}.
\end{align}
So, the configuration (\ref{assumption f_{n,k}}) and (\ref{assumption P_{n,k,l}}) are also correct for $k+1$.\par
The conclusion is
\begin{align}
  F_{l_1-n,-l_3+k}^{l_1,l_2,l_3}&=(-1)^{n+k}\sqrt{\frac{(2l_1-n)!(2l_3-k)!(l_2-l_1+l_3)!(l_2+l_1-l_3+k-n)!}{(2l_1)!n!(2l_3)!k!(l_2-l_1+l_3-k+n)!(l_2+l_1-l_3)!}}f_{n,k}F_{l_1,-l_3}^{l_1,l_2,l_3}\\
  f_{n,k}&=\sum_{l=0}^{k}P_{n,k,l}(-1)^{l}\frac{(l_2+l_1-l_3+k-l)!(l_2-l_1+l_3+n-k)!}{(l_2+l_1-l_3-n+k)!(l_2-l_1+l_3-k+l)!}\\
  P_{n,k,l}&=\frac{k!n!(2l_3-k+l)!}{l!(k-l)!(n-l)!(2l_3-k)!}.
\end{align}
\subsection{Direct integral calculation of three-point function}
In two-dimensional CFT, the direct integral calculation is straightforward because we can decompose it into holomorphic and antiholomorphic parts.
We can calculate them independently by a complex integral technique.
On the other hand, in three-dimensional CFT, we don't have such techniques, so we have to perform direct integral calculations straightforwardly. From the definition,
\begin{align}
\int\frac{\dd\Omega_1}{4\pi}\int\frac{\dd\Omega_2}{4\pi}\int\frac{\dd\Omega_3}{4\pi}&\langle\OO_3\OO_2\OO_1\rangle=\lambda_{321}r_1^{-\Delta_1}r_2^{-\Delta_2}r_3^{-\Delta_3}\sum_{p,q\in\mathbb{Z}_{\geq0}}F_{(p,q)}^{000}y_1^{\Delta_1+2p}y_3^{-\Delta_3-2q}\notag\\
&=\int\frac{\dd\Omega_1}{4\pi}\int\frac{\dd\Omega_2}{4\pi}\int\frac{\dd\Omega_3}{4\pi}\frac{\lambda_{321}}{x_{12}^{\Delta_1+\Delta_2-\Delta_3}x_{23}^{\Delta_2+\Delta_3-\Delta_1}x_{31}^{\Delta_3+\Delta_1-\Delta_2}},
\end{align}
where $x_{ij}^2=r_i^2+r_j^2-2r_ir_j\Phi_{ij}$. We can obtain $F_{(p,q)}^{000}$ by differentiating it.
\begin{align}
  F_{(p,q)}^{000}=&\int\frac{\dd\Omega_1}{4\pi}\int\frac{\dd\Omega_2}{4\pi}\int\frac{\dd\Omega_3}{4\pi}\frac{1}{(2p)!}\frac{1}{(2q)!}\left(\dv{z_1}\right)^{2p}\left(\dv{z_3}\right)^{2q}\notag\\
  &(1+z_1^2-2z_1\Phi_{12})^{g_{12}}(1+z_3^2-2z_3\Phi_{23})^{g_{23}}(1+(z_1z_3)^2-2z_1z_3\Phi_{31})^{g_{31}}|_{z_1=z_3=0}
\end{align}
To calculate it, we need the following relation.
\begin{equation}
\left.\frac{1}{(2q)!}\left(\dv{z}\right)^{2q}(1+z^2-2z\Phi)^g\right|_{z=0}=\sum_{k=0}^q\frac{2^{2k}}{(q-k)!(2k)!}\frac{g!}{(g-q-k)!}\Phi^{2k}
\end{equation}
We omit its derivation here. With this, calculate the amplitude $F_{(p,q)}^{000}$. First, consider the differentiation by $z_3$.
\begin{align}
  &\left(\dv{z_3}\right)^{2q}(1+z_3^2-2z_3\Phi_{23})^{g_{23}}(1+(z_1z_3)^2-2z_1z_3\Phi_{31})^{g_{31}}\notag\\
  =&\sum_{k=0}^q\sum_{j=0}^k\sum_{l=0}^{q-k}\frac{2^{2(j+l)}(2q)!}{(k-j)!(2j)!(q-k-l)!(2l)!}\frac{g_{31}!g_{23}!\Phi_{31}^{2j}\Phi_{23}^{2l}}{(g_{31}-k-j)!(g_{23}-q+k-l)!}z_1^{2k}\notag\\
  +&\sum_{k=0}^{q-1}\sum_{j=0}^k\sum_{l=0}^{q-k-1}\frac{2^{2(j+l+1)}(2q)!}{(k-j)!(2j+1)!(q-k-l-1)!(2l+1)!}\frac{g_{31}!g_{23}!\Phi_{31}^{2j+1}\Phi_{23}^{2l+1}}{(g_{31}-k-j-1)!(g_{23}-q+k-l)!}z_1^{2k+1}
\end{align}
Next, consider the differentiation by $z_1$.
\begin{equation}
  \left.\left(\dv{z_1}\right)^{2p}(1+z_1^2-2z_1\Phi_{12})^{g_{12}}z_1^{2k}\right|_{z_1=0}
\begin{cases}
0 & (p < k)\\
\sum_{n=0}^{p-k}\frac{2^{2n}(2p)!}{(p-k-n)!(2n)!}\frac{g_{12}!\Phi_{12}^{2n}}{(g_{12}-p+k-n)!} & (p \geq k)
\end{cases}
\end{equation}
\begin{equation}
  \left.\left(\dv{z_1}\right)^{2p}(1+z_1^2-2z_1\Phi_{12})^{g_{12}}z_1^{2k+1}\right|_{z_1=0}=
\begin{cases}
0 & (p \leq k)\\
\sum_{n=0}^{p-k-1}\frac{-2^{2n+1}(2p)!}{(p-k-n-1)!(2n+1)!}\frac{g_{12}!\Phi_{12}^{2n+1}}{(g_{12}-p+k-n)!} & (p > k)
\end{cases}
\end{equation}
With them, we get
\begin{align}
  F_{(p,q)}^{000}=&\sum_{k=0}^{\min(p,q)}\sum_{j=0}^{k}\sum_{n=0}^{p-k}\sum_{l=0}^{q-k}A_{k,j,n,l}\int\frac{\dd\Omega_1}{4\pi}\int\frac{\dd\Omega_2}{4\pi}\int\frac{\dd\Omega_3}{4\pi}\Phi_{12}^{2n}\Phi_{23}^{2l}\Phi_{31}^{2j}\notag\\
  &-\sum_{k=0}^{\min(p-1,q-1)}\sum_{j=0}^{k}\sum_{n=0}^{p-k-1}\sum_{l=0}^{q-k-1}B_{k,j,n,l}\int\frac{\dd\Omega_1}{4\pi}\int\frac{\dd\Omega_2}{4\pi}\int\frac{\dd\Omega_3}{4\pi}\Phi_{12}^{2n+1}\Phi_{23}^{2l+1}\Phi_{31}^{2j+1},
\end{align}
where
\begin{align}
  A_{k,j,n,l}=&\tfrac{2^{2(j+n+l)}}{(2j)!(2n)!(2l)!(k-j)!(p-k-n)!(q-k-l)!}\tfrac{g_{12}!g_{23}!g_{31}!}{(g_{31}-k-j)!(g_{12}-p+k-n)!(g_{23}-q+k-l)!}\\
  B_{k,j,n,l}=&\tfrac{2^{2(j+n+l)+3}}{(2j+1)!(2n+1)!(2l+1)!(k-j)!(p-k-n-1)!(q-k-l-1)!}\tfrac{g_{12}!g_{23}!g_{31}!}{(g_{31}-k-j-1)!(g_{12}-p+k-n)!(g_{23}-q+k-l)!}.
\end{align}

\newpage
\bibliographystyle{alphaurl}
\bibliography{biblio.bib}

\end{document}